\documentclass[twocolumn,trackchanges]{aastex701}

\newcommand{\OII}{{[O\,II]\,}}

\newcommand{\ho}{\textsc{Hom}e\,\,}
\graphicspath{{./}{figures/}}
\usepackage{amsmath}
\usepackage{tikz}
\usepackage{array}
\usepackage{graphicx}
\usepackage{multirow}
\usepackage{natbib}

\shorttitle{ELG$\times$LRG distribution through DM halo dynamics}
\shortauthors{Favole et al. (2026)}

\received{2025 December 3}
\revised{2026 March 19}
\accepted{2026 March 20}
\published{2026 April 28}

\begin{document}

\title{ELG$\times$LRG distribution through dark matter halo dynamics}

\correspondingauthor{Ginevra Favole}\email{gifavole@ull.edu.es}

\author[0000-0002-8218-563X,sname='Ginevra Favole']{Ginevra Favole}
\affil{Departamento de Astrof\'{\i}sica, Universidad de La Laguna, E-38206, La Laguna, Tenerife, Spain}
\affiliation{Instituto de Astrof\'{\i}sica de Canarias, c/ V\'ia L\'actea s/n, E-38205, La Laguna, Tenerife, Spain}
\email{gifavole@ull.edu.es}  

\author[0000-0002-9994-759X,sname='Francisco-Shu Kitaura']{Francisco-Shu Kitaura} 
\affiliation{Departamento de Astrof\'{\i}sica, Universidad de La Laguna, E-38206, La Laguna, Tenerife, Spain}
\affiliation{Instituto de Astrof\'{\i}sica de Canarias, c/ V\'ia L\'actea s/n, E-38205, La Laguna, Tenerife, Spain}
\email{fkitaura@ull.edu.es} 

\author[0000-0002-2312-3121,sname='Boryana Hadzhiyska']{Boryana Hadzhiyska}
\affiliation{Institute of Astronomy, University of Cambridge, Madingley Road, Cambridge CB3 0HA, UK}
\email{boryanah@berkeley.edu}

\author[0000-0002-2929-3121,sname='Daniel Eisenstein']{Daniel Eisenstein}
\affiliation{Center for Astrophysics, Harvard \& Smithsonian, 60 Garden Street, Cambridge, MA 02138, USA}
\email{deisenstein@cfa.harvard.edu}

\author[0000-0002-9853-5673,sname='Lehman Garrison']{Lehman H. Garrison}
\affiliation{Scientific Computing Core, Flatiron Institute, 162 Fifth Avenue, New York, NY 10010, USA}
\email{lgarrison@flatironinstitute.org}

\author[0000-0002-0974-5266,sname='Sownak Bose']{Sownak Bose}
\affiliation{Institute for Computational Cosmology, Department of Physics, Durham University, South Road, Durham, DH1 3LE, UK}
\email{sownak.bose@durham.ac.uk}

\begin{abstract}
We investigate the clustering and halo occupation distribution (HOD) of DESI Y1 emission-line (ELGs) and luminous red (LRGs) galaxies at $0.8<z<1.1$, including their cross-correlation (ELG$\times$LRG), using the \textsc{AbacusSummit} suite and a new Halo Occupation Model (\textsc{Hom}e) for galaxy multi-tracers. This integrates intra-halo dynamics, halo exclusion, and quenching, bridging insights from hydrodynamical, HOD, abundance-matching, and semi-analytic studies. Leveraging full phase-space information from the \textsc{Uchuu} $N$-body simulation, and sampling satellites from dark-matter particle positions via physically motivated prescriptions, 
\textsc{Hom}e reproduces the anisotropic clustering down to $s=200\,h^{-1}$kpc with unprecedented accuracy. Model parameters are inferred solely from two-point statistics using a two-level Bayesian framework, yielding high-fidelity ELG, LRG and cross-reference catalogs. We find that satellite ELGs behave as incoherent flows within their parent halos, dominating the clustering below $4\,h^{-1}$Mpc. The HOD from the best-fit \ho has the following properties: (i) 90.50\% (85.91\%) of ELGs (LRGs) are central galaxies without satellites, residing in halos of $M_{\rm vir}\sim6.6\times10^{11}\,(1.2\times10^{13})\,h^{-1}{\rm M}_\odot$; (ii) the ELG$\times$LRG cross-correlation is governed by central-central pairs and shaped by halo exclusion on $2-5\,h^{-1}$Mpc scales; (iii) 9.50\% (14.09\%) of ELGs (LRGs) are satellites, of which 1.09\% (3.52\%) inhabit halos with a central galaxy of the same species in a maximally conformal configuration, 7.02\% (0.005\%) orbit complementary hosts in a minimally conformal state, and 0.58\% (10.57\%) are orphans. 
The high sensitivity of \ho precisely captures the dynamics of satellites in different host environments, opening a promising avenue for understanding systematics, the dynamical nature of dark matter, potentially distinguishing gravity models.
\end{abstract}


\section{Introduction} \label{sec:intro}
A complete and physically motivated galaxy-halo connection scheme is crucial for building high-fidelity reference catalogs for mock calibration to achieve accurate covariance estimates for cosmological analyses. However, establishing a universal prescription is challenging due to the diversity of galaxy populations targeted by new-generation surveys, as the Dark Energy Spectroscopic Instrument \citep[DESI;][]{2013arXiv1308.0847L, 2016arXiv161100037D, 2022AJ....164..207D} or the {\textit{Euclid}} space mission \citep{2011arXiv1110.3193L}. These surveys observe a variety of galaxy types, each with different compositions, star formation histories, and physical properties. All these sources are complementary biased tracers of the same underlying dark matter field.

To capitalise on ongoing and upcoming observations, it is essential to generalise the standard galaxy-halo connection techniques, i.e. the sub-halo abundance matching \citep[SHAM;][]{2006ApJ...647..201C,2010ApJ...717..379B,2011ApJ...742...16T} and the halo occupation distribution \citep[HOD;][]{2002ApJ...575..587B,2005ApJ...630....1Z,2005ApJ...633..791Z,2007ApJ...667..760Z} models, to different surveys and galaxy tracers. This involves incorporating the effects of target selection, different bias schemes, intra-halo dynamics, halo exclusion and quenching in our clustering models.
 
Significant progress has been made in recent years with HOD \citep[e.g.,][]{2016MNRAS.462.2218F,2016MNRAS.459.3040G,2018MNRAS.474.4024G,2020MNRAS.499.5486A,2020MNRAS.497..581A,2021MNRAS.502.3599H,2021MNRAS.501.1603H,2022MNRAS.510.3301Y, 2023JCAP...10..016R,2025MNRAS.538.1216Y} and SHAM \citep[e.g.,][]{2016MNRAS.461.3421F,2017MNRAS.472..550F,2016MNRAS.460.1173R,2016MNRAS.460.3100C,2021MNRAS.504.5205C,2022MNRAS.516...57Y,2022MNRAS.509.1614F,2023MNRAS.520..489C,2024ApJ...961...74G, 2024A&A...689A..66O} models capable of reproducing, with different levels of precision, the clustering signals of galaxy multi-tracers and their evolution with redshift. 
However, these models still lack a comprehensive physical interpretation of the ELG configuration and its connection with LRGs, particularly in understanding the impact of peculiar motions and intra-halo dynamics on the small-scale anisotropic clustering and quenching mechanisms.

In this work we present a novel physically motivated Halo Occupation Model (``\textsc{Hom}e'', hereafter) to produce high-fidelity reference catalogs for DESI Y1 ELG, LRG and ELG$\times$LRG tracers based on the \textsc{AbacusSummit}\footnote{\url{https://abacussummit.readthedocs.io/en/latest/abacus.html}} simulation suite. This new approach includes a dynamical treatment of satellite halos---not tracked by the halo finder in the simulation---via dark-matter (DM) particle positions, and is \emph{applicable to any galaxy survey, galaxy tracer, and simulation}. 

Note that a key strength of \ho is its generality, as it can be applied to simulations that do not resolve (or output) substructures--such as \textsc{AbacusSummit}--as well as to those that do, such as \textsc{Uchuu} \citep{2021MNRAS.506.4210I}.

Our method reconciles different perspectives from current SHAM, HOD, hydrodynamical and semi-analytic model (SAM) studies about ELGs and LRGs within their host halos \citep[e.g.,][]{2016MNRAS.460.3100C,2021MNRAS.501.1603H,2021MNRAS.502.3599H,2022MNRAS.509.1614F,2018MNRAS.475.2530O}, and extends previous SHAM results \citep{2022MNRAS.516...57Y} down to $200\,h^{-1}$kpc scales, with $\sim20$ times better resolution, dramatically improving the accuracy of current clustering models \citep[e.g.,][]{2025A&A...698A.170P} on sub-Mpc scales. 

Our results demonstrate that the fraction and velocity bias of satellites with respect to their hosts, together with the exclusion mechanism happening between pairs of massive halos, and the quenching of ELG satellites in LRG hosts, are pivotal to accurately predict the anisotropic clustering of both tracers below $4\,h^{-1}$Mpc. 
The fine interplay between these model ingredients determines \textsc{Hom}e's unprecedented sensitivity, which opens new paths for understanding systematics, constraining the dynamical nature of DM, and breaking degeneracies between galaxy bias and cosmology.

We present \ho within a two-level Bayesian inference framework that allows us to constrain the physically meaningful parameters---those shaping the clustering and HOD of DESI galaxy multi-tracers---marginalizing over a set of nuisance parameters that drive analytic prescriptions used to build the latent properties of the halo input catalogs for \textsc{Hom}e. 

To guarantee the latent variables are properly informed by precise phase-space information and DM distribution, we calibrate the analytic prescriptions against independent predictions from external high-resolution $N$-body simulations. For our analysis we adopt \textsc{Uchuu}, that has the same volume of \textsc{AbacusSummit}.

In this way, the inference process is robust and computationally optimized at the same time, as we do not need to regenerate the full model at each point of the parameter space, which would be unfeasible.

The aim of this work is threefold: (i) presenting the \ho scheme, including full satellite treatment via \textsc{AbacusSummit} DM particles, as well as the two-level Bayesian inference framework; (ii) apply it to DESI Y1 observations to accurately predict the two-point ELG, LRG and ELG$\times$LRG correlation functions, and their halo occupation distribution down to $s=200\,h^{-1}$kpc; (iii) interpreting the \ho results in terms of intra-halo dynamics to shed light on the physics of peculiar motions, redshift-space distortions, halo exclusion and quenching, reconciling them with the current picture from SHAM, HOD, hydrodynamical and SAM studies.

In our analysis we adopt the \textsc{AbacusSummit} $\Lambda$CDM fiducial cosmology, which is consistent with \citet{2020A&A...641A...6P}: $\Omega_m=0.31519$, $h=0.6736$, $n_{\rm s}=0.9649$, $\sigma_8=0.807952$.

\section{Data} \label{sec:data}
\subsection{DESI Y1 observations} \label{sec:DESIdata}
We use the LSS catalog \citep{2025JCAP...01..125R,2025JCAP...07..017A} part of the DESI Y1 Data Release \citep{2025arXiv250314745D}, which includes observations collected between May 2021 and June 2022, following a survey validation phase \citep{2024MNRAS.534.3540L}.

DESI employs robotic fiber positioners to simultaneously capture spectra for 5,000 celestial targets organized in tiles, routing light through ten spectrographs. The survey observing time is classified into ``bright" and ``dark" programs based on conditions; both ELGs \citep{2023AJ....165..126R} and LRGs \citep{2023AJ....165...58Z} are targeted in dark time \citep[see][]{2025arXiv250314745D}. 

The DESI Y1 ELG sample comprises 243,2022 good redshifts in $0.8< z<1.6$, covering $5,914$ deg$^2$ \citep{2025JCAP...07..017A}. ELGs are the DESI targets with the lowest priority: their fiber assignment completeness is 35.5\% in Y1, then growing to 60\% in the final dataset of 14,000 deg$^2$. 
The LRG sample includes 2,138,600 good redshifts in $0.4 < z < 1.1$, over 5,840 deg$^2$, with 69.2\% fiber assignment completeness in Y1, then increasing to 90\% in the final release.

The ELG and LRG samples \citep{2025JCAP...07..017A} span different redshift ranges, with significant overlap only at $0.8<z<1.1$. We adopt this shared redshift window as the fiducial region where modeling both their auto- and cross-correlations using \textsc{Hom}e. 

Employing the same nomenclature as in \citet{2025JCAP...07..017A}, we analyze the clustering of the \texttt{ELG1} and \texttt{LRG3} samples, as well as their cross-correlation \texttt{ELG1}$\times$\texttt{LRG3}, at $0.8<z<1.1$. The number and density of targets in each sample is reported in Table\,\ref{tab:zrep}, and
we assume $z_{\rm eff}=0.95$ as the most representative redshift of the above range \citep[see][]{2025JCAP...04..012A}. 

\begin{deluxetable*}{lccccccc}
\tablecaption{Observed number ($N$) and number density ($n$) at the effective redshift ($z_{\rm eff}$) of the DESI Y1 \textsc{ELG1} and \textsc{LRG3} samples, as given in \citet{2025JCAP...04..012A}. We also show the redshift error ($z_{\rm err}$) affecting our velocity estimates, the \textsc{AbacusSummit} snapshot ($z_{\rm A}$) used for modeling, and the \textsc{Uchuu} snapshot ($z_{\rm U}$) chosen for calibration. \label{tab:zrep}}
\tablehead{
\colhead{DESI Y1 sample} & \colhead{observed $z$ range}& \colhead{$z_{\rm eff}$} & \colhead{$z_{\rm A}$}& \colhead{$z_{\rm err}\,$}&\colhead{$N(z_{\rm eff})$} &\colhead{$n(z_{\rm eff})\times10^{4}$} &$z_{\rm U}$\\
&&&&&&$[h^3\,{\rm Mpc^{-3}}]$
}
\startdata
{\texttt{ELG1}}&$0.8<z<1.1$&0.95& 0.8&-0.06&99575&$6.56$&0.78\\
{\texttt{LRG3}}&$0.8<z<1.1$& 0.95 & 0.8&-0.06&48491&$3.40$&0.78 \\
\enddata
\end{deluxetable*}

\subsection{The \textsc{AbacusSummit} simulation} \label{sec:simulation}
We analyze DESI Y1 observations (\S\,\ref{sec:DESIdata}) using the \textsc{AbacusSummit}\footnote{\url{https://abacussummit.readthedocs.io/en/latest/}} suite of 139 large-volume ($L_{\rm box}=2h^{-1}{\rm Gpc}$), high-resolution ($2\times10^9h^{-1}{\rm M_\odot}$ particle mass) ``base" $N$-body cosmological simulations \citep{2021MNRAS.508.4017M,2021MNRAS.508..575G}. We use the \textsc{CompaSO} halo catalogs \citep{2022MNRAS.509..501H} that include only central host halos (``h", hereafter). 

Here, the halo central core is robustly identified using a L2 halo finding algorithm, rather than simply defining quantities relative to the centre of mass of each \textsc{CompaSO} object. To circumvent the lack of satellites, which are not tracked by the halo finder, we assign them based on DM particle and host positions, as described in \S\,\ref{sec:satellites}.

To ensure reliable halo properties, we consider only halos with at least $N_{\rm p}\ge65$ DM particles, corresponding to a lower halo mass limit of $1.3\times 10^{11}h^{-1}{\rm M}_\odot$. This minimal cut returns a population of nearly $225$ million central halos---all potential hosts for satellites---including about $1.5\times10^{10}$ DM particles. 

We implement \ho on the closest \textsc{AbacusSummit} snapshot to the effective redshift of the DESI tracers to model (see \S\,\ref{sec:data}); this is $z_{\rm A}=0.8$, as shown in Table\,\ref{tab:zrep}. 
The impact of redshift errors in estimating halo velocities can be quantified as the ratio of the growth factors evaluated at the redshifts of interest: $z_{\rm err}=1-G(z_{\rm eff})/G(z_{\rm A})$. In our case the impact is minimal, as all deviations are within 6\% (see Table\,\ref{tab:zrep}). 

In addition to the $2\,h^{-1}$Gpc \textsc{AbacusSummit} base boxes, we also adopt 1800 $500\,h^{-1}$Mpc boxes (``small", hereafter) at the base mass resolution to estimate our model covariances.   

\section{Methodology} \label{sec:methodology}
In what follows, we describe the ingredients and implementation of our Halo Occupation Model (\textsc{Hom}e) for the DESI Y1 \texttt{ELG1}, \texttt{LRG3}, and \texttt{ELG1}$\times$\texttt{LRG3} samples, in the fiducial redshift range $0.8<z<1.1$, using the \textsc{AbacusSummit} simulation snapshot $z_{\rm A}=0.8$.

\subsection{Central halos} \label{sec:centrals}
We characterize central halos (``cen", hereafter) in \ho by extracting and assigning them the relevant properties of their \textsc{AbacusSummit} central hosts (\S\,\ref{sec:simulation}) at the snapshot of interest, $z_{\rm A}$ (Table\,\ref{tab:zrep}). These are: 
\begin{itemize}
\item \emph{Maximum circular velocity}: in the \textsc{AbacusSummit} notation, this is called {\texttt{vcirc\_max\_L2com}} and is computed relative to the center of mass position and velocity, based on the particles in the L1 CompaSO halo\footnote{In \textsc{AbacusSummit}, L0 groups are large sets of DM particles typically encompassing several L1 groups. These correspond to classical halos, while L2 groups identify with ``halo cores" or ``subhalos". For further details, see \url{https://abacussummit.readthedocs.io/en/latest/compaso.html}}. We assume $V_{\rm max}^{\rm cen}\equiv V_{\rm max}^{\rm h}$. 

\item \emph{Peak circular velocity}: it corresponds to the peak $V_{\rm max}$ value of the halo along its main progenitor branch. We compute it by matching the \texttt{HaloIndex} of each halo at the $z_{\rm A}$ of interest with the \texttt{HaloIndex\_mainprog} of all its progenitors from previous snapshots, both primary and seconday ones\footnote{\url{https://abacussummit.readthedocs.io/en/latest/data-products.html}}. We then assume $V_{\rm peak}^{\rm cen}\equiv V_{\rm peak}^{\rm h}$.

\item \emph{Cartesian positions}: we assume the 3-d spatial coordinates of the central halo center--\texttt{x\_L2com[:,i=0,1,2]}  in the \textsc{AbacusSummit} notation--to coincide with the center of mass of the largest L2 subhalo, meaning that: 
    \begin{equation}
    \vec{r}_{\rm cen}\equiv\vec{r}_{\rm h}=(x_{\rm h},\,y_{\rm h},\,z_{\rm h})\,.
    \label{eq:vzcen}
    \end{equation}
    
    \item \emph{Velocity dispersion}: we adopt the 3-d velocity dispersion ($\sigma_{\rm vd}$) of the inner 50\% of particles as a proxy for the entire halo. In \textsc{AbacusSummit}, this is called \texttt{sigmav3d\_r50\_L2com} and is estimated as the square root of the sum of eigenvalues of the second moment tensor of the velocities relative to the center of mass.  
 
    \item \emph{Peculiar velocity}: assuming that the line-of-sight (LOS) is parallel to the $\vec{z}$ direction, we adopt the center of mass LOS velocity of the largest L2 subhalo, which is \texttt{v\_L2com[:,2]} in the \textsc{AbacusSummit} notation. In this way, the peculiar velocity contribution to the central halo position in redshift space comes all from that of its host, $v_z^{\rm h}$. 
    
    We modulate it further using a constant velocity bias parameter, $b_{\rm cen}$ \citep[see][]{2023JCAP...10..016R}:  
    \begin{equation}
v_z^{\rm cen} = v_z^{\rm h} + b_{\rm cen}\,\sigma_{\rm vd}\,\mathcal{N}(0,1)\,,
\label{eq:vpeccen}
\end{equation}
where $\mathcal{N}(0,1)$ is a normal random realization with mean 0 and variance 1, and $\sigma_{\rm vd}$ is the halo velocity dispersion (see above). 

    \item \emph{Virial mass}: we calculate the host halo virial mass by multiplying the individual particle mass, \texttt{ParticleMassHMsun}, available in the header of the \textsc{AbacusSummit} catalogs, by the number \texttt{N} of its particles. We assign $M^{\rm cen}_{\rm vir}\equiv M^{\rm h}_{\rm vir}$.
    
    \item \emph{Virial radius}: we assign central halos their host virial radius, $R^{\rm cen}_{\rm vir}\equiv R^{\rm h}_{\rm vir}$. As a proxy for this quantity, we follow \citet{2023JCAP...10..016R} and adopt the radius enclosing 98\% of the halo mass relative to the L2 center, denoted \texttt{r98\_L2com} in the \textsc{AbacusSummit} catalog.

    \item \emph{Type}: we flag all central halos with {\texttt{type$\,=\,$0}} to distinguish them from satellites, which have  {\texttt{type$\,=\,$1}} (see \S\,\ref{sec:positions}).
\end{itemize}

\subsection{Satellite halos} \label{sec:satellites}
Since the \textsc{AbacusSummit} halo finder does not track satellite properties, these must be generated externally. This step is essential for \ho, which relies on abundance matching (AM) and therefore requires accurate satellite information. 

Our choice of using \textsc{AbacusSummit}, which does not output substructure information, reflects our goal of developing a fully general framework for the galaxy–halo connection. While simulations such as \textsc{Uchuu} provide resolved subhalo catalogs that could be directly used to model satellite galaxies, relying on them would restrict the applicability of the method to simulations with similar capabilities. By constructing satellite properties externally, \ho remains applicable to a broad class of cosmological simulations, including those that do not track substructure.

We construct satellites by combining the properties of the \textsc{AbacusSummit} DM particles and central host halos. Specifically, we build and assign each satellite the following key quantities for \textsc{Hom}e: $V_{\rm peak}^{\rm sat}$, $x_{\rm sat}$, $y_{\rm sat}$, $z_{\rm sat}$, $v_z^{\rm sat}$, $M^{\rm h}_{\rm vir}$, $R^{\rm h}_{\rm vir}$, {\texttt{type$\,=\,$1}}. Here, the superscripts ``sat" (``h") indicates that the property refers to the satellite (parent host) halo. 

Unless explicitly mentioned, in what follows we treat central and satellite $V_{\rm peak}$ values in the same manner, with no distinction, and we use the \texttt{type} flag whenever we need to discriminate them.

To assign satellite positions, we randomly select 30\% of the DM particles available in \emph{all} \textsc{AbacusSummit} host halos (see \S\,\ref{sec:simulation}), which provide about $4.5\times10^9$ potential locations. This large reservoir of satellites has an impact on the conformity level and orphan satellite predictions of our model; see discussion in \S\,\ref{sec:hodpred}.

By combining these particle properties with their hosts' ones, and applying physically motivated prescriptions that we calibrate against predictions from the \textsc{Uchuu} $N$-body simulation, we generate the satellite population required for \textsc{Hom}e. In the following, we describe how each satellite property is constructed.

\subsubsection{Peak circular velocities} \label{sec:Vmaxsampling}
Implementing the abundance matching scheme in \ho requires rank-ordering both the \textsc{AbacusSummit} halos and the DESI galaxy tracers by key properties. For halos, we adopt their peak circular velocity $V_{\rm peak}$ (\S\,\ref{sec:centrals}), as it has been proven to correlate tightly with galaxy luminosity and stellar mass, representing an accurate proxy for AM in general \citep[see e.g.,][]{2013ApJ...771...30R, 2016MNRAS.460.3100C}.

Since \textsc{AbacusSummit} does not provide satellite $V_{\rm peak}$ values, we sample them \emph{conditional} to their host peak circular velocities using a cumulative distribution function (CDF) that we build in six steps:
\begin{enumerate}
\item We fit the \textsc{AbacusSummit} host abundance, in bins of $V_{\rm peak}$, at the $z_{\rm A}$ of interest using the analytic formula by \citet{2011ApJ...740..102K}:
\begin{equation}
n(V_{\rm peak}, z)=A\,V_{\rm peak}^{-3}\exp{\left(-\left[\frac{V_{\rm peak}}{V_0}\right]^\gamma\right)}\,,
\label{eq:klypincen}
\end{equation}
which depends on 3 parameters and, implicitly, on redshift.
We obtain $n^{\rm A}_{\rm h}(V_{\mathrm{peak}}, z_{\rm A})$, whose optimal parameters $(A, V_0, \gamma)^{\rm A}_{\rm h}$ are given in Table \ref{tab:klypinpar}.
We explore their posterior distribution using a Monte Carlo Markov Chain (MCMC) coupled with an emcee sampler \citep{2013PASP..125..306F}; the results are discussed in \S\,\ref{sec:levelIresu}. 

\item Using Eq.\,\ref{eq:klypincen}, we fit the satellite and central halo abundances predicted by the \textsc{Uchuu} $N$-body simulation \citep{2021MNRAS.506.4210I}. This has $12800^3$ particles, as well as satellite properties, in the same volume as \textsc{AbacusSummit} ($L_{\rm box}=2\,h^{-1}$Gpc), resulting in higher mass resolution, i.e. $3.28\times10^8\,h^{-1}{\rm M}_\odot$\footnote{\url{https://www.skiesanduniverses.org/Simulations/Uchuu/}}. We adopt the closest available \textsc{Uchuu} snapshot ($z_{\rm U}=0.78$, hereafter) to the observed effective redshift. The \textsc{Uchuu} abundance fits and results, with optimal parameters $(A, V_0, \gamma)^{\rm U}_{\rm cen}$ and $(A, V_0, \gamma)^{\rm U}_{\rm sat}$, are shown in \S\,\ref{sec:levelIresu}.

\item From the central-to-satellite ratio of the \textsc{Uchuu} abundance fits, we calculate the correction factor:
\begin{equation}
    R(V_{\mathrm{peak}}, z_{\rm U}) = \frac{n_{\rm cen}^{\rm U}(V_{\mathrm{peak}}, z_{\rm U})}{n_{\rm sat}^{\rm U}(V_{\mathrm{peak}}, z_{\rm U})}\,.
    \label{eq:correction}
\end{equation}

\item We define the probability density function (PDF) to sample satellite $V_{\rm peak}$ values in \textsc{AbacusSummit} as that of their hosts, obtained from step 1 above, rescaled by the correction factor just computed:
\begin{equation}
    n^{\rm A}_{\rm sat}(V_{\mathrm{peak}}, z_{\rm A}) =\frac{ n^{\rm A}_{\rm h}(V_{\mathrm{peak}}, z_{\rm A}) }{\,R(V_{\mathrm{peak}}, z_{\rm U})}\,.
    \label{eq:pdfsats}
\end{equation}

\item By integrating Eq.\,\ref{eq:pdfsats} and normalizing it to unity, we obtain the satellite $V_{\rm peak}$ CDF, conditional to their hosts:
\begin{equation}
    \Psi_{\rm sat}(V_{\mathrm{peak}}, z_{\rm A}) = \frac{\int_{V_{\rm low}}^{V_{\mathrm{peak}}} n^{\rm A}_{\rm sat}(V', z_{\rm A}) \,dV'}{\int_{V_{\mathrm{low}}}^{V_{\rm up}} n^{\rm A}_{\rm sat}(V', z_{\rm A}) \,dV'}\,.
    \label{eq:cdfvmaxsat}
\end{equation}
Here, \( V_{\mathrm{low}}\equiv 0.2\,V_{\rm peak}^{\rm h} \) and \( V_{\rm up}\equiv V_{\rm peak}^{\rm h} \) are the lower and upper velocity limits set by the hosts. We choose the lower limit to be 20\% the host velocity to have an homogeneous $V_{\rm peak}$ distribution, when combining central and satellite halos. Lower than that, satellites are segregated to extremely low velocity values, which are unphysical.

\item 
Finally, we inverse transform sampling the satellite $V_{\rm peak}$ values from Eq.\,\ref{eq:cdfvmaxsat}, by drawing uniform random numbers $u_i \sim \mathcal{U}(0,1)$, and solving:
\begin{equation}
    V_{\rm peak}^{i,\,\rm{sat}} = \Psi_{\rm sat}^{-1}(u_i)\,,\,\,{\rm where}\,\,i=1,\cdots,N_{\rm sat}\,,
   \label{eq:inversesampling}
\end{equation}
\end{enumerate}
at fixed $z_{\rm A}$. In this way, the sampled satellite velocities follow the probability distribution set by the hosts.

\subsubsection{Satellite positions and placement in hosts}\label{sec:positions}
We randomly place satellites at the \textsc{AbacusSummit} DM particle locations using a normalized particle-level occupation scheme. 

First, for each host we compute its expected mean number of satellites using a power-law function of its virial mass \citep[see e.g.,][]{2002ApJ...575..587B,2004ApJ...609...35K,2005ApJ...633..791Z,2020MNRAS.499.5486A,2021MNRAS.504.4667A,2023JCAP...10..016R,2025MNRAS.538.1216Y}:
\begin{equation}
    \langle N_{\rm sat} \rangle = \left(\frac{M^{\rm h}_{\rm vir}-\kappa\,M_{\rm cut}}{M_1}\right)^\beta\,,
    \label{eq:powerlaw}
\end{equation}
to ensure that more massive hosts are assigned a larger number of satellites. Here, $M_1$ is the typical halo mass hosting one satellite, while $M_{\rm cut}$ ($\kappa\,M_{\rm cut}$) is the minimum halo mass to host a central (satellite) galaxy. 

The parameter $\kappa$ acts as a satellite–onset shifter: lowering $\kappa$ allows satellites to populate lower–mass halos, boosting the 1–halo term and steepening the small–scale monopole and quadrupole, while increasing $\kappa$ delays satellite formation to higher masses, reducing the satellite fraction and flattening small–scale clustering without affecting large–scale bias.

This expected number is then distributed across the halo’s particles by assigning each particle a selection probability equal to: 
\begin{equation}
p_{\rm sat}=\frac{\langle N_{\rm sat} \rangle}{N_{\rm p}}\,,
\label{eq:psat}
\end{equation}
where $N_{\rm p}$ is the total number of particles available per halo. 
We then perform a Bernoulli trial per particle, with retain probability $p_{\rm sat}$, to decide whether or not a satellite will occupy its position. In case of success, we flag the particle with {\texttt{type\,$=$\,1}} to distinguish satellites from central mocks in the resulting catalog (see \S\,\ref{sec:centrals}).

This particle-level approach, where we directly sample satellite positions from particle locations, has the great advantage of preserving the spatial correlations of the underlying DM field, providing \ho natural anisotropy and stochasticity, especially on small scales (further details in Appendix\,\ref{sec:occup}).

Dark matter particles in $N$-body simulations exhibit more centrally concentrated radial profiles within halos, compared to resolved satellite subhalos \citep[e.g.,][]{2005ApJ...618..557N,2008MNRAS.387..536G}. This difference arises since particles trace the full mass distribution, including regions where subhalos have been disrupted or stripped, while satellite catalogs only include resolved, surviving substructures. 

As a result, satellite profiles are typically suppressed in the inner halo due to physical effects such as tidal disruption, dynamical friction, or finite resolution \citep[e.g.,][]{2005MNRAS.359.1029V,2013ApJ...770...57B}, as opposed to cuspy DM profiles, such as Navarro-Frenk-White \citep[NFW;][]{1997ApJ...490..493N}.

\begin{figure}
\centering
\includegraphics[width=\linewidth]{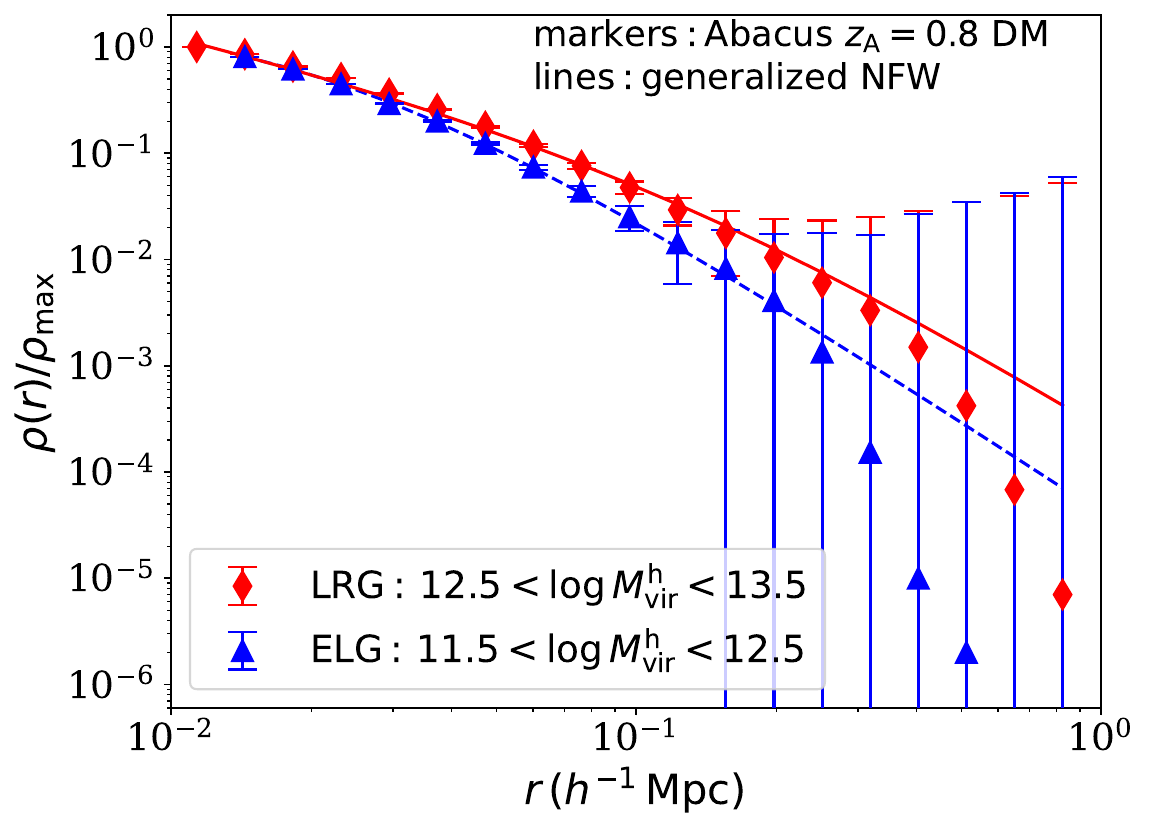}
\caption{Radial profile of the \textsc{AbacusSummit} DM particles in hosts with typical ELG (blue triangles, dashed line) and LRG (red diamonds, solid line) halo virial masses, in units of $(h^{-1}{\rm M}_\odot)$. The error bars are the standard deviation in each radial bin.}
\label{fig:abacusprofile}
\end{figure}

As shown in Figure\,\ref{fig:abacusprofile}, DM particles populating \textsc{AbacusSummit} halos with typical ELG ($3.2\times10^{11}-3.2\times10^{12}h^{-1}{\rm M}_\odot$) and LRG masses ($3.2\times10^{12}-3.2\times10^{13}h^{-1}{\rm M}_\odot$) naturally follow a generalized NFW profile:
\begin{equation}
    \rho_{\rm A}(r)=\frac{\rho_{\rm s}}{\left(\frac{r}{r_{\rm s}}\right)^\delta\left(1+\frac{r}{r_{\rm s}}\right)^{\eta-\delta}}\,,
\end{equation}
where $(\rho_{\rm s},\,r_{\rm s})$ are the scale parameters, and $(\delta,\,\eta)$ are the inner and outer slopes; their best-fit values are reported in Table\,\ref{tab:NFWpars}. 

\begin{deluxetable*}{cccccc}
\tablecaption{Best-fit parameters of the generalized NFW profiles for \textsc{AbacusSummit} DM particles in hosts with typical ELG and LRG masses, as shown in Figure\,\ref{fig:abacusprofile}.\label{tab:NFWpars}}
\tablehead{
   \colhead{$M_{\rm h}^{\rm vir}$\,($h^{-1}$Mpc)}&\colhead{$\rho_{\rm s}\,(h^2 {\rm M}_\odot \,{\rm Mpc^{-3}})$}& \colhead{$r_{\rm s}\,(h^{-1}{\rm Mpc})$}&\colhead{$\delta$}&\colhead{$\eta$}& \colhead{$\chi^2/{15\,\rm dof}$}
  }
\startdata
$10^{11.5}-10^{12.5}$&$(1.02\pm0.28)\times 10^{-2}$&$25.03\pm10.44$&$-1.65\pm0.51$&$2.90\pm0.16$&1.7\\
$10^{12.5}-10^{13.5}$&$0.17\pm0.18$&$0.11\pm0.18$&$0.71\pm0.07$&$6.06\pm4.66$&$2.3$
\enddata
\end{deluxetable*} 

For the above reasons, especially in high-resolution simulations, sampling satellites directly from the particle distribution, or from a pure NFW profile, systematically leads to an excess of clustering in projection on scales $r_{\rm p}\lesssim1\,h^{-1}$Mpc. In redshift space, this is smeared by peculiar velocities, in a way that its impact on the anisotropic clustering becomes negligible.

In terms of intra-halo dynamics, such an excess has two contributions: (i) satellites too concentrated around the host in the halo core; (ii) satellite-satellite pairs where the two objects are too close to each other. The latter has a stronger impact on the clustering.

Whereas traditional HOD models often assumed that satellite galaxies follow a pure NFW profile---adequate on intermediate and large scales---modern analyses extend to sub-Mpc and kpc scales, where NFW tends to overpredict the number of close galaxy pairs. 

To mitigate this excess, either one combines a radial suppression mechanism with a probabilistic core-suppression downsampling (see Appendix\,\ref{sec:core_supp}) or, more easily and with less parameters, fine tunes the inner halo profile by modulating the relative distance between pairs, while preserving the outer halo profile. We choose this second strategy and explain it later on. 

After directly sampling satellite positions from the \textsc{AbacusSummit} particle locations---so that our model inherits the full DM realism and structure---we modulate the cuspy particle profile to account for the observational evidence that ELGs preferentially occupy the outskirts of their host halos \citep[e.g.,][]{2007ApJ...664..791B,2012MNRAS.424..232W, 2017MNRAS.466.1880C,2018MNRAS.474..547K,2021ApJ...918...53G,2024A&A...691A.236D}, while LRGs prefer the central regions \citep[e.g.,][]{2012ApJ...751L...5T,2015MNRAS.452..998H}. 

In line with semi-analytic \citep[e.g.][]{2018MNRAS.475.2530O} and HOD models \citep[e.g.,][]{2020MNRAS.499.5486A, 2023JCAP...10..016R, 2021MNRAS.501.1603H,2021MNRAS.504.4667A}, that have successfully reproduced the ELG clustering properties by placing satellites in the outskirts of their hosts, we redefine the satellite positions in our model as: 
\begin{equation}
\vec{r}_{\rm sat} \equiv \vec{r}_{\rm h} + K_{\rm out}(\vec{r}_{\rm dm}-\vec{r}_{\rm h})\,.
\label{eq:outskirt}
\end{equation}
Here, $\vec{r}=(x,\,y,\,z)$ are the vectors of cartesian real-space\footnote{i.e., before applying redshift-space distortions due to peculiar velocities (\S\,\ref{sec:vpec})} positions, and $K_{\rm out}$ is a model parameter that changes the halo density profile acting as a scaling factor on the radial offset from the halo center. This parameter allows the ELG and LRG density profiles to be modeled independently. 
When DM particles are randomly selected, for $K_{\rm out}=1$ ($K_{\rm out}=0$), satellites occupy the particle (host) position, while $0<K_{\rm out}<1$ pulls them towards the host center, and $K_{\rm out}>1$ pushes them outward into the halo outskirts. However, when DM particles go through non-random selection processes---such as the AM (\S\,\ref{sec:AM}) or the joint-occupation (\S\,\ref{sec:jointoccup}) down-samplings---$K_{\rm out}$ compensates for the bias introduced by over-correcting the spatial profile (see also \S\,\ref{sec:hodpred}).

To prevent the rescaling from unphysically ejecting satellites outside the virial boundary, we impose a physical cap on the effective displacement such that
\begin{equation}
\left|\vec{r}_{\rm sat}-\vec{r}_{\rm h}\right|
\le R_{\rm vir} \,,
\label{eq:outskirt_cap}
\end{equation}
ensuring satellites remain bound within their parent halo. 

By redefining the satellite positions through Eq.\,\ref{eq:outskirt}, \ho gains additional leverage to place satellites in different regions of their host halos depending on the galaxy population. This flexibility is key to match the distinct 1–halo clustering signatures of ELGs and LRGs.

\subsubsection{Peculiar motions}\label{sec:vpec}
Inspired by \citet{2018MNRAS.475.2530O}, we model satellites as {\textit{coherent flows}} with a velocity bias with respect to their host halos. 

 Assuming that the line-of-sight is parallel to the $\vec{z}$ direction (see \S\,\ref{sec:centrals}), we define the satellite peculiar motions as the sum of a radial and a tangential component---modulated by velocity bias parameters $b_{\rm r}$ and $b_{\rm t}$---relative to their host, plus a random contribution proportional to the host velocity dispersion. 
 
 With the particle-host radial vector and unit direction given by:
\begin{equation}
    \vec{r}=\vec{r}_{\rm dm}-\vec{r}_{\rm h},\hspace{0.5cm}\hat{r}=\frac{\vec{r}}{||\vec{r}||}\,,
\end{equation}
the radial and tangential velocity components are:
\begin{equation}
\begin{aligned}
    &v_{\rm rad}=(\vec{v}_{\rm dm}-\vec{v}_{\rm h})\cdot\hat{r}\,,\\
    &v_{\rm tan}=(\vec{v}_{\rm dm}-\vec{v}_{\rm h}) - \vec{v}_{\rm rad}\,\,\,{\rm with}\,\,\vec{v}_{\rm rad}=v_{\rm rad}\cdot\hat{r}\,.
    \end{aligned}
\end{equation}

Combining all these ingredients, the satellite peculiar velocity along the LOS is:
\begin{equation}
v_z^{\rm sat} = v_z^{h}+ b_{\rm r}\,v_{{\rm rad},z}+b_{\rm t}\,v_{{\rm tan},z}+ \mathcal{N}(0,f_{\sigma}\,\sigma_{\rm vd})\,,
\label{eq:vzsat}
\end{equation}
where the host peculiar velocity, $v_z^{\rm h}$, represents the coherent component, and the DM particle velocity $v_z^{\rm DM}$ the dispersed one. 

The last term in Eq.\,\ref{eq:vzsat} is a Gaussian random realization with zero mean and dispersion proportional to the DM velocity dispersion $\sigma_{\rm vd}$ (\S\,\ref{sec:centrals}), rescaled by the $f_\sigma$ parameter \citep[][]{2023JCAP...10..016R}. This term maintains the model realistic, avoiding artificial correlations.

The radial velocity bias, $b_r$, allows us to control the infall of satellites towards their hosts: $b_r<1$ ($b_r>1$) translates into slow (fast) infall motions, resulting in weaker (stronger) compression on $5-10\,h^{-1}$Mpc scales \citep{1987MNRAS.227....1K}.

As discussed in \S\,\ref{sec:appvz}, peculiar intra-halo motions have strong impact on the satellite dynamics and are responsible of shaping the small-scale anisotropic clustering.

\subsection{Halo Occupation Model (\textsc{Hom}e)} \label{sec:AM}
To connect DESI ELGs and LRGs with their \textsc{AbacusSummit} host halos, we revisit the abundance matching prescriptions by \citet{2016MNRAS.460.1173R} and \citet{2016MNRAS.461.3421F, 2017MNRAS.472..550F,2022MNRAS.509.1614F}, and generalize them to galaxy multi-tracers. All these methods modify the standard AM \citep[e.g.,][]{2006ApJ...647..201C, 2010ApJ...717..379B} recipe in different ways to account for the luminosity or stellar mass incompleteness of the galaxy sample to model, hence they are suitable for multi-tracer analyses.

The standard AM assumes that more massive (or luminous) galaxies occupy more massive halos, with deeper gravitational potential wells. In its basic version, the method rank-orders halos and galaxies based on some primary properties, and puts them in correspondence allowing some scatter between them.

On the halo side, we adopt $V_{\rm peak}$ as proxy for the halo mass, and use it to rank-order the \textsc{AbacusSummit} halos (see \S\,\ref{sec:Vmaxsampling}). Note that, at this stage of our model pipeline, the input halo catalog for \ho is composed of central (\texttt{type\,$=$\,0}) and satellite (\texttt{type\,$=$\,1}) halos including all the relevant properties to perform abundance matching and clustering analysis---i.e., $V_{\rm peak}$, positions, peculiar velocities---as built in the previous sections. 

On the galaxy side, we consider as main property the stellar mass ($M_\star$) for both ELGs and LRGs. One could alternatively use the \OII \,line ($r-$band) luminosity for ELGs (LRGs), but it is preferable to choose a unique consistent proxy for all tracers, so $M_\star$ is more adecuate.

We sample the $M_\star$ values from the current fits (see Appendix\,\ref{sec:MF}) to the star-forming (ELG) and quiescent (LRG) populations observed in COSMOS \citep{2023A&A...677A.184W}. Besides the matching, we do not employ the stellar masses in our analysis.

Besides the minimal cut in the number of particles per halo (see \S\,\ref{sec:simulation}), we also impose a lower velocity threshold, $V_{\rm peak}^{\rm min}$, that we fix for each galaxy sample to match its large-scale bias. This threshold is not free parameter in our model, and its values for ELGs and LRGs are reported in Table\,\ref{tab:clusteringresu}.

Between $V_{\rm peak}$ and $M_\star$, we allow a constant scatter in order to be physical. This is introduced by perturbing $V_{\rm peak}$ using a random Gaussian realization ($\mathcal{N}$) with zero mean and dispersion $\sigma_{\rm AM}$, which is a model parameter:
\begin{equation}
V_{\rm peak}^{'}=V_{\rm peak}\,[1+\mathcal{N}(0,\,\sigma_{\rm AM})]\,.
\end{equation}

In \textsc{Hom}e, the abundance matching procedure establishes the galaxy–halo connection without distinguishing between central and satellite halos. However, during the construction of the mock catalog, each central (satellite) mock is flagged with \texttt{type=0}\,(\texttt{1}), enabling separation by type in what follows. 

Once the basic catalog is assembled, further refinement can be achieved by matching secondary halo and galaxy properties conditioned on $M_\star$ \citep[e.g.,][]{2022MNRAS.509.1614F}, enhancing the predictive power of the high-fidelity mocks; we defer such extensions to follow-up work.

We then incorporate the ELG and LRG incompleteness effects by downsampling the final number of objects in the mock catalog to match the observed DESI Y1 ELG and LRG number densities \citep{2025JCAP...07..017A}. As discussed in Appendix\,\ref{sec:MF}, this downsampling alters the shape of the stellar mass functions in by selecting ELG and LRG populations that are incomplete with respect to the original ones.

The downsampling is applied separately to central and satellite mocks---identified through the above \texttt{type} flag---using probability values computed in bins of $V_{\rm peak}$, as \citep[see][]{2016MNRAS.460.1173R, 2017MNRAS.472..550F}:
\begin{equation}
p_{\rm cen\,(sat)}=\frac{N_{\rm tot}^{\rm cen\,(sat)}(V_{\rm peak})}{N_{\rm Gauss}^{\rm cen\,(sat)}(V_{\rm peak})}\,,
\end{equation}
each one normalized to unity. 

Here, $N_{\rm tot}^{\rm cen\,(sat)}(V_{\rm peak})$ are the total number of central (satellite) halos in bins of $V_{\rm peak}$, meaning their velocity functions computed from the simulation. The  $N_{\rm Gauss}^{\rm cen\,(sat)}(V_{\rm peak})$ counterparts are central (satellite) Gaussian velocity functions that we compute from $V_{\rm peak}$ values sampled using a multi-tracer Gaussian PDF, which considers ELG and LRG as complementary populations. This is:
\begin{equation}
\begin{aligned}
\phi(&V_{\rm peak},V_{\rm elg},\,\sigma_{\rm elg},V_{\rm lrg},\,\sigma_{\rm lrg}) = \\ 
    &w_{\rm elg}(z)\left[\mathcal{G}^{\rm cen}_{\rm elg}(V_{\rm peak},V_{\rm elg},\,\sigma_{\rm elg})+ \mathcal{G}^{\rm sat}_{\rm elg}(V_{\rm peak},V_{\rm elg},\,\sigma_{\rm elg})\right] \,+\\
    &w_{\rm lrg}(z)\left[\mathcal{G}^{\rm cen}_{\rm lrg}(V_{\rm peak},V_{\rm lrg},\,\sigma_{\rm lrg})+\mathcal{G}^{\rm sat}_{\rm lrg}(V_{\rm peak},V_{\rm lrg},\,\sigma_{\rm lrg})\right]\,,
\label{eq:PDFdown}
\end{aligned}
\end{equation}
where $\mathcal{G}^{\rm cen}_{\rm elg\,(lrg)}$ and $\mathcal{G}^{\rm sat}_{\rm elg\,(lrg)}$ are \emph{independent} Gaussian realizations for central and satellite ELG (LRG) mocks defined, in the same way, as a function of 4 model parameters: 
\begin{equation}
\begin{aligned}
    \mathcal{G}^{\rm cen}_{\rm elg\,(lrg)}&(V_{\rm peak},V_{\rm elg\,(lrg)},\,\sigma_{\rm elg\,(lrg)})=&\\
    &\mathcal{G}^{\rm sat}_{\rm elg\,(lrg)}(V_{\rm peak},V_{\rm elg\,(lrg)},\,\sigma_{\rm elg\,(lrg)})=&\\
    &\frac{1}{\sqrt{2\,\pi}\,\sigma_{\rm elg\,(lrg)}}\exp\left[{-\frac{1}{2}\left(\frac{V_{\rm peak} - V_{\rm elg\,(lrg)}}{\sigma_{\rm elg\,(lrg)}}\right)^2}\right]\,.
    \label{eq:realiz}
\end{aligned}
\end{equation}
Here, the parameters $V_{\rm elg,(lrg)}$ and $\sigma_{\rm elg,(lrg)}$ represent the typical mean maximum circular velocities of the tracers, and the corresponding scatter. 

We emphasize that the two Gaussian realizations---one for centrals and one for satellites---yield different outcomes, as the sampling is performed independently for each population.

The contribution of each tracer to the PDF in Eq.\,\ref{eq:PDFdown} is weighted by its observed number as a function of redshift \citep[][]{2025JCAP...07..017A}, i.e. $N_{\rm elg\,(lrg)}(z)$. The ELG (LRG) weights are defined as:  
\begin{equation}
\begin{aligned}
    &w_{\rm elg\,(lrg)}(z)=\frac{N_{\rm elg\,(lrg)}(z)}{N_{\rm elg}(z)+N _{\rm lrg }(z)}\,,
    \end{aligned}
\end{equation}
so that, when modeling their auto-correlation functions, the complementary tracer does not contribute, since $N_{\rm lrg\,(elg)}(z)=0$.

Note that the full redshift evolution is accounted for only when working with a light-cone. For a fixed simulation snapshot, as in our case, we evaluate $N_{\rm elg\,(lrg)}(z_{\rm eff})$, as reported in Table\,\ref{tab:zrep}. 

The Gaussian realizations in Eq.\,\ref{eq:realiz} are normalized to match the observed ELG and LRG numbers in terms of 4 model parameters as:

\begin{equation}
\begin{aligned}
\int \left(w_{\rm elg}\,\mathcal{G}_{\rm elg}^{\rm cen} + w_{\rm lrg}\,\mathcal{G}_{\rm lrg}^{\rm cen}\right)\,dV_{\rm peak} &=\\
\left(1 - f^{\rm elg}_{\rm sat}\right)\,N_{\rm elg}(z) +&
\left(1 - f^{\rm lrg}_{\rm sat}\right)\,N_{\rm lrg}(z)\,, \\\\
\int \left(w_{\rm elg}\,\mathcal{G}_{\rm elg}^{\rm sat} + w_{\rm lrg}\,\mathcal{G}_{\rm lrg}^{\rm sat}\right)\,dV_{\rm peak} &=\\
 f^{\rm elg}_{\rm sat}&\,N_{\rm elg}(z)
+  f^{\rm lrg}_{\rm sat}\,N_{\rm lrg}(z)\,.
\end{aligned}
\label{eq:normaliz}
\end{equation}
This guarantees \ho sufficient flexibility to precisely and simultaneously model the \texttt{ELG1} and \texttt{LRG3} auto- and cross-correlation functions below $4\,h^{-1}$\,Mpc, where the $1-$halo contribution to the clustering is particularly strong \citep[][]{2023JCAP...10..016R}.

\subsubsection{Orphan treatment}
\label{sec:orphans}
The $f^{\rm elg}_{\rm sat}$ and $f^{\rm lrg}_{\rm sat}$ terms in Eq.\,\ref{eq:normaliz} represent the \emph{total} fractions of satellite ELGs and LRGs calculated with respect to the total number of objects in the final mock catalog. Note that, in principle, these fractions can include:
\begin{itemize}
    \item ELG (LRG) satellites orbiting a ELG (LRG) host, meaning a maximally conformal configuration;
    \item ELG (LRG) satellites living in a host of the complementary species, meaning minimal conformity;
    \item Orphans, i.e. satellites whose central host halos are not part of the resulting mock catalog. Note that here the term ``orphans" does not refer to disrupted subhalos, but simply to satellites whose corresponding centrals are not selected as part of the ELG or LRG samples.
\end{itemize}

Contrarily to SHAM models, standard HOD frameworks typically omit orphan satellites by construction because satellites are only populated in halos that already host a central. Only a few modified HOD schemes have explicitly allowed for orphans \citep[see e.g.][]{2017MNRAS.469..749P}, despite growing evidence that their inclusion can be important for reproducing the small-scale clustering signal. This point has been emphasized in semi-analytic models, where disrupted or unresolved centrals naturally generate satellite populations without an identified host \citep[e.g.,][]{2013ApJ...771...30R,2019MNRAS.488.3143B}.

In \textsc{Hom}e, orphans are not parametrized nor constrained by the likelihood. Rather, the relative ELG and LRG orphan abundances emerge naturally and self-consistently within our forward model as pure predictions, as a direct consequence of applying the AM down-sampling of satellites to match the observed number densities of tracers (\S,\ref{sec:AM}), halo exclusion, and the joint-occupation condition (\S\,\ref{sec:jointoccup}). 
The resulting orphan fractions (see Table\,\ref{tab:clusteringresu}) are relatively high by construction, since potential satellite candidates are drawn from the full pool of \textsc{AbacusSummit} DM particles---including those belonging to host halos that are not included in the final mock catalog (\S\,\ref{sec:satellites}).


\subsubsection{Halo exclusion}\label{sec:halo_exclusion}
Halo exclusion refers to the physical constraint that DM halos cannot overlap in space due to their finite sizes \citep[e.g.,][]{2023OJAp....6E..39A}. This leads to a suppression of clustering power on small scales and introduces a scale-dependent bias, particularly relevant in the transition between the 1-halo and 2-halo regimes. 

In observational data, halo exclusion manifests as a deficit of close galaxy pairs residing in distinct halos, visible in the two-point correlation functions at small separations. 

In $N$-body simulations, exclusion arises naturally from halo finding algorithms---such as Friends-of-Friends \citep[e.g.,][]{2011ApJS..195....4M}, Spherical Overdensity \citep[e.g.,][]{2011MNRAS.415.2293K}, or Rockstar \citep{2013ApJ...762..109B}---where halos are defined as non-overlapping entities. Its effect is especially prominent for massive halos, where the exclusion scale can reach up to several Mpc.

From the modeling perspective, halo exclusion has been successfully incorporated in fast methods for covariance mock generation \citep[see][]{2016MNRAS.456.4156K}. Here, the \textsc{hadron} \citep{2015MNRAS.451.4266Z} approach probabilistically assigns halo masses while accounting for both local and non-local environmental effects, including minimum distance constraints to suppress unphysical clustering among massive halos.

Analytical halo models have incorporated this effect by modifying the 2-halo term with exclusion corrections, using either step-function or soft-exclusion schemes \citep[see e.g.,][]{2013PhRvD..88h3507B,2013MNRAS.430..725V}.

HOD frameworks naturally incorporate exclusion by enforcing one central galaxy per halo and restricting the distribution of satellites accordingly. This approach has been successful in modeling SDSS and DES data \citep[see e.g.,][]{2005ApJ...633..791Z,2011MNRAS.410..210M,2012ApJ...745...16T,2018MNRAS.476.1637Z}. Recent observational models incorporate refined exclusion treatments to improve parameter inference and consistency with lensing measurements \citep[e.g.,][]{2017MNRAS.467.3024L}.

Hydrodynamical simulations, as IllustrisTNG, EAGLE, or Horizon-AGN, confirm that halo exclusion persists even when baryonic physics is included. Baryonic processes can alter halo profiles and sizes, but the fundamental exclusion scale remains tied to the DM distribution. These simulations show that models lacking exclusion cannot accurately reproduce galaxy–galaxy lensing or clustering measurements \citep[e.g.,][]{2018MNRAS.480.3978A}.

Halo exclusion has been recently integrated also in the modeling of line-intensity mapping, together with non-linear bias, in a physically motivated and simulation-validated framework \citep{2025arXiv250603015J}.

In light of these results, we incorporate halo exclusion in \textsc{Hom}e by suppressing the probability, for pairs of massive halos (i.e., more massive than a given threshold, $M_{\rm excl}$), to be closer than a given (small) exclusion radius, $r_{\rm excl}$. This translates into the condition:
\begin{align}
&{\rm If}\,\,\,
\begin{cases}
&M^{{\rm cen},\,i}_{\rm vir}>M_{\rm excl}\\
&M^{{\rm cen},\,j}_{\rm vir}>M_{\rm excl} \\
& |r_i-r_j|<r_{\rm excl}
\end{cases}\\
&\Rightarrow \,\,{\rm randomly\,remove\,}i\,{\rm or}\,j\,{\rm with\,probability\,}p_{\rm excl}\,,
\end{align}
where $r_{\rm excl}$, $M_{\rm excl}$ and $p_{\rm excl}$ are model parameters.

In practice, we apply the exclusion condition to the \ho input catalog---composed of central hosts and satellites---just before sampling the $V_{\rm peak}$ values from the Gaussian PDF (Eq.\,\ref{eq:PDFdown}). In this way, we ensure that close, massive pairs are excluded prior to galaxy assignment. 
This approach naturally embeds exclusion physics into the mock-building process, while preserving the normalization of the PDF above. 

Because our exclusion is keyed to the host mass, dropping a massive host automatically drops all of its bound (non-orphan) satellites, i.e. no extra satellite cut is needed there. Hence, the only place where satellite exclusion matters is for orphans, which carry the mass tag of their original hosts from the \textsc{AbacusSummit} simulation, but these hosts are missing in the \ho input catalog.
If left unmitigated, orphans from massive hosts behave like low-mass centrals, adding central--central--like pairs and inflating $w_p$ on $\sim8-10\,h^{-1}$ Mpc scales. Therefore, excluding satellites besides centrals, primarily serves to regulate further the presence of orphans in the final mock catalog (i.e., besides mitigating it via $f_{\rm orph}$) and its impact on non-orphans is redundant once host-based exclusion is in place.

Note that, although satellites are placed within the virial radius of their hosts, halo exclusion is not automatically guaranteed in the resulting galaxy population. The reason is that the galaxy catalog is constructed by sampling halos according to the $V_{\rm peak}$ distribution through abundance matching, which effectively reselects a subset of halos from the original simulation. This selection can produce configurations where two massive hosts that both pass the abundance matching threshold lie at separations smaller than physically expected for the galaxy populations under consideration. While the underlying simulation enforces non-overlapping halos at the halo-finder level, it does not guarantee that the subset of halos selected to host galaxies preserves the appropriate exclusion scale for massive systems. The explicit exclusion condition implemented here therefore acts as a corrective step that suppresses such unphysical close pairs, ensuring that the resulting clustering signal remains consistent with the expected transition between the 1-halo and 2-halo regimes.

\begin{figure}
\centering
\includegraphics[width=\linewidth]{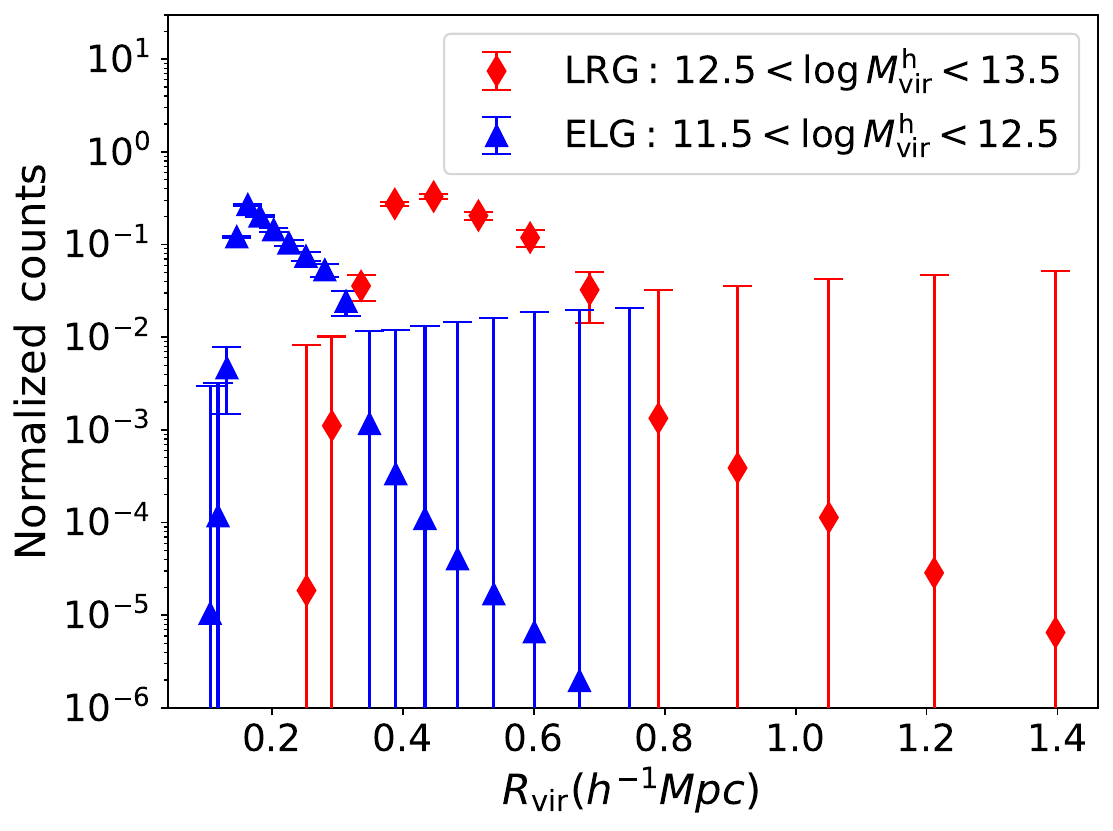}
\caption{Normalised $R_{\rm vir}$ distribution of the \textsc{AbacusSummit} central host halos of ELG and LRG.}
\label{fig:abacusRvir}
\end{figure}

Figure\,\ref{fig:abacusRvir} presents the normalized distribution of virial radii ($R_{\rm vir}$), for \textsc{AbacusSummit} halos selected within mass ranges typical of ELGs ($3.2\times10^{11}-3.2\times10^{12}h^{-1}{\rm M}_\odot$) and LRGs ($3.2\times10^{12}-3.2\times10^{13}h^{-1}{\rm M}_\odot$), following the same selection as in Figure\,\ref{fig:abacusprofile}. While ELG host halos typically peak at $R_{\rm vir}\sim0.2\,h^{-1}$Mpc, and extend up to $\sim 0.7\,h^{-1}$Mpc, LRG hosts are substantially larger, peaking around $\sim 0.5\,h^{-1}$Mpc and reaching up to $ \sim 1.4\,h^{-1}$Mpc. 

These distributions imply that, for most ELG (LRG) halos, the 2-halo regime starts beyond $R_{\rm vir}\sim0.4\,h^{-1}$Mpc ($ 0.8\,h^{-1}$Mpc), the latter marking the scale where exclusion between pairs of more massive central halos becomes significant.

\subsubsection{ELG and LRG class assignment}\label{sec:class}

We stochastically classify galaxies into ELG or LRG tracers while constructing the mock catalog, using the two-component Gaussian mixture model in Eq.\,\ref{eq:PDFdown}. The classification is performed on the fly---separately for central and satellite halos--- during the mock assembly, enabling us to track how the combined ELG and LRG HOD is progressively constructed within \textsc{Hom}e.

For centrals, the assignment is mutually exclusive by construction. We first assign LRG centrals following their selection probability. Once a halo is flagged as an LRG host, it is removed from the pool of available central halos to prevent reassignment. ELG centrals are then selected from the complementary halo population. This sequential procedure guarantees that a halo hosts at most one central galaxy in the final catalog, preserving physical consistency.

For satellites, however, the assignment is not exclusive to the host tracer type, since any satellite selected from the particle reservoir may inhabit a halo regardless of whether its central is an ELG or LRG. This design naturally allows:
\begin{itemize}
\item ELG satellites in LRG hosts, characterizing the environmental quenching regime; 
\item LRG satellites in ELG hosts, an observationally rare but permitted configuration,
\end{itemize}
thereby enabling the forward model to predict cross–tracer satellite populations and reveal potential central–satellite conformity.

In prectice, for each sampled $V_{\rm peak}$ value\footnote{Remember that the sampling is performed separately for centrals and satellites, hence classes are assigned accordingly.}, we evaluate the unnormalized posterior probabilities:

\begin{equation}
    p_{\rm elg\,(lrg)}(V_{\rm peak})=w_{\rm elg\,(lrg)}(z)\,\mathcal{G}(V_{\rm peak}|V_{\rm elg\,(lrg)},\,\sigma_{\rm elg\,(lrg)})\,,
    \label{eq:probpost}
\end{equation}
which define the normalized ELG-class probability:
\begin{equation}
    P_{\rm elg}(V_{\rm peak})=\frac{p_{\rm elg}(V_{\rm peak})}{p_{\rm elg}(V_{\rm peak})+p_{\rm lrg}(V_{\rm peak})}\,.
    \label{eq:probnorm}
\end{equation}
A uniform draw $u\sim\mathcal{U}(0,1)$ then determines the label:
\begin{equation}
{\rm \texttt{class}}(V_{\rm peak})=
\begin{cases}
&{\rm \texttt{ELG},}\,\,\,\,{\rm if}\,\,u<P_{\rm elg}(V_{\rm peak}) \\
&{\rm \texttt{LRG},}\,\,\,\,{\rm otherwise}
\label{eq:classes}
\end{cases}
\end{equation}
Thus, \ho performs probabilistic Bayesian classification for centrals and satellites, while enforcing central exclusivity and preserving the freedom for physically motivated cross–tracer satellite configurations.
\subsubsection{Modeling environmental quenching through the joint halo occupation of ELGs and LRGs}\label{sec:jointoccup}

Understanding the mechanisms that drive galaxy quenching, i.e. the cessation of star formation, remains one of the central challenges in galaxy evolution. 

Satellite galaxies, in particular, are observed to quench rapidly upon infall into dense environments. Observations show that ELGs cease star formation shortly after entering the potential wells of massive hosts \citep[e.g.][]{2021ApJ...918...53G, 2013MNRAS.432..336W, 2014MNRAS.444.2938H, 2024ApJ...971..111R}.

Large cluster surveys support a delayed-then-rapid scenario in which satellites continue forming stars for roughly 1-2 Gyr after infall before undergoing a rapid shutdown \citep[e.g.][]{2013MNRAS.432..336W, 2015ApJ...806..101H}.
\citet{2023MNRAS.526.3716B} further identify two dominant pathways: core quenching, acting swiftly on satellites reaching the inner regions of their hosts, and starvation, a slower process driven by the exhaustion of gas in the outskirts.

Recent IFU surveys, such as K-CLASH, also show evidence for environmental quenching induced by ram-pressure stripping and strangulation, i.e. the halt of cosmic gas inflow \citep{2020MNRAS.496.3841V}.

On the theoretical side, semi-analytic and hydrodynamical studies find that strangulation dominates quenching in low-mass galaxies, while ram-pressure stripping or overconsumption are more efficient at high redshift or in dense environments \citep{2015Natur.521..192P, 2014MNRAS.442L.105M}.

Numerical simulations, such as AREPO, confirm a two-phase picture in which gas depletion via starvation precedes rapid stripping during pericentric passages \citep{2016A&A...591A..51S}.

Machine-learning analyses of cosmological simulations also identify black-hole mass and central potential depth as key predictors of quenching in massive systems, highlighting AGN feedback as the principal internal mechanism \citep{2022MNRAS.512.1052P, 2023ApJ...944..108B}.
Taken together, these results support a dual-channel view of galaxy quenching: internal AGN feedback dominates in massive centrals, while environment-driven processes govern the shutdown of star formation in satellites.

In \textsc{Hom}e, we emulate environmental quenching statistically, through a joint-occupation condition that regulates the co-existence of ELGs and LRGs within the same halos.
Instead of explicitly evolving star-formation histories, we modulate the ELG occupation probability by the local density field traced by massive LRG hosts, effectively suppressing the presence of star-forming galaxies in dense environments.

For each ELG halo $i$, we compute a kernel-weighted suppression field:
\begin{equation}
S_i = \sum_j
\left[\frac{M_{{\rm vir},j}^{\rm lrg}}{M_0}\right]^{m_{\rm slope}}
\exp\left[-\left(\frac{r_{ij}}{R_0}\right)^{\delta}\right]\,,
\end{equation}
where $r_{ij}$ is the comoving separation between the ELG candidate and the $j$-th LRG host. Here, $M_{{\rm vir},j}^{\rm lrg}$ is the LRG halo virial mass, while the $M_0$ is the threshold mass above which a LRG halo is considered massive enough to influence ELG quenching in its vicinity.

The kernel decays exponentially with distance over a characteristic scale $R_0$, while the exponent $\delta$ controls its steepness, and the mass-weighting term, $m_{\rm slope}$, enhances the contribution of massive hosts.
This effectively encodes the environmental influence of nearby LRG halos, i.e. those most likely to drive quenching through tidal interactions, ram-pressure stripping, or starvation.

In practice, we apply the joint-occupation condition by rescaling the baseline abundance-matching probability for a halo to be populated by an ELG---i.e., $p_{\rm elg}(V_{\rm peak})$ in Eq.\,\ref{eq:probpost}---via the above suppression field, as:
\begin{equation}
p_{\rm elg}^{\rm joint}(V_{\rm peak}) = p_{\rm elg}(V_{\rm peak})\,\exp(-\alpha\,S_i)\,,
\end{equation}
where the parameter $\alpha$ controls the strength of environmental suppression.
Here, high-density regions (i.e., large $S_i$ values), correspond to LRG-dominated environments where quenching is efficient, yielding a smaller $p_{\rm elg}^{\rm joint}$, while isolated regions retain high ELG occupation probabilities.

This exponential coupling therefore emulates quenching statistically, that is reducing the abundance of central and satellite ELGs near massive hosts, without the need for an explicit treatment of gas physics or feedback.

As such, the joint occupation model serves as a phenomenological but physically motivated proxy for environmental quenching, connecting the observed scarcity of star-forming ELGs in cluster cores with the underlying DM halo distribution.

\section{Hierarchical Bayesian inference framework}
\label{sec:inference}
We adopt a \emph{two-level hierarchical Bayesian inference model} designed to account for the uncertainty and physical realism in the generation of mock catalogs for galaxy multi-tracers based on precise $N$-body simulations.

Our aim is to forward model the observed clustering statistics (\( \mathbf{D} \), hereafter) using mock catalogs generated coupling the \textsc{AbacusSummit} host halo properties with the latent properties that we build for satellites.

These latent satellite variables are constructed---conditional to their host and DM particle properties---using analytic prescriptions based on a set of physically motivated nuisance parameters (\( \boldsymbol{\theta} \), hereafter).

Incorporating \( \boldsymbol{\theta} \) in the likelihood used for cosmological inference would require regenerating the full set of latent variables for all halos at every point in parameter space---a task that is computationally prohibitive.
To overcome this, we structure our inference strategy in two steps:
\begin{itemize}
\item {\bf Level-I inference:} we generate the latent variables by calibrating the nuisance parameters \( \boldsymbol{\theta} \) against independent predictions from the \textsc{Uchuu} $N$-body simulation \citep{2021MNRAS.506.4210I}; this has the same volume as \textsc{AbacusSummit}, but higher resolution, and it includes substructures. We then use the \textsc{Uchuu} posterior distribution to sample Gaussian priors to build the latent catalogs we need as inputs for \ho in the next step.
\item {\bf Level-II inference:} we build high-fidelity mock catalogs by coupling level-I posterior samples (i.e., the latent catalogs above) with our Halo Occupation Model, which depends on a set of physically motivated parameters, \(\boldsymbol{\phi}\). Using these mocks, we forward-model the clustering statistics to constrain \( \boldsymbol{\phi} \).
\end{itemize}

The above separation allows us to decouple the expensive generative step---i.e., producing and assigning the latent satellite variables to the \textsc{AbacusSummit} hosts) from the cosmological inference---at the cost of approximating the marginalization over \( \boldsymbol{\theta} \) using a finite number of latent sample realizations.

\begin{figure*}
\centering
\includegraphics[width=\linewidth]{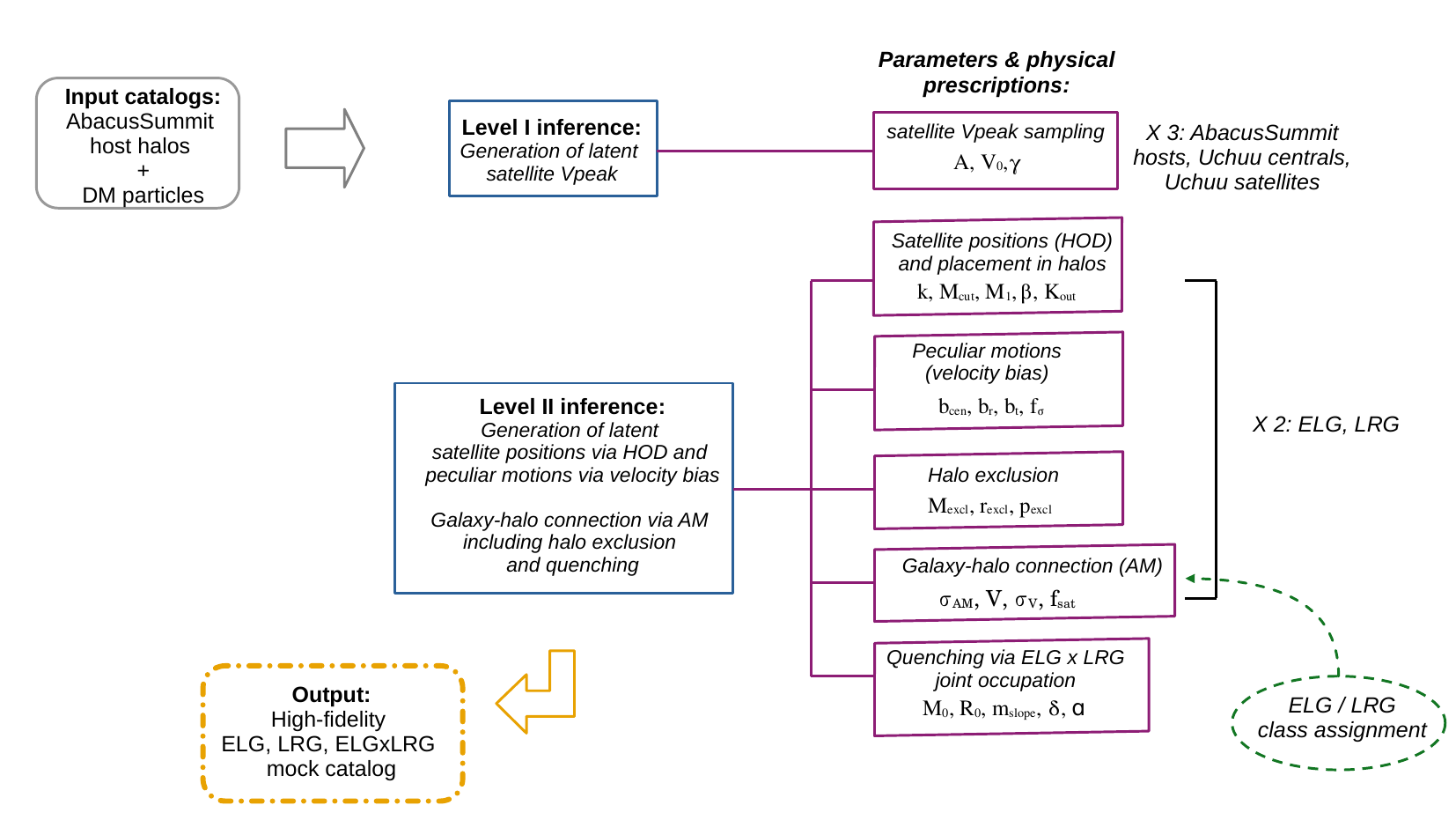}
\caption{Schematic illustration of the \ho workflow. We start from the \textsc{AbacusSummit} input halo and DM particle catalogs and generate the latent satellite $V_{\rm peak}$ values in level-I inference. In level II, we apply the \ho HOD prescription to assign satellite positions from DM particles, and the velocity bias model to generate their peculiar motions. Then, we perform galaxy-halo connection via abundance matching, probabilistically assigning ELG/LRG classes to the halos. Just prior to the matching, in the \ho input catalog we model halo exclusion. We then mimic the environmental quenching of ELG satellites in massive LRG hosts statistically, through the joint occupation condition. Finally, by constraining \ho physical parameters (37 in total) of our level-II inference, we obtain the high-fidelity mock catalog for DESI multi-tracers.}
\label{fig:workflow}
\end{figure*}

The \ho workflow, including all its parameters and the physical processes they drive, is schematically illustrated in Figure\,\ref{fig:workflow}.

\subsection{Posterior formulation}
\label{sec:posterior}
The full posterior distribution over all parameters is:
\begin{equation}
\begin{aligned}
P(\boldsymbol{\phi}, &\boldsymbol{\theta} \mid \mathbf{D}) \propto \\
&P(\mathbf{D} \mid \boldsymbol{\phi}, \mathcal{C}(\boldsymbol{\phi}, \boldsymbol{\theta})) \,
P(\mathcal{C}(\boldsymbol{\phi}, \boldsymbol{\theta}) \mid \boldsymbol{\theta}) \,
P(\boldsymbol{\phi}) \,
P(\boldsymbol{\theta}),
\end{aligned}
\label{eq:posterior}
\end{equation}
where:
\begin{itemize}
    \item \( P(\mathcal{C}(\boldsymbol{\phi}, \boldsymbol{\theta}) \mid \boldsymbol{\theta}) \) encodes the generation of the mock catalog $\mathcal{C}(\boldsymbol{\phi}, \boldsymbol{\theta})$ based on both the nuisance ($\boldsymbol{\theta}$) and the \ho ($\boldsymbol{\phi}$) parameters;
    \item \( P(\mathbf{D} \mid \boldsymbol{\phi}, \mathcal{C}) \) represents the likelihood of the observations, given a forward model evaluated on the mock catalog \( \mathcal{C} \);
    \item \( P(\boldsymbol{\theta}) \) and \( P(\boldsymbol{\phi}) \) are priors on the nuisance and model parameters, respectively.
\end{itemize}

\subsection{Likelihood structure}
\label{sec:likelihoodavg}
Since jointly sampling the posterior in Eq.\,\ref{eq:posterior} is computationally unfeasible, we split the inference in two levels, that we detail here.

\subsubsection{Level-I: nuisance parameter calibration}
\label{sec:levelI}
We group our 9 nuisance parameters, which are defined in \S\,\ref{sec:Vmaxsampling}, in the vector:
\begin{equation}
\boldsymbol{\theta} = \{(A,\,V_0,\,\gamma)^{\rm A}_{\rm h},(A,\,V_0,\,\gamma)^{\rm U}_{\rm cen},
(A,\,V_0,\,\gamma)^{\rm U}_{\rm sat}\}\,,
\label{eq:theta}
\end{equation} 
where the 3 sets of $(A,\,V_0,\,\gamma)$ values---one for the \textsc{AbacusSummit} hosts, and the others for \textsc{Uchuu} central and satellite halos---govern the satellite $V_{\rm peak}$ sampling conditional to their host velocities.
 
To marginalize over $\boldsymbol{\theta}$, we calibrate the analytic prescriptions in \S\,\ref{sec:Vmaxsampling} against external predictions ($\mathcal{D}_{\text{sim}}$, hereafter) from the \textsc{Uchuu} $N$-body simulation \citep{2021MNRAS.506.4210I}; the results are discussed in \S\,\ref{sec:levelIresu}. 

In this way, the analytic prescription used to sample satellite $V_{\rm peak}$ values in \textsc{AbacusSummit}, that shapes our galaxy-halo connection model, is informed by precise DM distribution, ensuring that the latent satellite variables are consistent with the gravitational dynamics. 
 
As a result, we infer the level-I posterior distribution:
\begin{equation}
    P(\boldsymbol{\theta} \mid \mathcal{D}_{\text{sim}})\propto P(  \mathcal{D}_{\text{sim}} \mid \boldsymbol{\theta})\,P(\boldsymbol{\theta})\,,
    \label{eq:levelIposterior}
\end{equation}
from which we sample \( N=30\) sets of parameters \( \{ \boldsymbol{\theta}_i \}_{i=1}^{N} \) that we use as Gaussian priors to generate corresponding realizations of the latent catalog (see below).


\subsubsection{Level-II: forward modeling and likelihood averaging over latent realizations}
\label{sec:levelII}
In the second stage of our inference pipeline, we construct high-fidelity mock galaxy catalogs, denoted as $\mathcal{C}_i(\boldsymbol{\phi}, \boldsymbol{\theta}_i)$, by coupling the $N$ latent realizations \( \{ \boldsymbol{\theta}_i \}_{i=1}^{N} \) obtained from level I with our physically motivated \ho prescriptions driven by 37 parameters grouped in the vector:
\begin{equation}
\begin{aligned}
\boldsymbol{\phi} =&\{(\kappa,\,M_{\rm cut},\,M_1,\,\beta)^{\rm elg},\,K_{\rm out}^{\rm elg},\,(b_{\rm cen},\,b_r,\,b_t,\,f_\sigma)^{\rm elg},\\
&(\sigma_{\rm AM},\,V,\,\sigma_{V})^{\rm elg},\,f^{\rm elg}_{\rm sat},\,(M_{\rm excl}, \,r_{\rm excl},\,p_{\rm excl})^{\rm elg},\\
&(\kappa,\,M_{\rm cut},\,M_1,\,\beta)^{\rm lrg},\,K_{\rm out}^{\rm lrg},\,(b_{\rm cen},\,b_r,\,b_t,\,f_\sigma)^{\rm lrg},\\
&(\sigma_{\rm AM},\,V,\,\sigma_{V})^{\rm elg},\,f^{\rm lrg}_{\rm sat},\,(M_{\rm excl}, \,r_{\rm excl},\,p_{\rm excl})^{\rm lrg},\\
&\,(M_0,\,R_0,\,m_{\rm slope},\,\delta,\,\alpha)\}\,,
\label{eq:phi}
\end{aligned}
\end{equation}
where, for both tracers, we have:
\begin{itemize}
\item $(\kappa,\,M_{\rm cut},\,M_1,\,\beta):$ HOD parameters determining the number of satellites assigned per host at their DM particle locations (see \S\,\ref{sec:positions});
   \item $K_{\rm out}:$ regulates the spatial distribution of satellites within halos (\S\,\ref{sec:positions});
    \item $(b_{\rm cen},\,b_{\rm r},\,b_{\rm t},\,f_\sigma):$ control central and satellite peculiar motions, modulating redshift-space distortions (\S\,\ref{sec:centrals}, \ref{sec:vpec});
    \item $(\sigma_{\rm AM}$,\,$V$,\,$\sigma_{V})$: parametrize abundance matching  (\S\,\ref{sec:AM});
    \item $f_{\rm sat}:$ normalize the final halo occupation based on the observed number density of tracers (\S\,\ref{sec:AM}).
    \item $(M_{\rm excl},\,r_{\rm excl},\,p_{\rm excl}):$ drive the halo exclusion mechanism in massive tracers (\S\,\ref{sec:halo_exclusion});
    \end{itemize}
and for the joint occupation condition:
\begin{itemize}
\item $(M_0,\,R_0,\,m_{\rm slope},\,\delta,\,\alpha):$ regulate the coexistence of ELGs and LRGs in the same halos mimicking the effect of environmental quenching (\S\,\ref{sec:jointoccup}).
\end{itemize}

To constrain the full parameter space of $\boldsymbol{\phi}$, we explore its posterior distribution using MCMC sampling. For each proposed set of $\boldsymbol{\phi}$, we generate $N=30$ mock realizations $\mathcal{C}_i(\boldsymbol{\phi}, \boldsymbol{\theta}_i)$ by applying \ho to each of the 30 latent halo configurations $\boldsymbol{\theta}_i$.
The model likelihood is then approximated by averaging over the $N$ mocks as:
\begin{equation}
P(\boldsymbol{\phi} \mid \mathbf{D}) \propto P(\boldsymbol{\phi}) \cdot \frac{1}{N} \sum_{i=1}^{N} P(\mathbf{D} \mid \mathcal{C}_i(\boldsymbol{\phi}, \boldsymbol{\theta}_i))\,,
\label{eq:likelihood_avg}
\end{equation}
where each term in the average likelihood above is a Gaussian likelihood. 

Assuming the data are divided in $N_{\rm b}$ bins, normally distributed around the model prediction $\mathcal{C}_i$, and that the total uncertainty---encompassing both data (d) and mock (m) contributions---is captured by the covariance matrix $\mathbf{C}_{\text{tot}}$ (see \S\,\ref{sec:measurements} for details on its computation),
\begin{equation}
\mathbf{C}_{\rm tot}=\mathbf{C}_{\rm d}+\mathbf{C}_{\rm m}\,,
\end{equation}
each individual likelihood in Eq.\,\ref{eq:likelihood_avg} takes the form:
\begin{equation}
\begin{aligned}
&\log P\left(\mathbf{D} \mid \mathcal{C}_i(\boldsymbol{\phi}, \boldsymbol{\theta}_i)\right) =\\
&- \frac{1}{2} \left[ \mathbf{D} - \mathcal{C}_i(\boldsymbol{\phi}, \boldsymbol{\theta}_i) \right]^T \mathbf{C}_{\text{tot}}^{-1} \left[ \mathbf{D} - \mathcal{C}_i(\boldsymbol{\phi}, \boldsymbol{\theta}_i) \right] \\
&- \frac{1}{2} \log \det \mathbf{C}_{\text{tot}} - \frac{N_{\rm b}}{2} \log(2\pi)\,,
\end{aligned}
\end{equation}
where $N_{\rm b}=26$ is the number of spatial bins we use to measure the clustering statistics $\mathbf{D}$ (see \S\,\ref{sec:measurements}).

This Monte Carlo marginalization provides a tractable way to account for uncertainty in the latent halo properties \( \boldsymbol{\theta}_i \) without regenerating them for every step in parameter space. As a result, we are able to propagate cosmological and nuisance uncertainties from the halo level to the final galaxy statistics in an efficient and robust way.


\section{Clustering measurements and fits} 
\label{sec:measurements}
We employ the two-point correlation function (2PCF) measurements of the DESI Y1 \texttt{ELG1}, \texttt{LRG3} and \texttt{ELG1$\times$LRG3} samples \citep{2025JCAP...01..125R,2025JCAP...07..017A}. Specifically, we consider the first two even multipoles, $\xi_{l=0,\,2}(s)$, computed as \citep[][]{2013MNRAS.431.2634C}:
\begin{equation}
\xi_l(s)=\frac{2l+1}{2}\int_{-1}^{1} \xi(s,\mu)\,L_l(\mu)\,d\mu\,,
\label{eq:multipoles}
\end{equation}
where $L_l$ is the $l-$th order Legendre polynomial, and the projected function \citep{1983ApJ...267..465D}:
 \begin{equation}
w_{\rm p}(r)=2\int_0^{\rm \pi_{\rm max}} \xi(r_{\rm p},\pi)\,d\pi\,.
\label{eq:projected}
\end{equation}
The correlation functions $\xi(s,\mu)$ in Eq.\,\ref{eq:multipoles} and $\xi(r_{\rm p},\pi)$ in Eq.\,\ref{eq:projected} are evaluated in $N_{\rm b}=26$ logarithmic bins of $s$ $(r_{\rm p})$, where $s=\sqrt{r_{\rm p}^2+\pi^2}$, between $0.17\,h^{-1}$Mpc and $32\,h^{-1}$Mpc, 200 linear bins of $\mu$ in $[-1,1]$, and 40 linear bins of $\pi$ between zero and $\pi_{\rm max}=40\,h^{-1}$Mpc.  

Together with the DESI Y1 measurements, we adopt the jackknife covariance matrices ($\mathbf{C}_{\rm d}$) estimated from $N_{\rm res}=128$ re-samplings on the Y1 survey footprint \citep{2025JCAP...01..125R}. 

Using the same binning scheme, we calculate the multipole and projected 2PCFs of the \ho high-fidelity mock catalog using the \textsc{FCFC} code \citep{2023A&A...672A..83Z} coupled with the natural estimator \citep{1983ApJ...267..465D}.

To estimate the mock covariances ($\mathbf{C}_{\rm m}$), we impose the \ho best-fit configuration to the available $N_{\rm r}=1800$ realizations of the small \textsc{AbacusSummit} boxes (\S\,\ref{sec:simulation}), and compute corresponding 2PCFs for each tracer. The model covariance for each tracer is then computes as:
\begin{equation}
\mathbf{C}_{\rm m} = \frac{1}{N_{\rm res} - 1} \sum_{i=1}^{N_{\rm res}} \left( \mathbf{X}_{\rm m}^{(i)} - \bar{\mathbf{X}}_{\rm m} \right) \left( \mathbf{X}_{\rm m}^{(i)} - \bar{\mathbf{X}}_{\rm m} \right)^{\rm T}\,,
\label{eq:covformula}
\end{equation}
where $\mathbf{X}_{\rm m} = (\xi_0(s),\,\ \xi_2(s),\,\ w_{\rm p}(r_{\rm p}))$ is the grouped vector of the model clustering statistics, stacked on the same binning used in the likelihood, and $\bar{\mathbf{X}}_{\rm m}$ its average over the $N_{\rm res}=1800$ realizations given by:
\begin{equation}
\bar{\mathbf{X}}_{\rm m} = \frac{1}{N_{\rm res}} \sum_{i=1}^{N_{\rm res}} \mathbf{X}_{\rm m}^{(i)}\,.
\end{equation}

We jointly fit the monopole, quadrupole and projected correlation functions of each galaxy tracer under study (i.e., \texttt{ELG1}, \texttt{LRG3}, and \texttt{ELG1$\times$LRG3}), and estimate its $\chi^2$ as \citep[e.g.,][]{2021MNRAS.505.5833F,2023JCAP...10..016R}:
\begin{equation}
    \chi^{2}=(\mathbf{X}_{\rm d}-\mathbf{X}_{\rm m})^T\,\mathbf{\hat{\Psi}}\,(\mathbf{X}_{\rm d}-\mathbf{X}_{\rm m})\,,
    \label{eq:chi2}\,
\end{equation}
where $\mathbf{\hat{\Psi}}$ is the total assembled precision matrix:
\begin{equation}
    \mathbf{\hat{\Psi}}=\left[\frac{\mathbf{C}_{\rm d}}{(1-H_{\rm d})}+\frac{\mathbf{C}_{\rm m}}{(1-H_{\rm m})}\right]^{-1}\,.
    \label{eq:precisionm}\,
\end{equation}
Here, $\mathbf{X}_{\rm m\,(d)}$ is the grouped vector of the model (data) clustering statistics, and $H_{\rm d\,(m)}$ their Hartlap factors computed as \citep{2007A&A...464..399H}: 
\begin{equation}
    H_{\rm d\,(m)}=(N_{\rm res}-N_{\rm b}-2)/(N_{\rm res}-1)\,,
\end{equation}
with $N_{\rm b}=26$, and $N_{\rm res}=128\,(1800)$ in $H_{\rm d}\,(H_{\rm m})$.

Following \citet{2023JCAP...10..016R} and \cite{2024MNRAS.530..947Y}, in the likelihood estimation we reduce the noise in the jackknife covariances by replacing their off-diagonal terms with those of the covariances from the 1800 small \textsc{AbacusSummit} mocks (see \S\,\ref{sec:simulation}). In this way, the resulting \ho covariances are more stable, their inversion is well-behaved, and the likelihood estimation more robust.


\section{Results}\label{sec:results}
In what follows, we present the results of our DESI Y1 ELG and LRG clustering analysis structured as a two-level Bayesian inference scheme (see \S\,\ref{sec:inference}). 

First, we discuss level-I results, that is the validation of the posterior distributions of the nuisance parameters through independent $N$-body predictions (\S\,\ref{sec:levelIresu}). 

Then, we present and discuss our main findings from level-II inference: the \ho galaxy clustering predictions (\S\,\ref{sec:clustering}) and halo occupation distribution (\S\,\ref{sec:hodpred}) for DESI Y1 galaxy multi-tracers.

\subsection{Nuisance parameter posterior validation}
\label{sec:levelIresu}

\begin{deluxetable}{ccccccc}
\tablecaption{\textit{Top:} best-fit parameters \label{tab:klypinpar} of the \textsc{AbacusSummit} host $V_{\rm peak}$ functions at the redshift snapshots of interest, compared with the \textsc{Uchuu} central and satellite results shown in Figure\,\ref{fig:VFfits}. The analytic function employed for fitting is given in Eq.\,\ref{eq:klypincen}. Next to each parameter, we report the Gaussian priors ($\mathcal{N}$) used to explore the posterior distributions shown in Figure\,\ref{fig:VFposterior}.}
\tablehead{
   \colhead{$z_{\rm snap}$}&\colhead{halo type} &\colhead{$A\,(h^{-1} {\rm M}_\odot/s^{-1}\,{\rm km})^{-3}$}& \colhead{$V_0\,(s^{-1}\rm km)$}& \colhead{$\gamma$}& \colhead{$\chi^2/{\rm dof}$}& \colhead{dof}
  }
\startdata
\multicolumn{1}{c}{$z_{\rm A}$}&\multicolumn{6}{c}{{\bf \textsc{AbacusSummit}}}\\
$0.8$& hosts &$52804^{+21}_{-21}\,\,\,\mathcal{N}(60000,15000)$&$407.38_{-0.15}^{+0.14}\,\,\,\mathcal{N}(550,200)$&$1.418_{-0.001}^{+0.001}\,\,\,\mathcal{N}(1.0,0.5)$&1.79&25 \\
\hline
\multicolumn{1}{c}{$z_{\rm U}$}&\multicolumn{6}{c}{{\bf Uchuu}}\\
0.78& satellites &$14585^{+10}_{-10}\,\,\,\mathcal{N}(18000,8000)$&$297.36^{+0.09}_{-0.09}\,\,\,\mathcal{N}(450,200)$&$1.777^{+0.002}_{-0.002}\,\,\,\mathcal{N}(1.5,0.5)$&2.05&21\\
0.78& centrals &$39831^{+16}_{-16}\,\,\,\mathcal{N}(40000,10000)$&$497.76^{+0.16}_{-0.16}\,\,\,\mathcal{N}(550,200)$&$1.665^{+0.001}_{-0.001}\,\,\,\mathcal{N}(1.5,0.5)$&2.19&25\\
\enddata
\end{deluxetable} 

We fit the abundance formula in Eq.\,\ref{eq:klypincen} to the $V_{\rm peak}$ function of the \textsc{AbacusSummit} central hosts at $z_{\rm A}=0.8$. Then, we fit the same formula to the $V_{\rm peak}$ functions of the central and satellite halos in the \textsc{Uchuu} $N$-body simulation \citep{2021MNRAS.506.4210I} at $z_{\rm U}=0.78$ (see Table\,\ref{tab:zrep}). From the \textsc{Uchuu} fits we calculate the correction factor (Eq.\,\ref{eq:correction}), which we use to build the CDF to sample $V_{\rm peak}$ values for the \textsc{AbacusSummit} satellites.

For consistency with \textsc{AbacusSummit}, we impose \textsc{Uchuu} a minimal $M_{\rm h}^{\rm vir}\ge2.1\times 10^{10}\,h^{-1}{\rm M}_{\odot}$ cut, corresponding to $N_{\rm p}\ge65$ particles (\S\,\ref{sec:simulation}). 

\begin{figure}
\centering
\includegraphics[width=\linewidth]{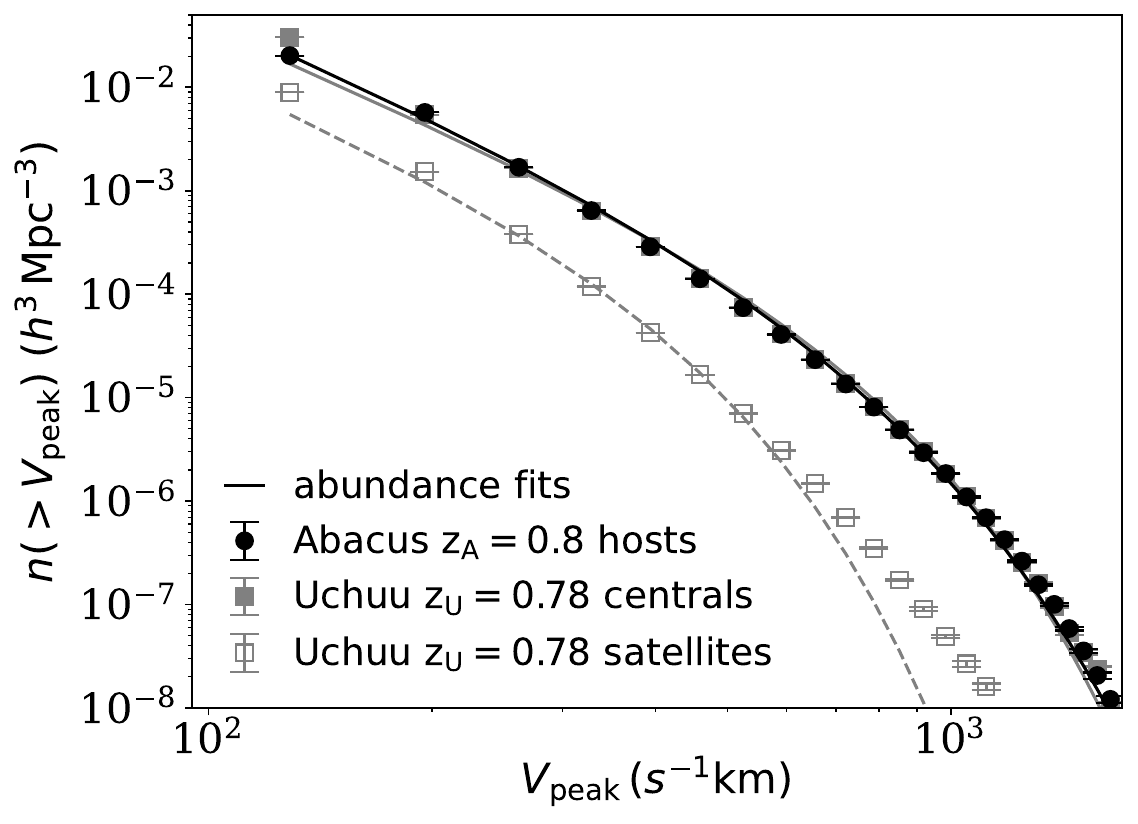}
\caption{$V_{\rm peak}$ functions (markers), and corresponding best fits (lines) based on Eq.\,\ref{eq:klypincen}, for the \textsc{AbacusSummit} host halos (black dots), compared to the \textsc{Uchuu} central (grey full squares almost perfectly overlapping with the black dots) and satellite (grey empty squares) halos at the fiducial redshift. For \textsc{Uchuu} the closest snapshot to $z_{\rm A}=0.8$ is $z_{\rm U}=0.78$. The uncertainties are computed from 30 bootstrap re-samplings. All the results are normalized to the simulation volume, that is $(2\,h^{-1}$Gpc)$^3$ for both \textsc{AbacusSummit} and Uchuu. The optimal parameters are reported in Table\,\ref{tab:klypinpar}. The posterior distributions are shown in Figure\,\ref{fig:VFposterior}}.
\label{fig:VFfits}
\end{figure}

Figure\,\ref{fig:VFfits} compares the \textsc{AbacusSummit} and \textsc{Uchuu} $V_{\rm peak}$ functions (note that the central contributions in both simulations almost perfectly overlap) with their best fits, whose optimal parameters are in Table\,\ref{tab:klypinpar}.

\begin{figure}
\centering
\includegraphics[width=0.85\linewidth]{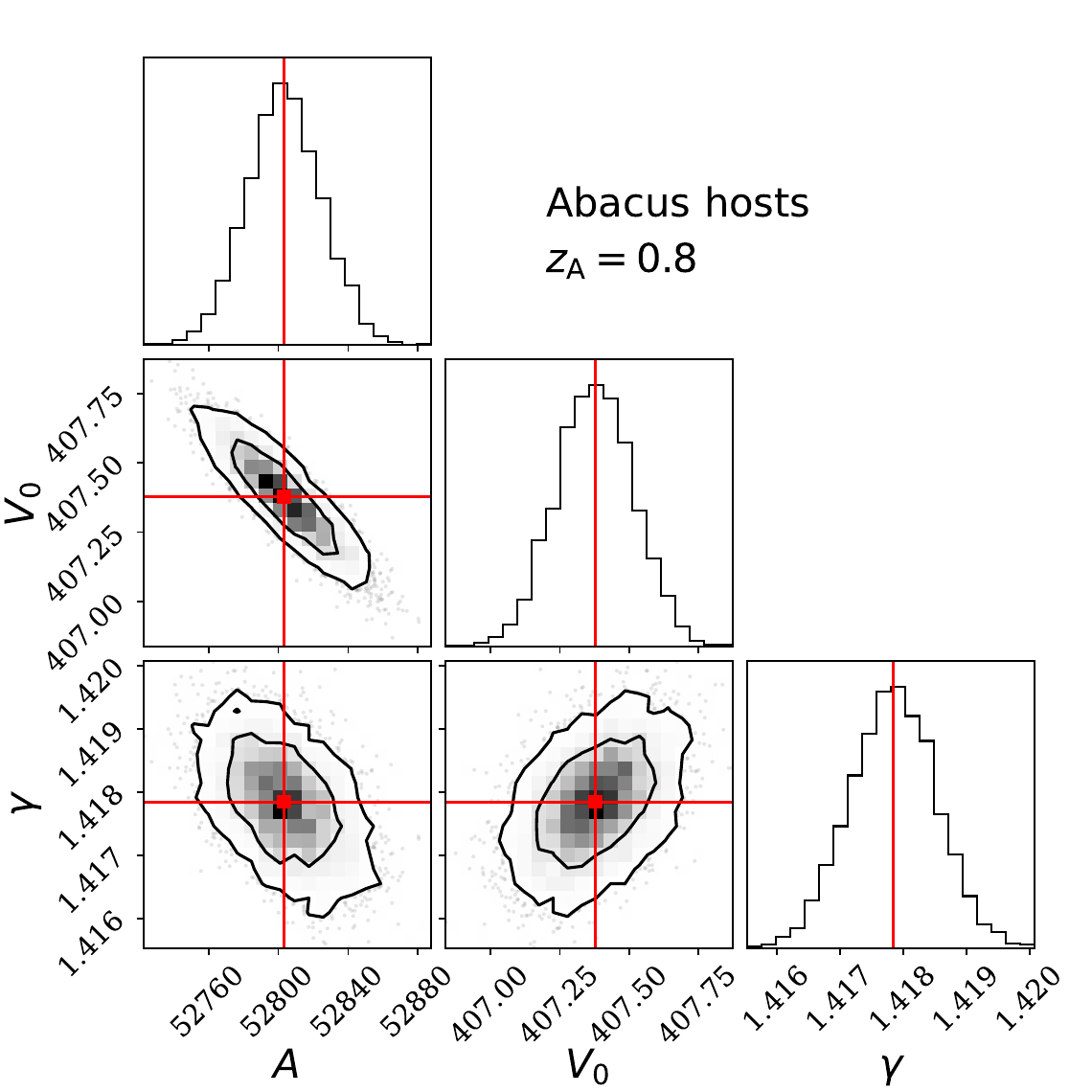}
\includegraphics[width=0.85\linewidth]{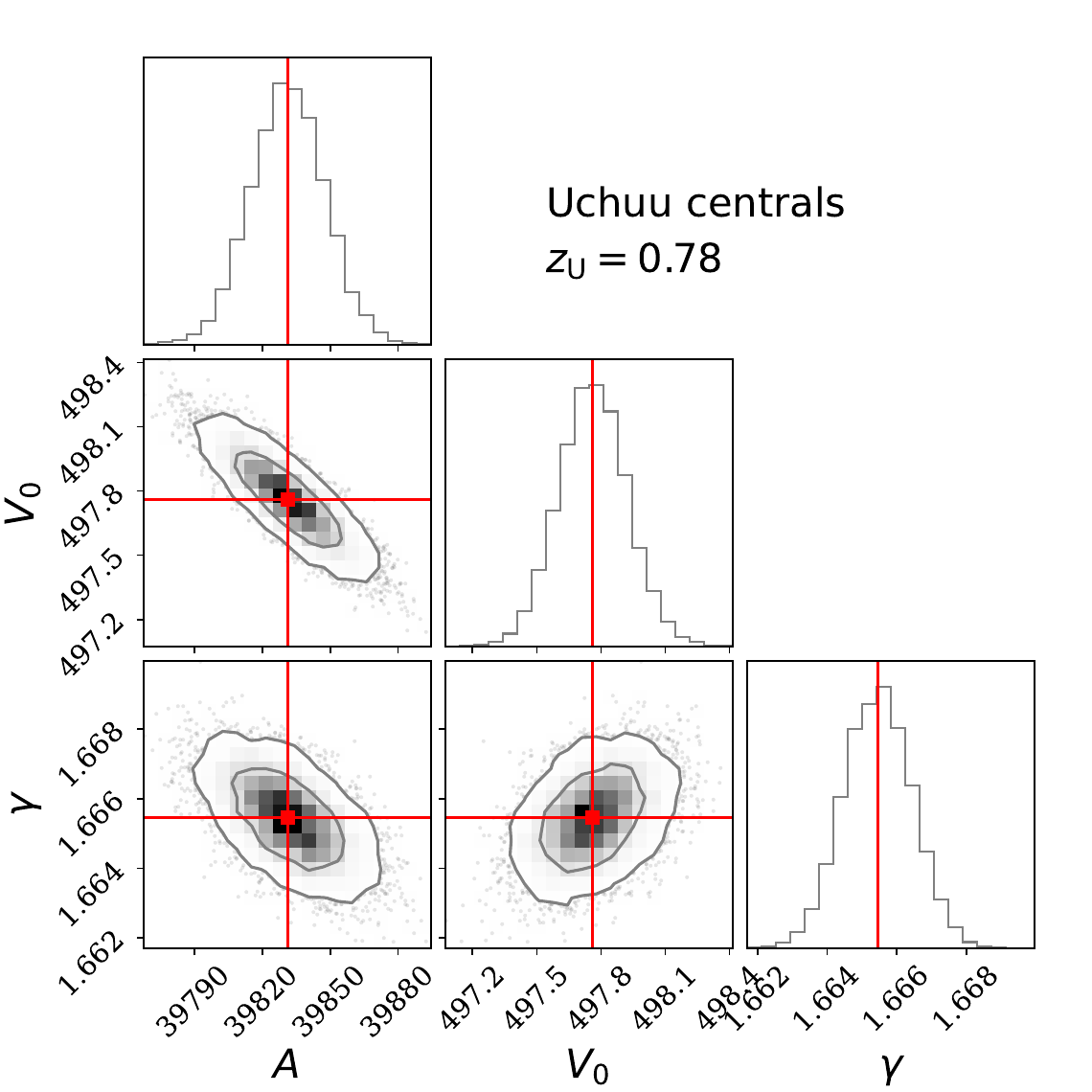}
\includegraphics[width=0.85\linewidth]{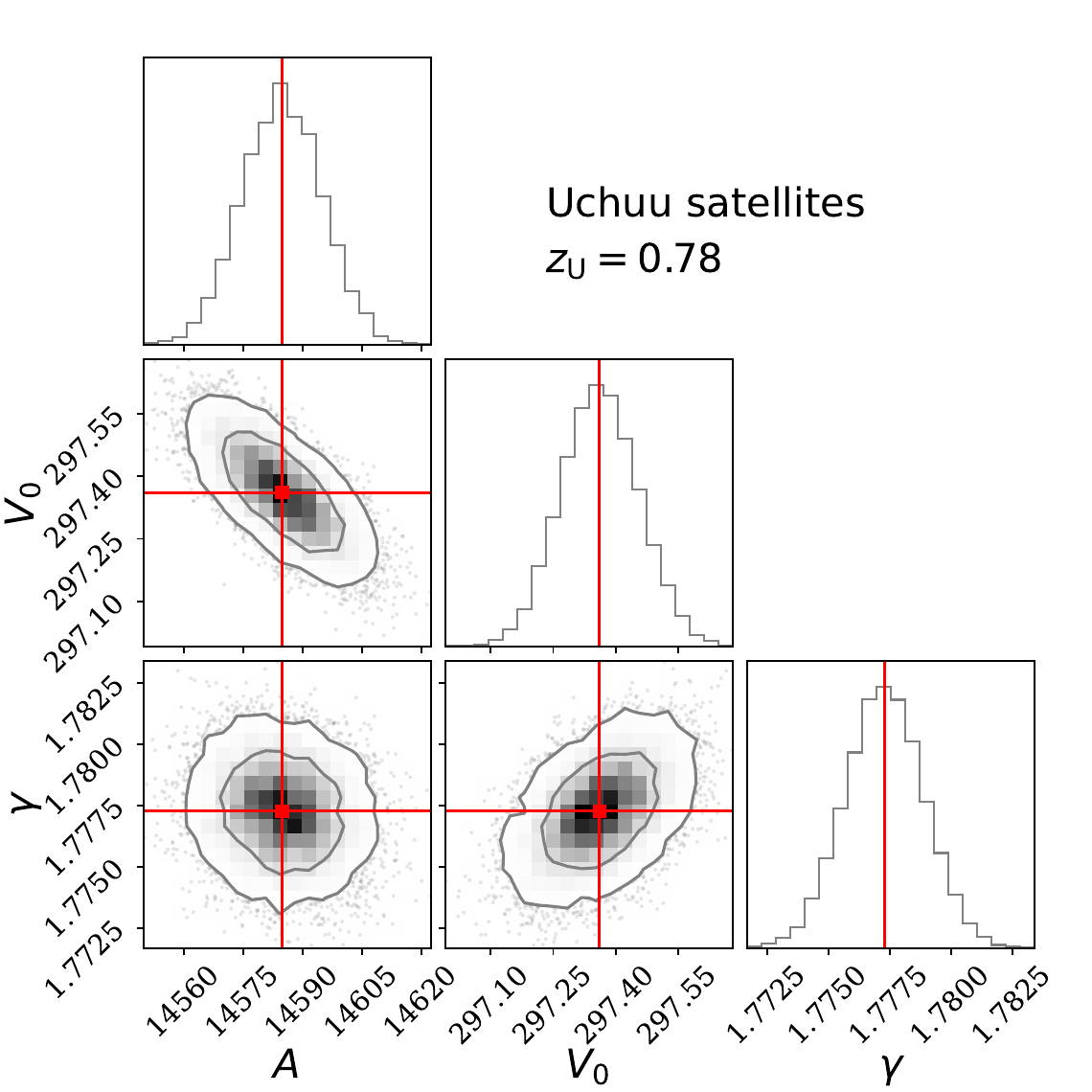}
\caption{Posterior distributions of the halo abundance fit in Eq.\,\ref{eq:klypincen} to the \textsc{AbacusSummit} hosts (top panel), \textsc{Uchuu} central (middle) and satellite (bottom) halos, as shown in Figure\,\ref{fig:VFfits}. The optimal parameters are reported in Table \ref{tab:klypinpar}.}
\label{fig:VFposterior}
\end{figure}

We employ an emcee sampler to explore the full posterior distributions of both simulations; the results are presented in Figure\,\ref{fig:VFposterior}.

From these posteriors, we sample Gaussian priors and use them to build the latent satellite catalogs needed as inputs for \ho in the second level of our inference process (\S\,\ref{sec:levelII}). This approach enables us to validate the physical plausibility and predictive power of the nuisance parameters in Eq.\,\ref{eq:theta}.

\subsection{\ho predictions} 
\label{sec:levelIIresu}
In what follows we present our main findings: the galaxy clustering and HOD predictions for the DESI Y1 \texttt{ELG1}, \texttt{LRG3}, and \texttt{ELG1$\times$LRG3} samples.

\subsubsection{Galaxy clustering results} 
\label{sec:clustering}
\begin{figure}
\centering
\includegraphics[width=\linewidth]{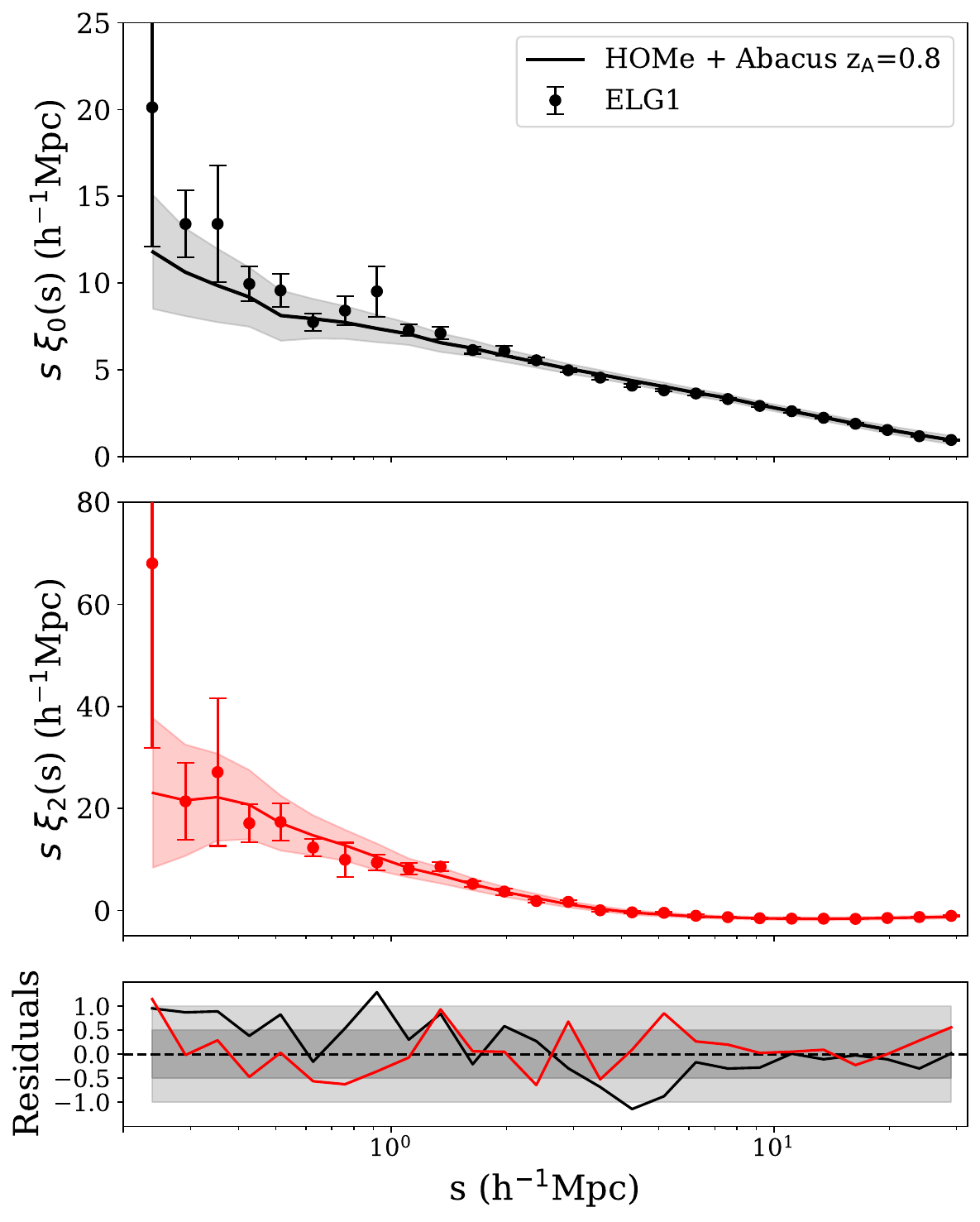}\\
\includegraphics[width=\linewidth]{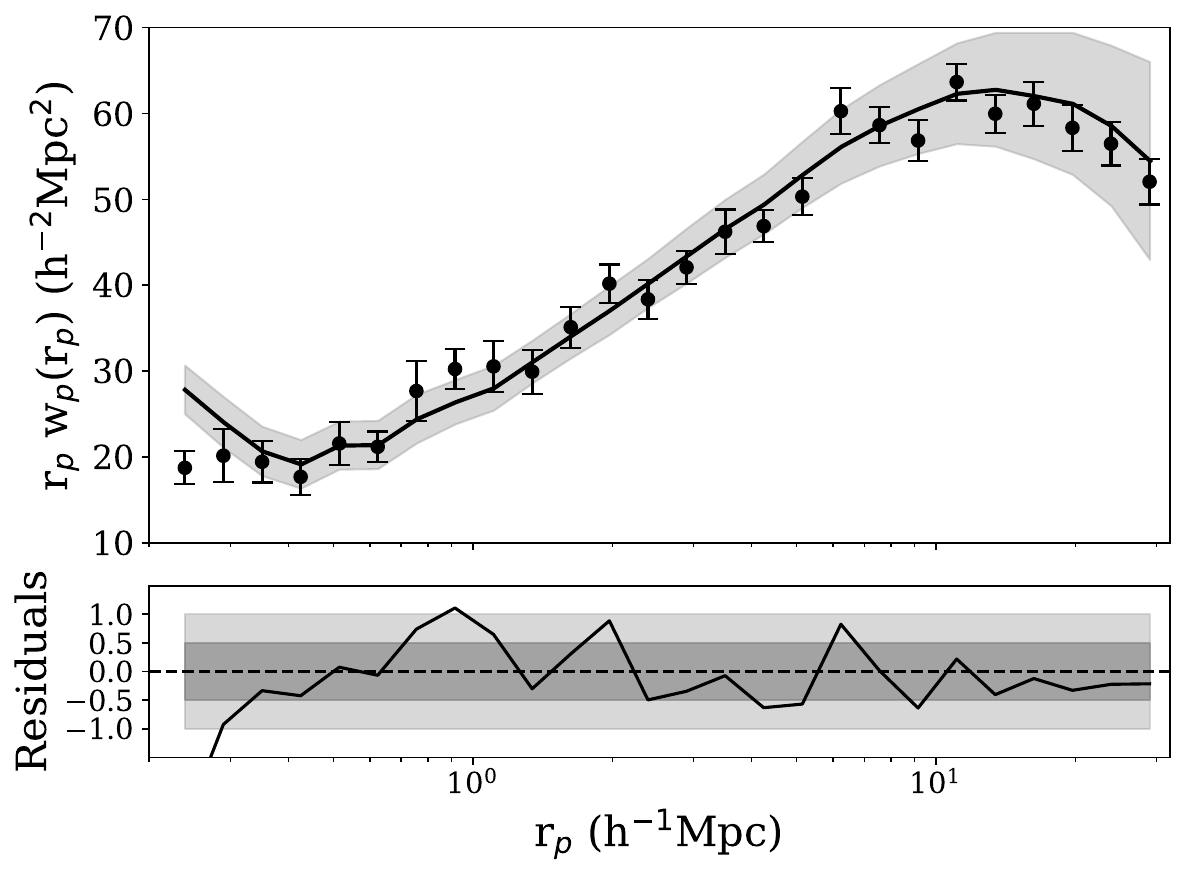}
\caption{\texttt{ELG1} monopole (\textit{top panel}, black markers), quadrupole (\textit{top panel}, red) and projected (\textit{bottom panel}, black) auto-correlation functions, compared with the \ho high-fidelity mock (lines color-coded as the markers). The observational uncertainties are computed from 128 jackknife re-samplings on the data; those on the models from 1800 \textsc{AbacusSummit} small boxes (details in the text). The lower panel below of each set of figures shows the residuals between data (d) and models (m), computed as $\rm{Residuals} = (\rm{d-m})/\sqrt{\sigma_{\rm d}^2+\sigma_{\rm m}^2}$. The grey shades represent the 5\% (inner) and 10\%  (outer) confidence regions. The best-fit parameters are reported in Table\,\ref{tab:clusteringresu}.}
\label{fig:clusteringELG}
\end{figure}

\begin{figure}
\centering
\includegraphics[width=\linewidth]{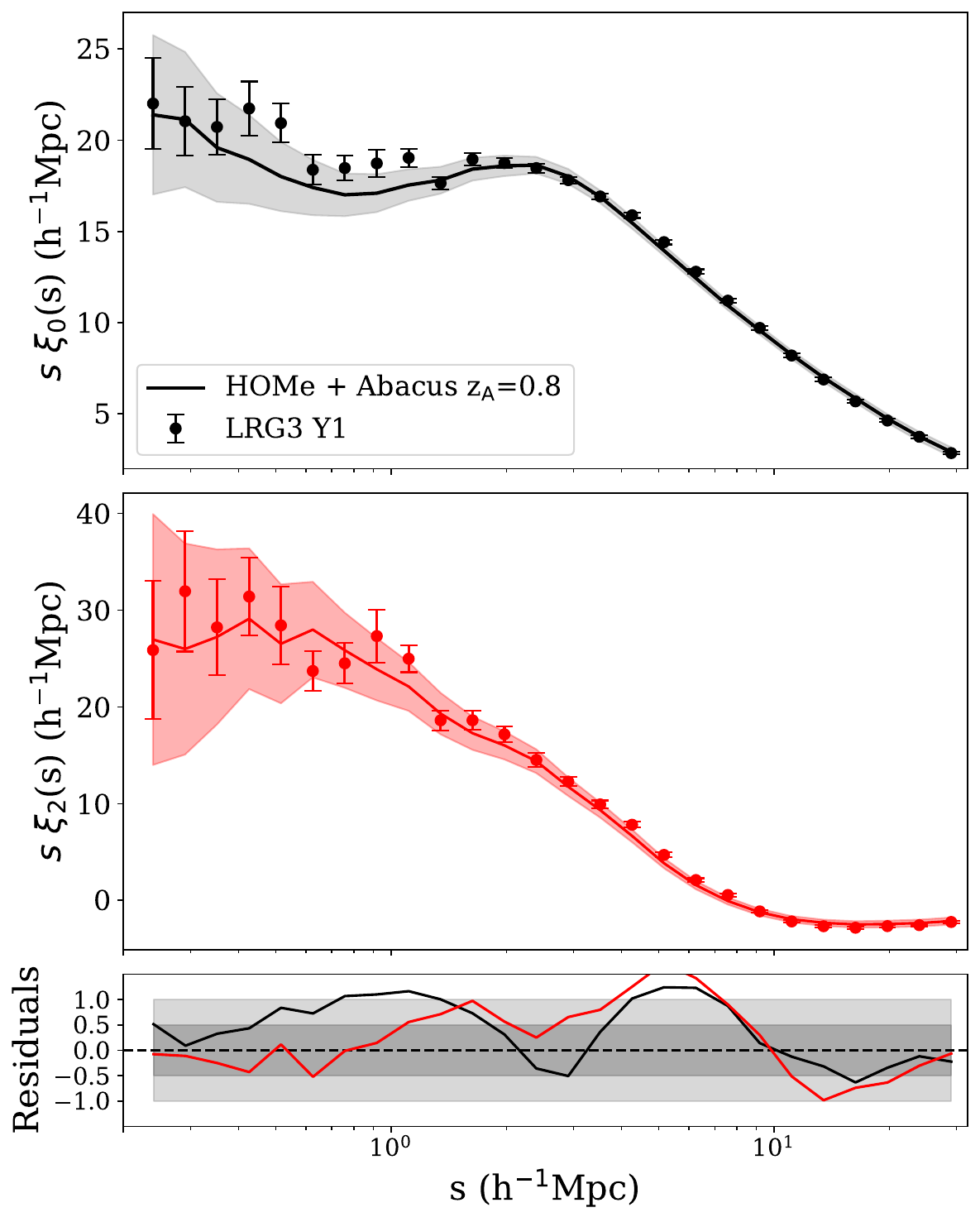}\\
\includegraphics[width=\linewidth]{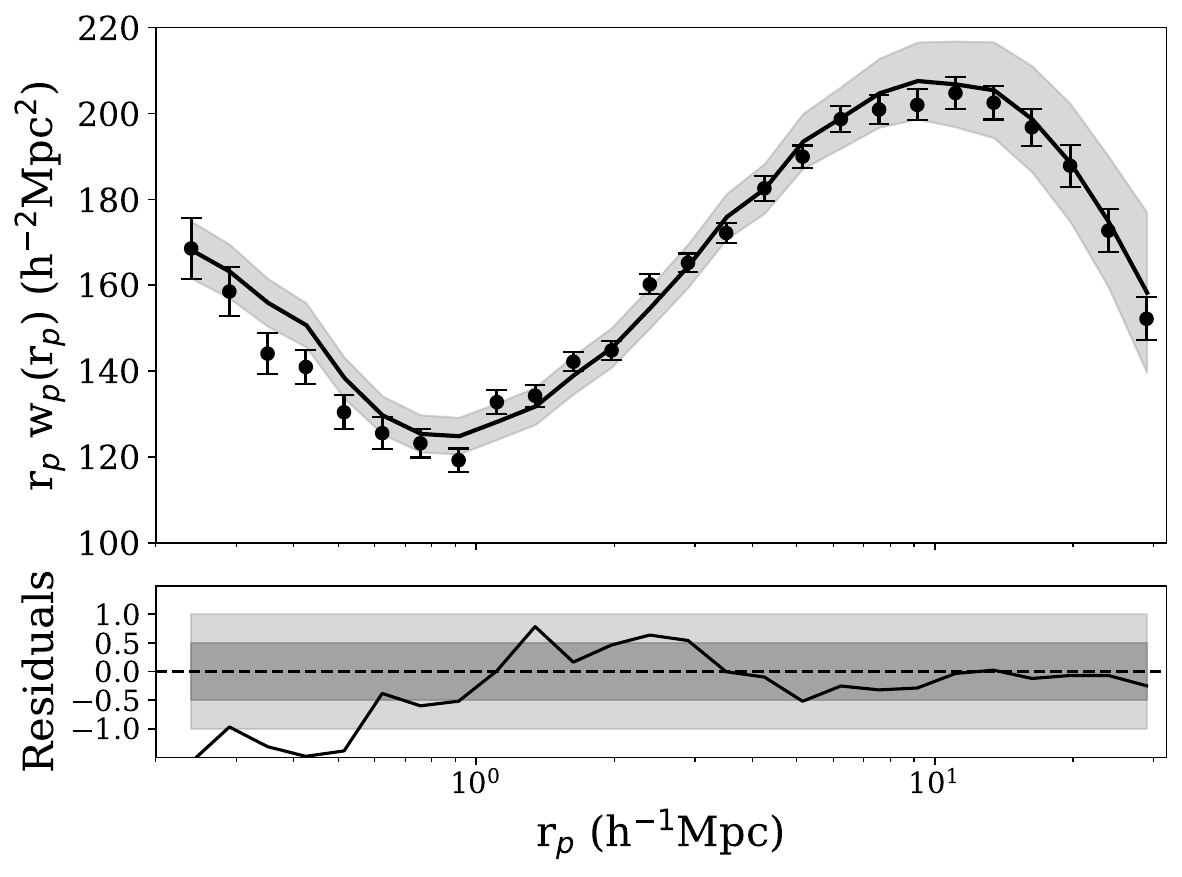}
\caption{Same result as Figure\,\ref{fig:clusteringELG}, but for \texttt{LRG3}.}
\label{fig:clusteringLRG}
\end{figure}

\begin{figure}
\centering
\includegraphics[width=\linewidth]{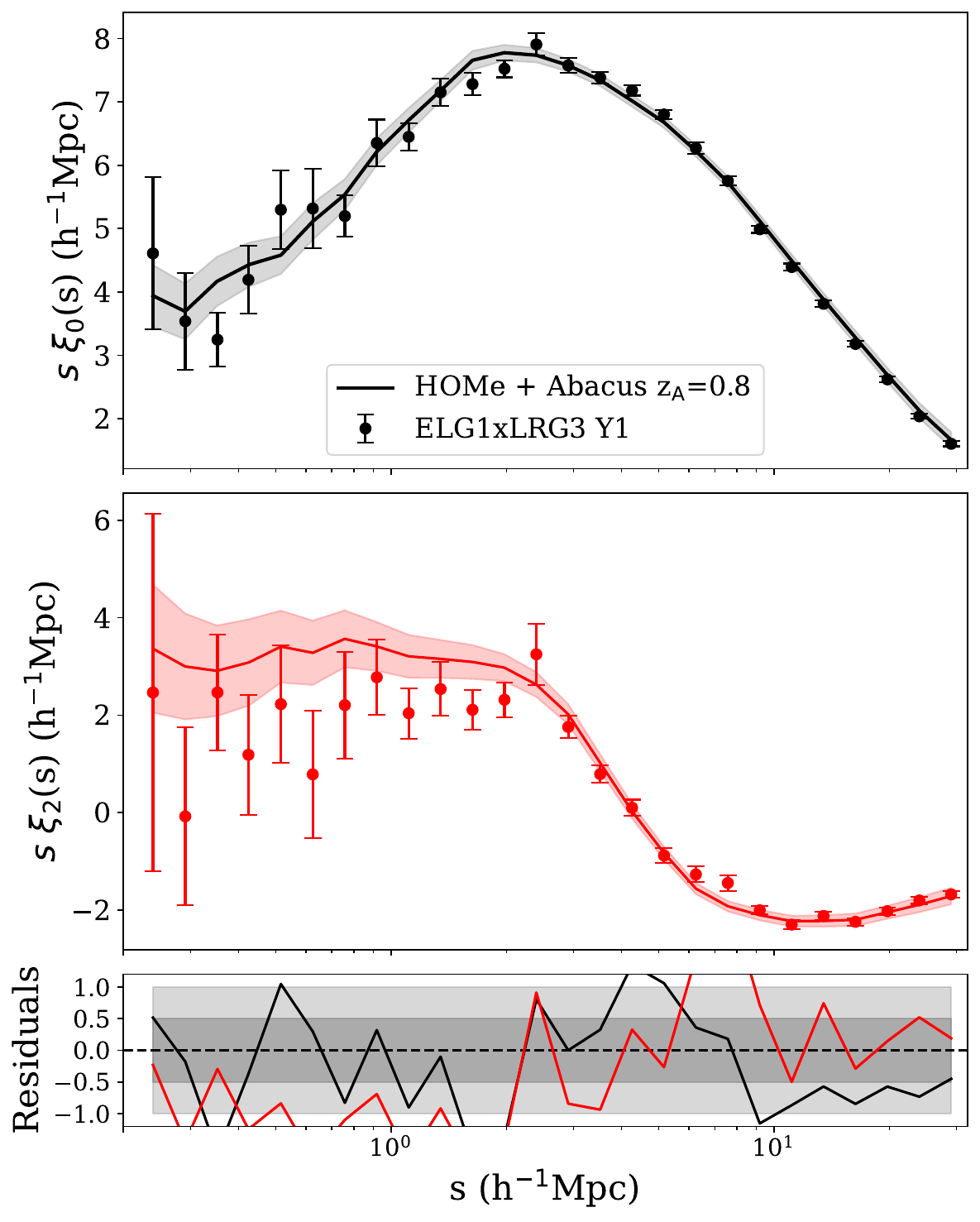}\\
\includegraphics[width=\linewidth]{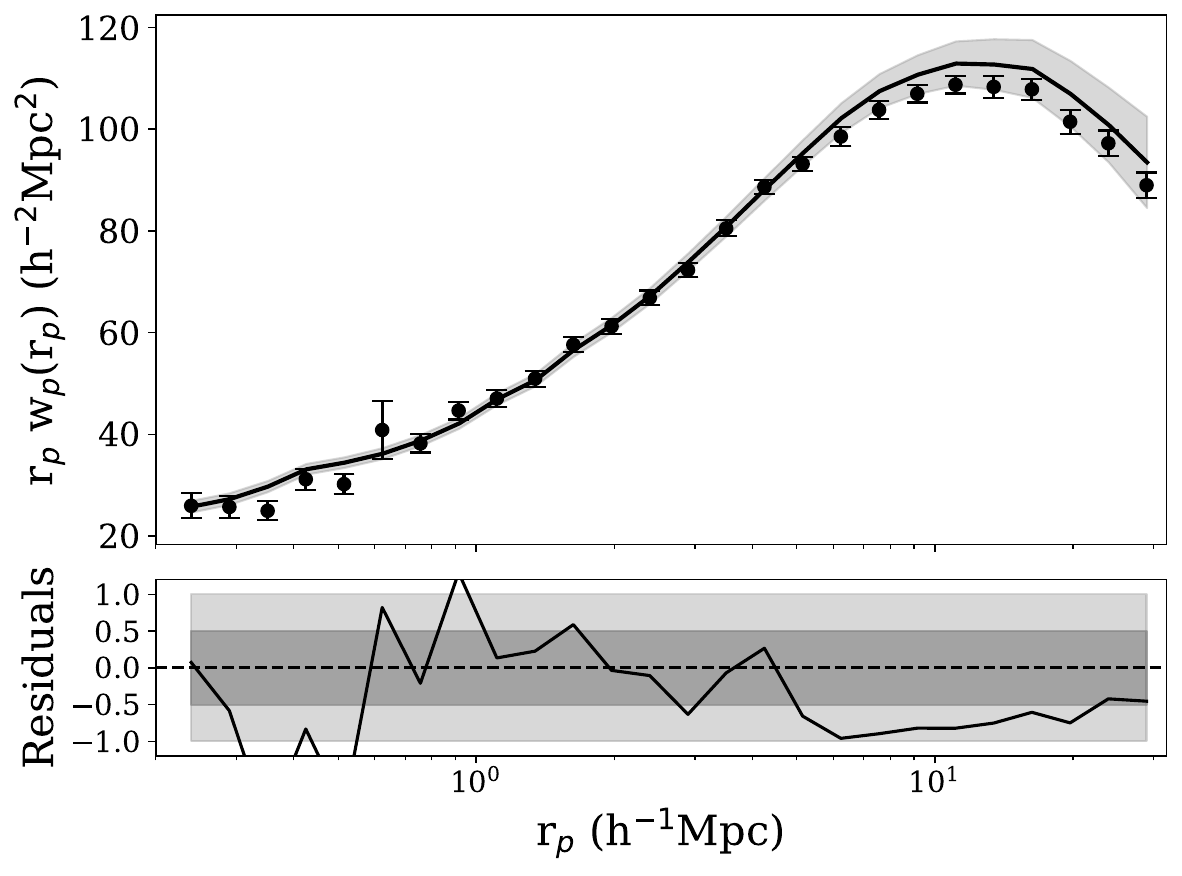}
\caption{Same result as Figures\,\ref{fig:clusteringELG}-\ref{fig:clusteringLRG}, but for \texttt{ELG1$\times$LRG3}.}
\label{fig:clusteringELGLRG}
\end{figure}

Figures\,\ref{fig:clusteringELG}, \ref{fig:clusteringLRG}, and \ref{fig:clusteringELGLRG} compare the DESI Y1 \texttt{ELG1}, \texttt{LRG3}, \texttt{ELG1$\times$LRG3} monopole, quadrupole and projected correlation functions with our \ho high-fidelity mock. 

\emph{We emphasize that a single model run outputs a unique  mock catalog simultaneously fitting the ELG and LRG auto- and cross-correlation functions, both multipoles and projected.}

\begin{figure*}
\centering
\includegraphics[width=0.48\linewidth]{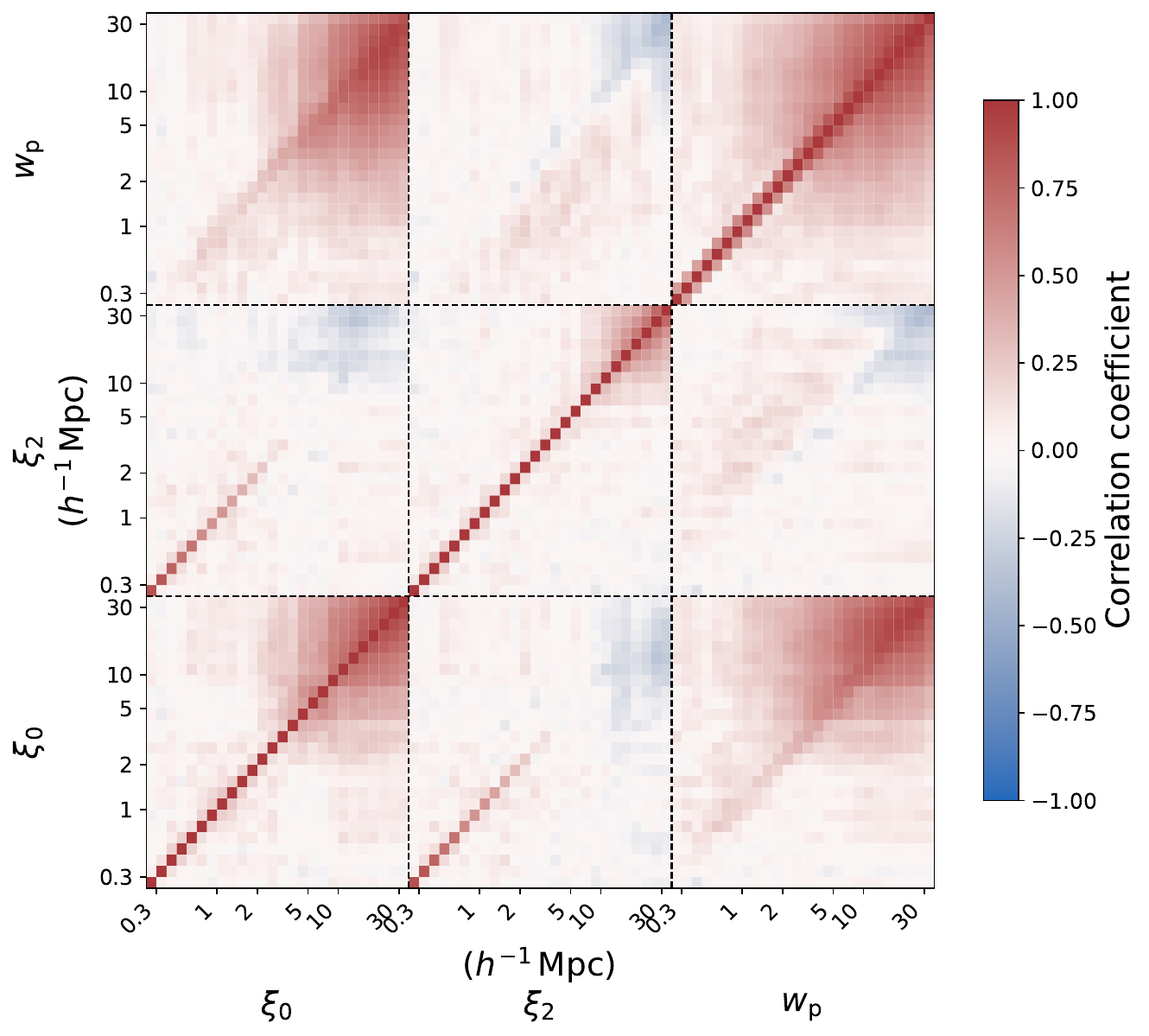}
\includegraphics[width=0.48\linewidth]{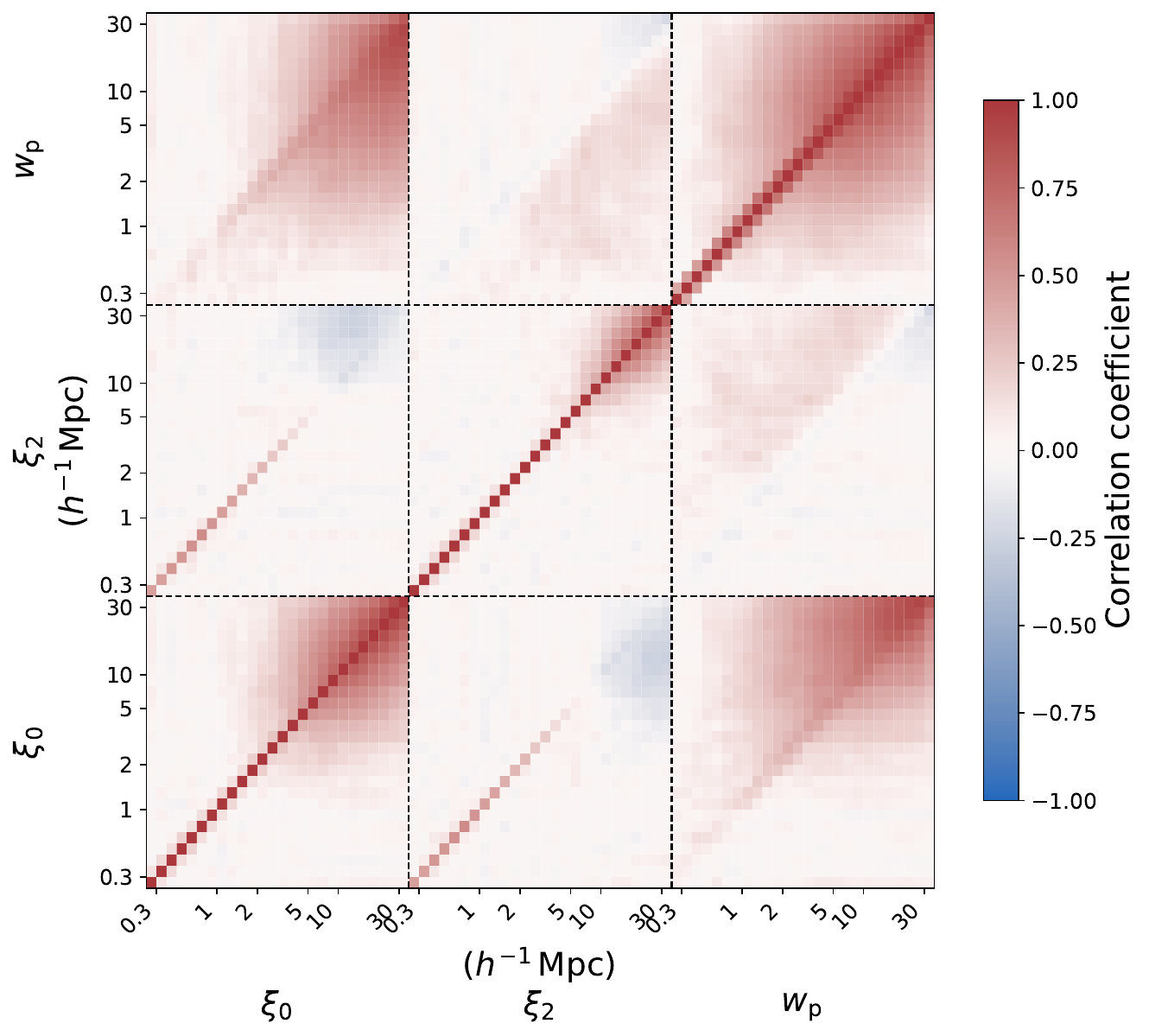}\\
\includegraphics[width=0.48\linewidth]{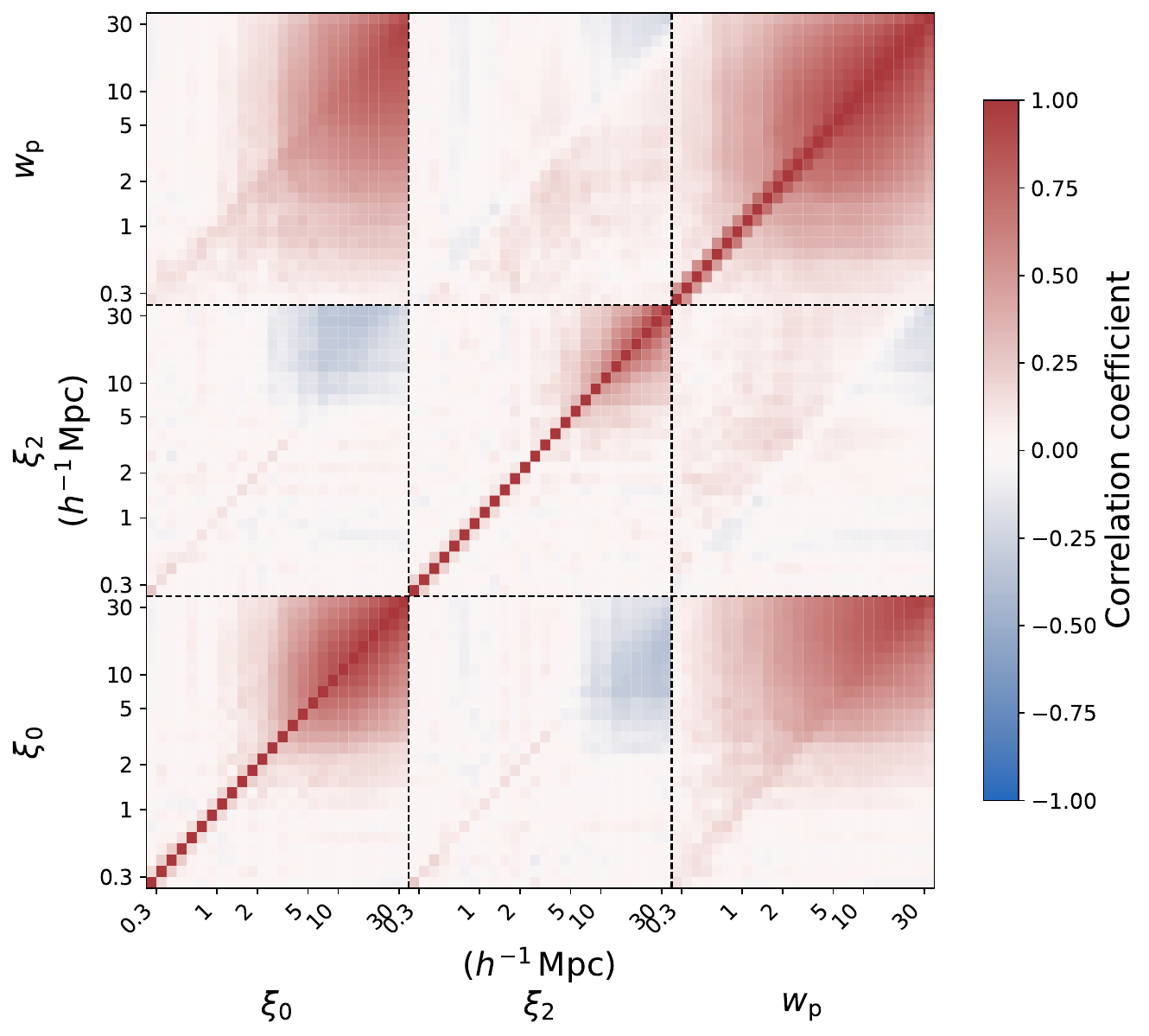}
\caption{Normalised covariance matrices of the \texttt{ELG1} (top left), \texttt{LRG3} (top right), and \texttt{ELG1$\times$LRG3} (bottom) \ho high-fidelity mocks computed from 1800 $500\,h^{-1}$Mpc \textsc{AbacusSummit} boxes.}
\label{fig:covsmall}
\end{figure*}

\begin{figure*}
\centering
\includegraphics[width=0.3\linewidth]{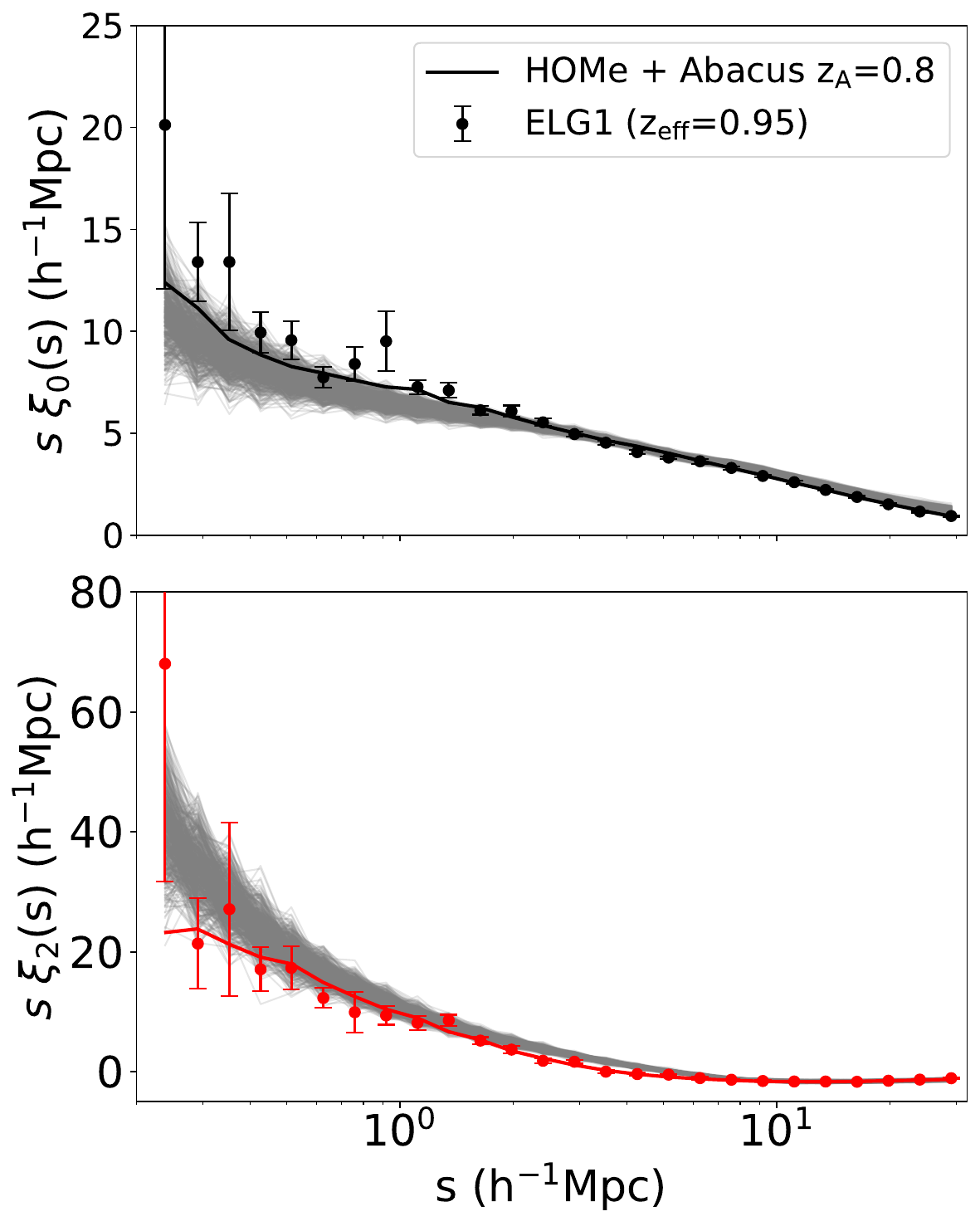}
\includegraphics[width=0.3\linewidth]{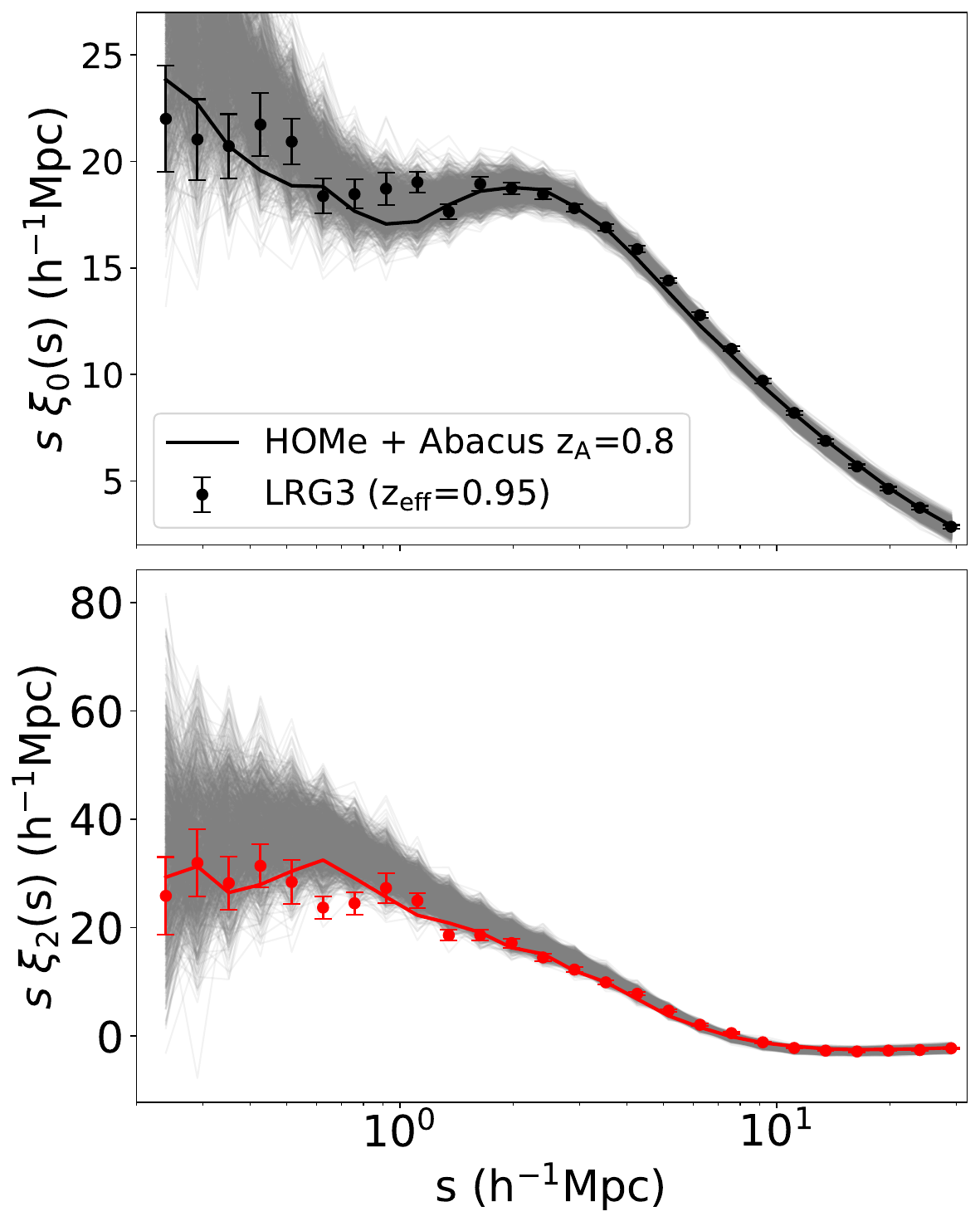}
\includegraphics[width=0.3\linewidth]{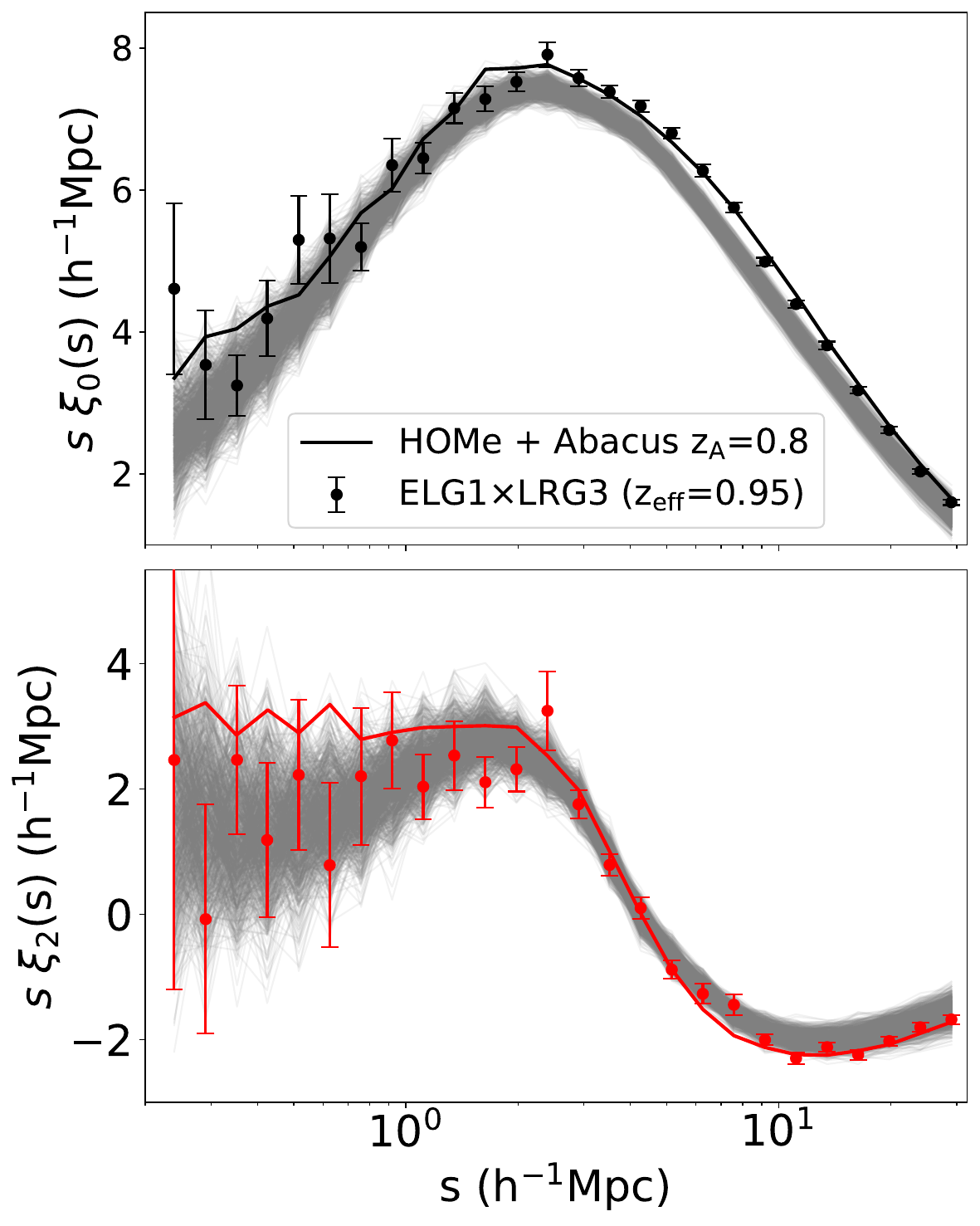}\\
\includegraphics[width=0.3\linewidth]{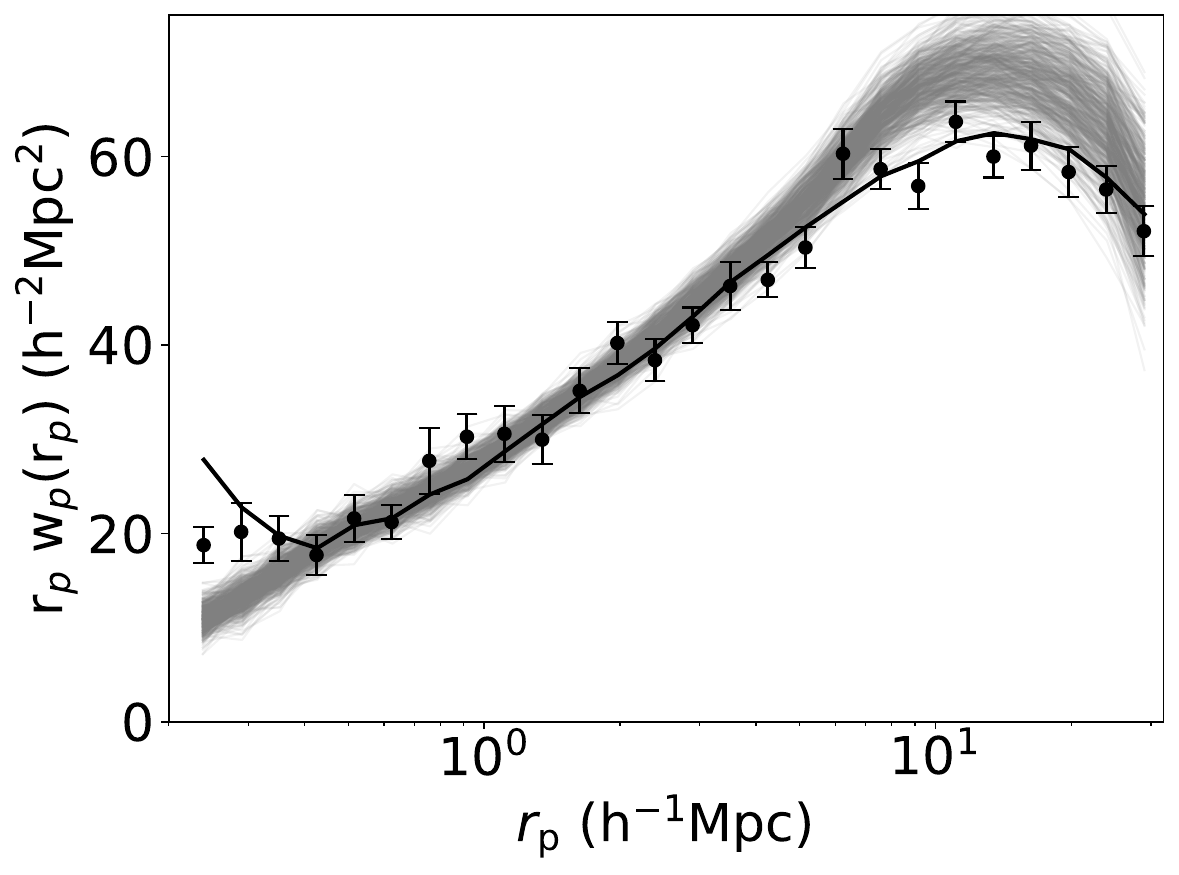}
\includegraphics[width=0.3\linewidth]{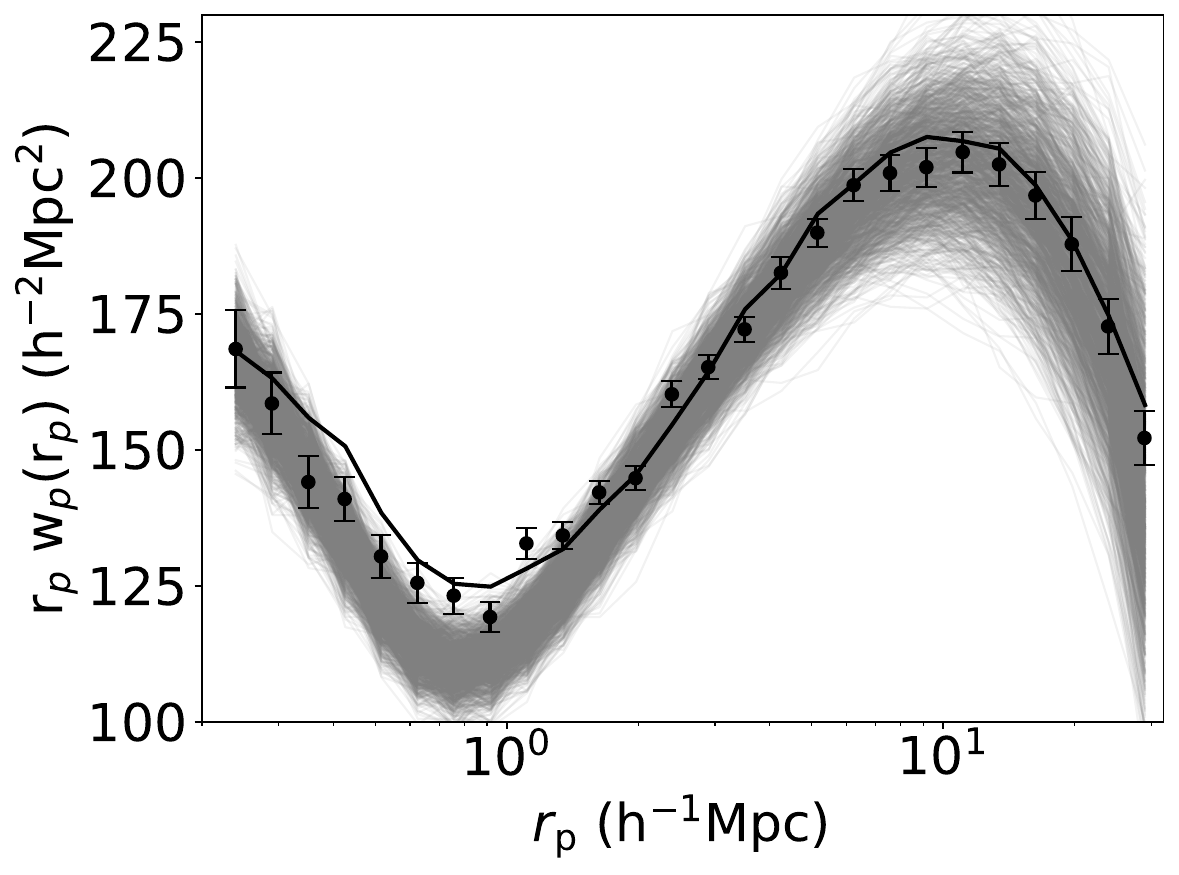}
\includegraphics[width=0.3\linewidth]{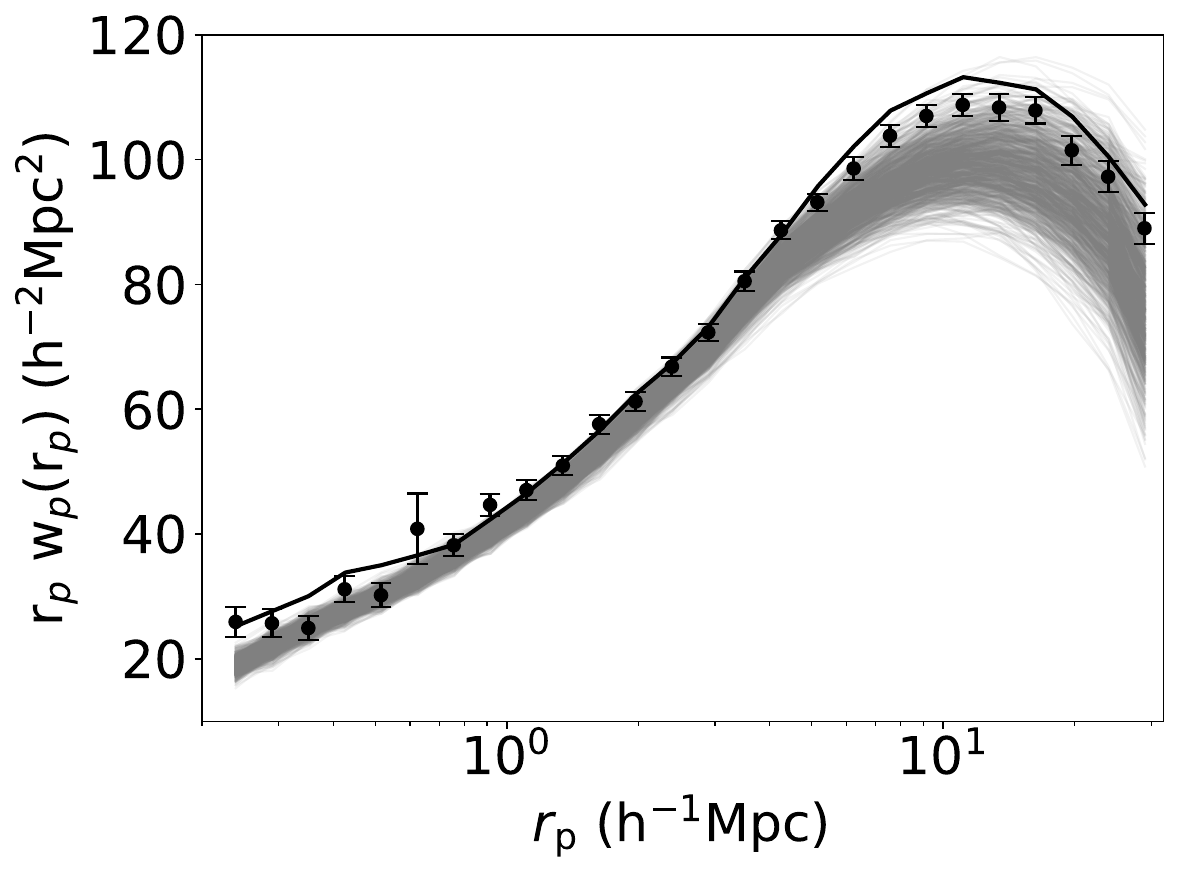}
\caption{\texttt{ELG1} (left column), \texttt{LRG3} (middle), and \texttt{ELG1$\times$LRG3} (right) correlation functions (markers) with the \ho high-fidelity mocks (thick solid lines), compared with the 1800 2PCFs (grey thin lines) from the $500\,h^{-1}$Mpc \textsc{AbacusSummit} boxes used to compute the correlation matrix in Figure\,\ref{fig:covsmall}. Note that all the 1800 2PCFs for the 3 tracers are obtained by imposing the \ho best-fit configuration obtained from the $2\,h^{-1}$Gpc boxes (see Table\,\ref{tab:clusteringresu}) to the small boxes, with minor adjustments only for \texttt{ELG1}. Here, the discrepancies observed are mainly due to the limited volume of the small boxes, which enhances sample variance and suppresses the contribution of massive halos and long-wavelength modes. }
\label{fig:clusteringcov}
\end{figure*}

The observational uncertainties are obtained from 128 jackknife re-samplings on the Y1 survey footprint \citep{2025JCAP...01..125R}. The errors on the models are obtained as the standard deviation of the 2PCFs measured from 1800 mocks constructed by applying the \ho best-fit configuration to 1800 realizations of the \textsc{AbacusSummit} $500\,h^{-1}$Mpc boxes (see \S\,\ref{sec:simulation}). The correlation matrices of the 1800 mocks from these small boxes are shown in Figure\,\ref{fig:covsmall}, while Figure\,\ref{fig:clusteringcov} compares their clustering results with those of the \ho $2\,h^{-1}$Gpc high-fidelity mocks. 

Note that we directly apply \ho with the parameters given in Table\,\ref{tab:clusteringresu} to the 1800 small boxes, with only minor adjustments in the satellite fractions. For both tracers and their cross-correlation, the agreement between the small-box mocks and the high-fidelity results is very good across all scales. The small residual differences is mostly due to the limited volume of the $500\,h^{-1}$Mpc boxes, which amplifies sample variance and reduces the contribution of massive halos and long-wavelength modes, penalizing the strong 1-halo term in the ELG clustering (for which we need some retuning in the satellite fractions). Importantly, resolution is not a limiting factor here, since both the $500\,h^{-1}$Mpc and the $2\,h^{-1}$Gpc boxes have identical resolution.

As previously done by \citet{2023JCAP...10..016R} and \cite{2024MNRAS.530..947Y}, in estimating the model likelihood, we reduce the noise of the jackknife covariances by replacing their off-diagonal terms by those of the 1800 small-box covariances. The diagonal elements remain unaltered to preserve the jackknife variance.


The residuals between the \ho high-fidelity mock catalog and the observations, weighted by both uncertainties, are displayed in the lower panels of Figures\,\ref{fig:clusteringELG}--\ref{fig:clusteringELGLRG}. These show remarkable agreement (mostly within 5\%) on all scales for all tracers and clustering statistics. The best-fit parameters and $\chi^2$ values are in Table\,\ref{tab:clusteringresu}.  

These results show that we accurately model the observed anisotropy in the Universe, capturing both the complex quadrupole shape on all scales and the strong 1-halo observed both in the monopole and in $w_p(r_p)$. 

The model accuracy is particularly remarkable below $2\,h^{-1}$Mpc, where achieving this level of agreement requires \ho to precisely characterize the dynamics of satellites within their host halos---which is highly tracer-dependent.
This has strong impact on the conformity level required by DESI observations, which emerges from \ho as a pure prediction (see \S\,\ref{sec:orphans} and \ref{sec:hodpred}).

The unprecedented accuracy we achieve in modeling the multi-tracer anisotropic clustering down to $s=200\,h^{-1}$kpc enables us to impose stringent constraints on the ELG$\times$LRG halo occupation distribution, as well as the presence and contribution of orphan satellites to the clustering signal, and the ELG quenching mechanism.

This precision hinges primarily on the physical prescriptions driven by the satellite fraction parameters, $f_{\rm sat}^{\rm elg\,(lrg)}$, the velocity biases $(b_{\rm r}$, $b_{\rm t}$, $b_{\rm cen})^{\rm elg\,(lrg)}$, the halo exclusion levers $(M_{\rm excl},\,r_{\rm excl},\,p_{\rm excl})^{\rm elg\,(lrg)}$, and the joint-occupation variables that emulate quenching ($M_0,\,R_0,\,m_{\rm slope},\,\delta,\,\alpha$).

A key feature of our forward modeling framework is that the $f_{\rm sat}^{\rm elg,(lrg)}$ entering the MCMC (Table\,\ref{tab:clusteringresu}, top part) are input parameters of the model, not direct predictors of the actual number of satellites in the final mock catalog. The \emph{realized satellite fractions} (Table\,\ref{tab:clusteringresu}, bottom part) instead emerge self-consistently from the coupled AM down-sampling (\S\ref{sec:AM}), the stochastic class assignment scheme, the halo exclusion condition, and the joint–occupation quenching model (\S\ref{sec:jointoccup}), which redistribute galaxies between central and satellite roles.

Our best-fit \ho predicts that $9.50\%$ of \texttt{ELG1} and $14.09\%$ of \texttt{LRG3} galaxies are satellites, while the remainder are centrals with no satellites. Among these satellites, $1.09\%$ (ELGs) and $3.52\%$ (LRGs) occupy halos whose central is of the same type (maximal conformity), whereas $7.02\%$ (ELGs) and only $0.005\%$ (LRGs) reside in halos hosting a central galaxy of the opposite type (minimal conformity). The remaining $0.58\%$ (ELGs) and $10.57\%$ (LRGs) are classified as orphans, i.e. satellites whose parent halos do not appear as centrals in the final realization.

According to our model, orphans tend to populate low-mass, low-bias halos that lack resolved substructure. On large scales, they redistribute galaxies toward lower-bias environments, suppressing both monopole and quadrupole amplitudes; on small scales, they retain satellite-like velocity dispersions and thus contribute to FoG damping.

From the above fractions, we infer that ELG satellites strongly favor minimally conformal environments, orbiting predominantly in LRG-host halos. Conversely, LRG satellites are mostly maximally conformal, orbiting around LRG centrals. The substantially larger orphan fraction among LRGs reflects the halo-selection hierarchy encoded in \textsc{HOM}e: LRGs are assigned first and independently, whereas ELG satellites are conditioned on the already-populated LRG halo field.

The stark difference in orphan incidence between the two tracers has both a physical and modeling motivation. For ELGs, the joint–occupation prescription (\S\ref{sec:jointoccup}) suppresses ELG satellites in massive LRG hosts, such that many satellites that would otherwise be orphans are instead placed into LRG halos---consistent with observed environmental quenching. For LRGs, however, no such conditioning is imposed: their orphan fraction is therefore a pure forward-model prediction, reflecting the subhalo survival rates associated with their massive hosts and the resolution limits of the merger tree.

Understanding the astrophysical interpretation of this non-negligible LRG orphan population---including subhalo disruption and selection incompleteness---is an exciting direction for follow-up work, which we plan to pursue using state-of-the-art hydrodynamical simulations, such as IllustrisTNG.

The realized satellite fractions in \ho lie between those obtained in current SHAM analyses \citep{2022MNRAS.516...57Y, 2025A&A...698A.170P}.
They govern the entire small-scale anisotropy budget, especially for ELGs, whose clustering below $\sim 4\,h^{-1}$Mpc is dominated by 1-halo contributions \citep{2023JCAP...10..016R}.

\begin{deluxetable*}{l c c }
\tablecaption{Best-fit values of the model input parameters for the DESI \texttt{ELG1} and \texttt{LRG3} tracers, producing the clustering signal in Figures\,\ref{fig:clusteringELG}--\ref{fig:clusteringELGLRG}. In parentheses we report the Gaussian priors adopted in the \texttt{emcee} sampler. Here the satellite fractions are total ones, i.e., computed with respect to the total number of mocks in the final catalog. We show the reduced $\chi^2$ obtained by jointly fitting the multipoles and the projected auto- and cross-correlation functions of both tracers. We also report the realized values of total satellite and orphan fractions, together with the fraction of satellites that are maximally (minimally) conformal, i.e., occupying central hosts of the same (complementary) species. Note that these are pure predictions of our forward model resulting from the interplay of the AM down-sampling, halo exclusion, and joint occupation condition. The $V_{\rm peak}$ minimal selection cuts are shown at the bottom of the table; these are fixed values---i.e., not free parameters constrained in the likelihood---chosen to match the large-scale bias of each tracer.
\label{tab:clusteringresu}}
\tablehead{
  \multicolumn{3}{c}{\bf Best-fit values (model inputs)}\\
  \colhead{Parameter / prescription} & \colhead{ELG} & \colhead{LRG}
}
\startdata
\multicolumn{3}{l}{\bf HOD (satellite positions)} \\
$\kappa$ & $0.721^{+0.046}_{-0.049}$\,\,\,$\mathcal{N}(0.7,0.2)$ & $0.473^{-0.105}_{+0.096}$\,\,\,$\mathcal{N}(0.7,0.20)$ \\
$\log{(M_{\rm cut}/h^{-1}{\rm M}_\odot)}$ & $11.651^{+0.137}_{-0.138}$\,\,\,$\mathcal{N}(11.6,0.3)$ & $13.151^{+0.129}_{-0.090}$\,\,\,$\mathcal{N}(12.7,0.3)$ \\
$\log{(M_{1}/h^{-1}{\rm M}_\odot)}$ & $11.801^{+0.071}_{-0.088}$\,\,\,$\mathcal{N}(11.6,0.3)$ & $13.402^{+0.145}_{-0.108}$\,\,\,$\mathcal{N}(13.6,0.4)$  \\
$\beta$ & $0.634_{-0.086}^{+0.110}$\,\,\,$\mathcal{N}(0.6,0.2)$ & $0.992_{-0.128}^{+0.109}$\,\,\,$\mathcal{N}(0.8,0.3)$ \\
\multicolumn{3}{l}{\bf Satellite placement in halos} \\
$K_{\rm out}$ & $1.028_{-0.117}^{+0.124}$\,\,\,$\mathcal{N}(1.2,0.3)$ & $1.105_{-0.114}^{+0.098}$\,\,\,$\mathcal{N}(1.5,0.3)$\\
\multicolumn{3}{l}{\bf Velocity bias (peculiar motions)} \\
$b_{\rm cen}$ & $0.116_{-0.069}^{+0.070}$\,\,\,$\mathcal{N}(0.1,0.3)$ & $0.134_{-0.021}^{+0.028}$\,\,\,$\mathcal{N}(0.1,0.3)$ \\
$b_{\rm r}$ & $0.873_{-0.082}^{+0.093}$\,\,\,$\mathcal{N}(0.7,1.2)$ & $0.959_{-0.130}^{+0.095}$\,\,\,$\mathcal{N}(0.9,0.4)$ \\
$b_{\rm t}$ & $0.802_{-0.106}^{+0.146}$\,\,\,$\mathcal{N}(0.7,1.3)$ & $1.183_{-0.107}^{+0.098}$\,\,\,$\mathcal{N}(0.8,1.4)$\\
$f_\sigma$ & $0.128_{-0.057}^{+0.050}$\,\,\,$\mathcal{N}(0.2,0.2)$ & $0.053_{-0.059}^{+0.075}$\,\,\,$\mathcal{N}(0.5,0.2)$\\
\multicolumn{3}{l}{\bf Halo exclusion} \\
$\log{(M_{\rm excl}/h^{-1}{\rm M}_\odot)}$ & $13.452_{-0.125}^{+0.129}$ \,\,\,$\mathcal{N}(13.5,0.3)$ & $13.417_{-0.126}^{+0.128}$ \,\,\,$\mathcal{N}(14.0,0.3)$\\
$r_{\rm excl}\,(h^{-1}{\rm Mpc})$ & $2.306_{-0.323}^{+0.489}$\,\,\,$\mathcal{N}(2.2,1.0)$ & $3.226_{-0.259}^{+0.268}$\,\,\,$\mathcal{N}(4.5,0.5)$\\
$p_{\rm excl}$ & $0.631_{-0.123}^{+0.088}$ \,\,\,$\mathcal{N}(0.6,0.3)$ & $0.480_{-0.067}^{+0.081}$ \,\,\,$\mathcal{N}(0.3,0.3)$\\
\multicolumn{3}{l}{\bf AM (galaxy--halo connection)} \\
$\sigma_{\rm AM}$ & $0.294_{-0.053}^{+0.065}$\,\,\,$\mathcal{N}(0.3,0.2)$ & $0.301_{-0.070}^{+0.056}$\,\,\,$\mathcal{N}(0.3,0.2)$\\
$V\,\,(s^{-1}{\rm km})$ & $185.903_{-10.786}^{+82.474}$\,\,\,$\mathcal{N}(190,100)$ &
$347.927_{-4.294}^{+27.715}$\,\,\,$\mathcal{N}(370,100)$  \\
$\sigma_{V}\,\,(s^{-1}{\rm km})$ & $88.135_{-12.407}^{+29.430}$\,\,\,$\mathcal{N}(100,90)$ &
$134.890_{-8.835}^{+20.4284}$\,\,\,$\mathcal{N}(170,60)$\\
$f_{\rm sat}\,\,(\%)$ total & $6.921_{-0.452}^{+0.370}$ \,\,\,$\mathcal{N}(7.8,1.5)$ &
$9.311_{-0.557}^{+0.629}$ \,\,\,$\mathcal{N}(10.4,1.5)$\\
\multicolumn{3}{l}{\bf Quenching (via joint occupation)} \\
$\alpha$ & $0.501_{-0.057}^{+0.065}$ \,\,\,$\mathcal{N}(0.4,0.2)$&  --\\
$R_0\,\,(h^{-1}{\rm Mpc})$ &$4.498_{-0.237}^{+0.255}$ \,\,\,$\mathcal{N}(5.0,2.0)$ & -- \\
$\delta$ & $2.987_{-0.308}^{+0.325}$ \,\,\,$\mathcal{N}(3.0,1.5)$ & --\\
$\log{(M_{0}/h^{-1}{\rm M}_\odot)}$ & $13.232_{-0.060}^{+0.055}$ \,\,\,$\mathcal{N}(13.0,0.4)$ & --\\
$m_{\rm slope}$ &$0.997_{-0.072}^{+0.075}$ \,\,\,$\mathcal{N}(1.0,0.4)$ &-- \\
\tableline
$\chi^2_{\rm joint}/{\rm dof}$ & 1.37 & 1.61 \\
dof & \multicolumn{2}{c}{119\hspace{1cm}(26 data points $\times$ 3 observables $\times$ 2 tracers $\rightarrow$ 156 data points $-$ 37 parameters)} \\
\tableline
\tableline
\multicolumn{3}{c}{\bf Realized values (pure predictions)}\\
$f_{\rm sat}\,\,(\%)$ total & 9.50  & 14.09 \\
$f_{\rm sat}\,\,(\%)$ max conformal & 1.09 & 3.52  \\
$f_{\rm sat}\,\,(\%)$ min conformal & 7.02 & 0.005  \\
$f_{\rm orph}\,\,(\%)$ total & 0.58  & 10.57  \\
\tableline
\multicolumn{3}{l}{Minimal selection cuts}\\
$V_{\rm peak}^{\rm min}(s^{-1}{\rm km})$ & 110 & 370 \\
\enddata
\end{deluxetable*}

The satellite fraction and velocity bias (\S\,\ref{sec:appvz}) exert complementary and competing effects on the anisotropic clustering on small and intermediate scales. While $f_{\rm sat}^{\rm elg\,(lrg)}$ primarily control the clustering amplitude, the radial and tangential velocity biases, $b_{\rm r}$ and $b_{\rm t}$, govern redshift-space distortions. These effects manifest both in the so-called finger-of-god (`FoG') regime below $1\,h^{-1}$Mpc, where random peculiar motions dominate, and on larger scales, around $10\,h^{-1}$Mpc, where satellite coherent infall towards overdensities induces the Kaiser squashing.

In the 2-halo regime, a third effect becomes significant: the mutual exclusion between close pairs of massive hosts. This mechanism, particularly relevant to shape the 1-to-2-halo transition in the \texttt{ELG1}, \texttt{LRG3} and \texttt{ELG1$\times$LRG3} monopole correlation functions.

Specifically, from our best-fit \ho we find that \texttt{ELG1} (\texttt{LRG3}) halos with $\log{(M_{\rm vir}/(h^{-1}{\rm M}_\odot))}\gtrsim 13.45$ ($\log{(M_{\rm vir}/(h^{-1}{\rm M}_\odot))}\gtrsim 13.4$) are segregated by a minimum radial distance of about $2.3\,h^{-1}$Mpc ($3.2\,h^{-1}$Mpc) from any other central tracer of the same mass.   
These exclusion scales mark the onset of nonlinear halo interactions, which are essential for accurately modeling the 1-to-2-halo transition in the clustering.

Another key ingredient in reproducing the \texttt{ELG1$\times$LRG3} cross-correlation is the joint–occupation condition, which acts as a phenomenological model for environmental quenching. In this prescription, each ELG candidate receives a kernel-weighted suppression factor that depends on the masses and distances of nearby LRG hosts, such that ELGs are progressively disfavored in high-density, LRG–dominated environments. 

Our best-fit parameters in Table\,\ref{tab:clusteringresu} indicate a strongly local and mass-dependent effect: massive LRG halos within a few Mpc exert the strongest suppression, while more distant or lower-mass systems contribute negligibly. This interaction naturally reduces the ELG occupation around LRG hosts and thereby controls the 1-halo contribution to the \texttt{ELG1$\times$LRG3} cross-correlation, yielding the level of conformity required by the data without invoking explicit baryonic physics. The resulting modulation of ELG centrals is mild on large scales but crucial for matching the small-scale anisotropic signal.

In conclusion, the non-trivial interplay among satellite fraction, velocity biases, halo exclusion and environmental quenching (via joint occupation) underpins the sensitivity and accuracy of our forward modeling approach. The individual impact of each model parameter on the clustering is shown and discussed in Appendix\,\ref{sec:impact_params}.

\subsubsection{\ho posterior distribution} 
\label{sec:home_posterior}

\begin{figure*}
\centering
\includegraphics[width=0.8\linewidth]{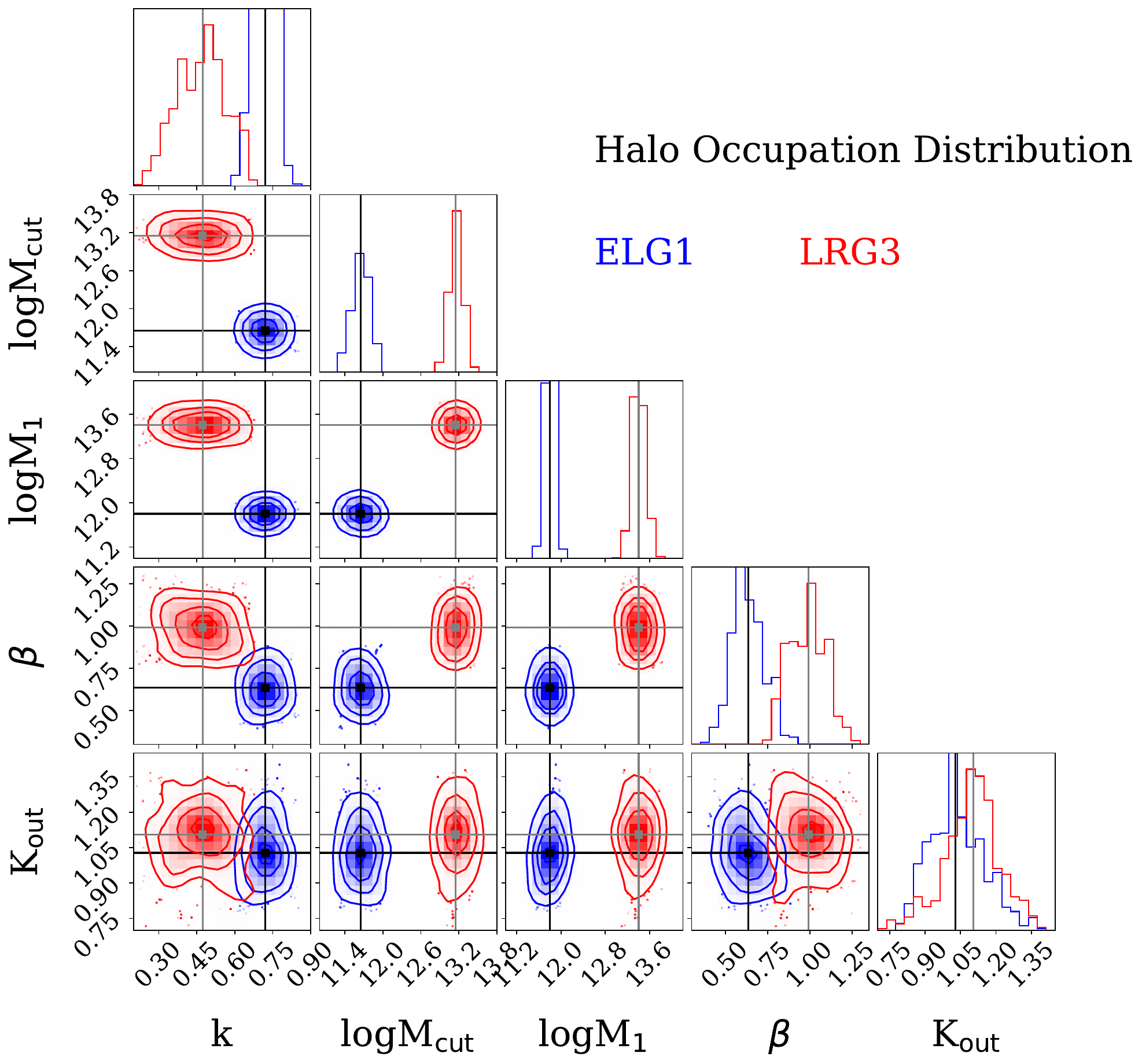}
\caption{Posterior distribution of the HOD model parameters for the \texttt{ELG1} (blue contours and black straight lines) and \texttt{LRG3} (red contours and grey lines) samples showing the HOD parameters. The best-fit parameter values are are written above each panel as well as in Table \ref{tab:clusteringresu}, and they are highlighted by the intersection of the straight lines; the resulting models are shown in Figure\,\ref{fig:clusteringELG}.}
\label{fig:cornerHOD}
\end{figure*}

\begin{figure*}
\centering
\includegraphics[width=0.65\linewidth]{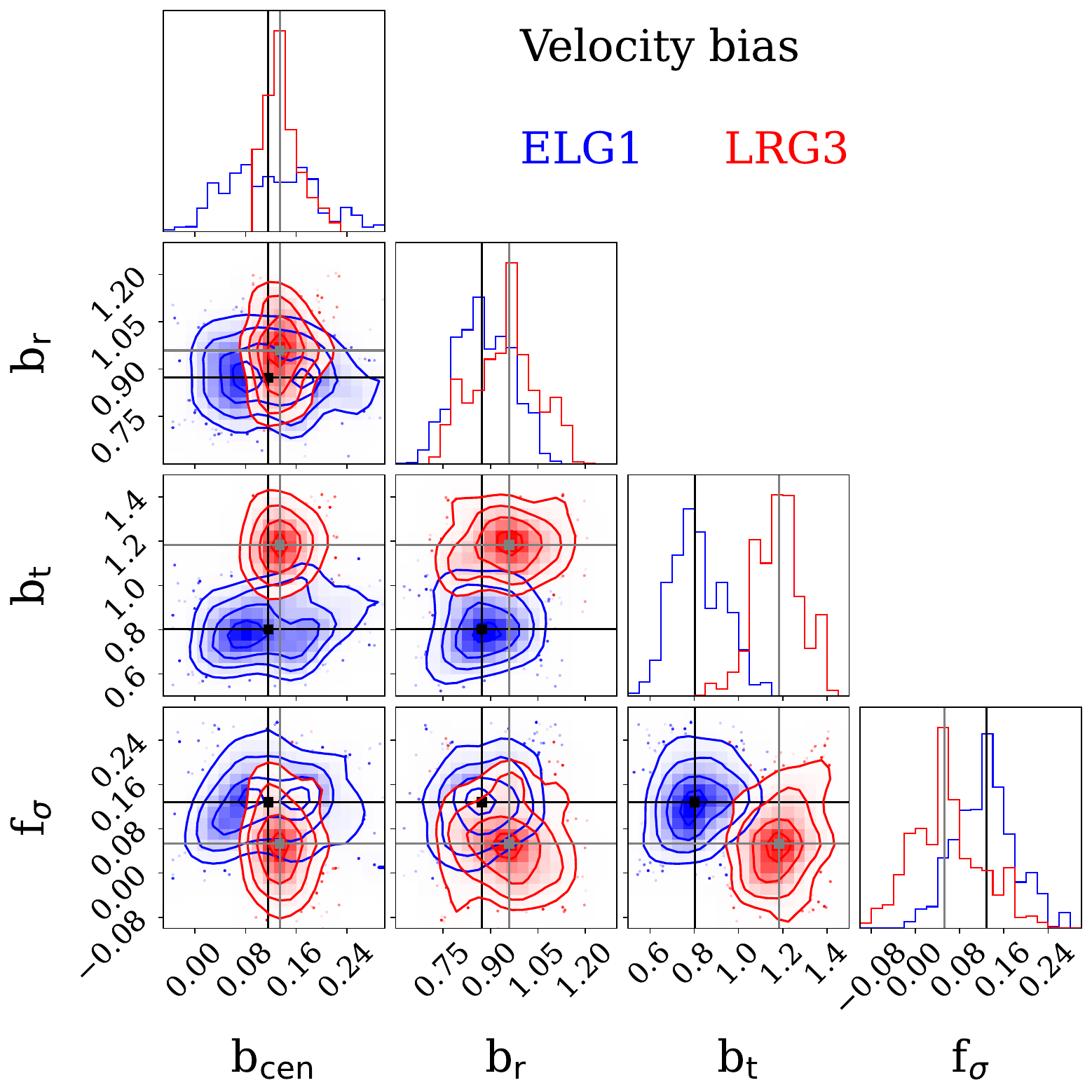}
\caption{Posterior distribution of the ELG and LRG velocity-bias parameters.}
\label{fig:cornervbias}
\end{figure*}

\begin{figure*}
\centering
\includegraphics[width=0.6\linewidth]{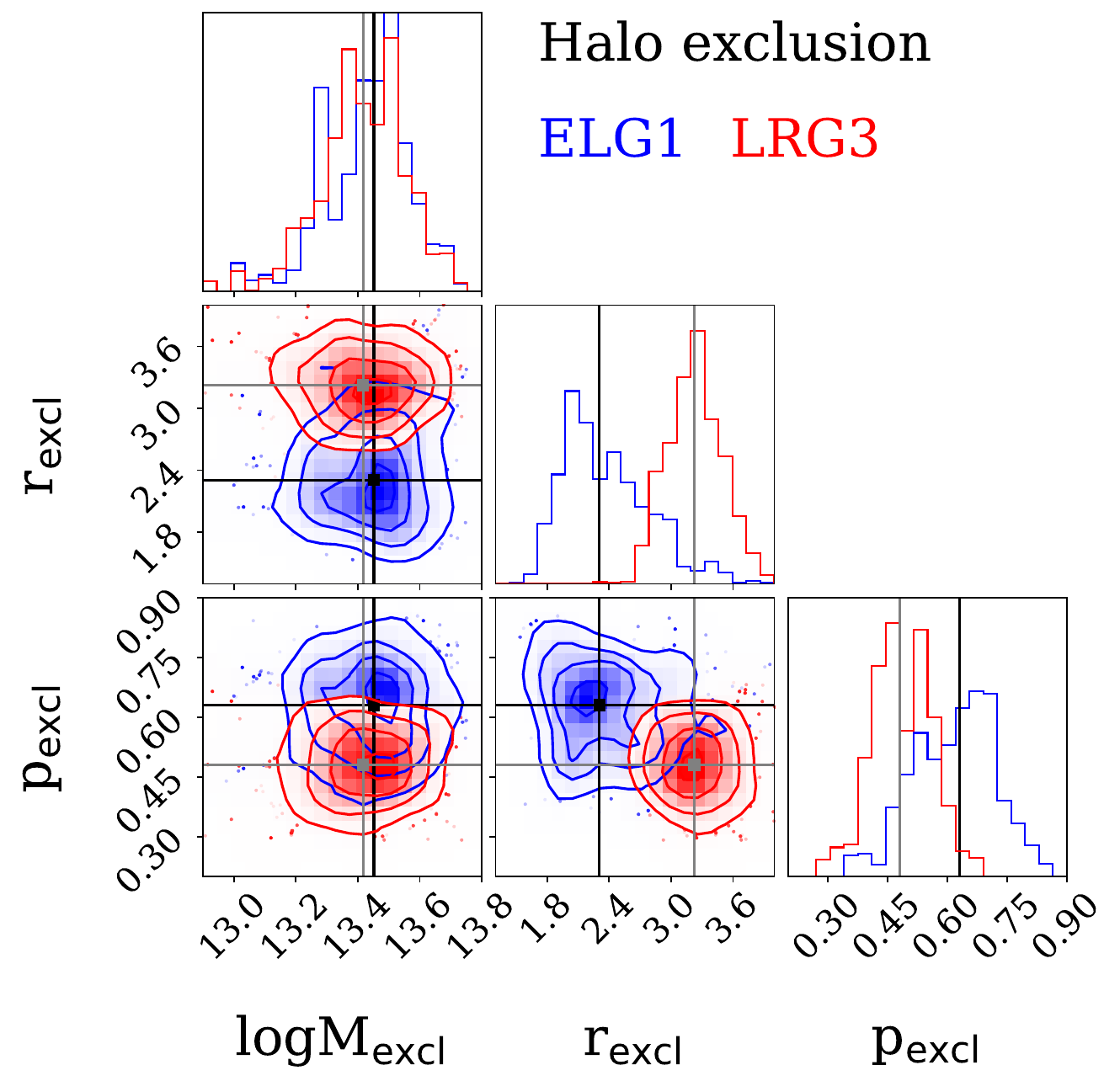}
\caption{Posterior distribution of the ELG and LRG halo-exclusion parameters.}
\label{fig:cornerexclusion}
\end{figure*}

\begin{figure*}
\centering
\includegraphics[width=0.7\linewidth]{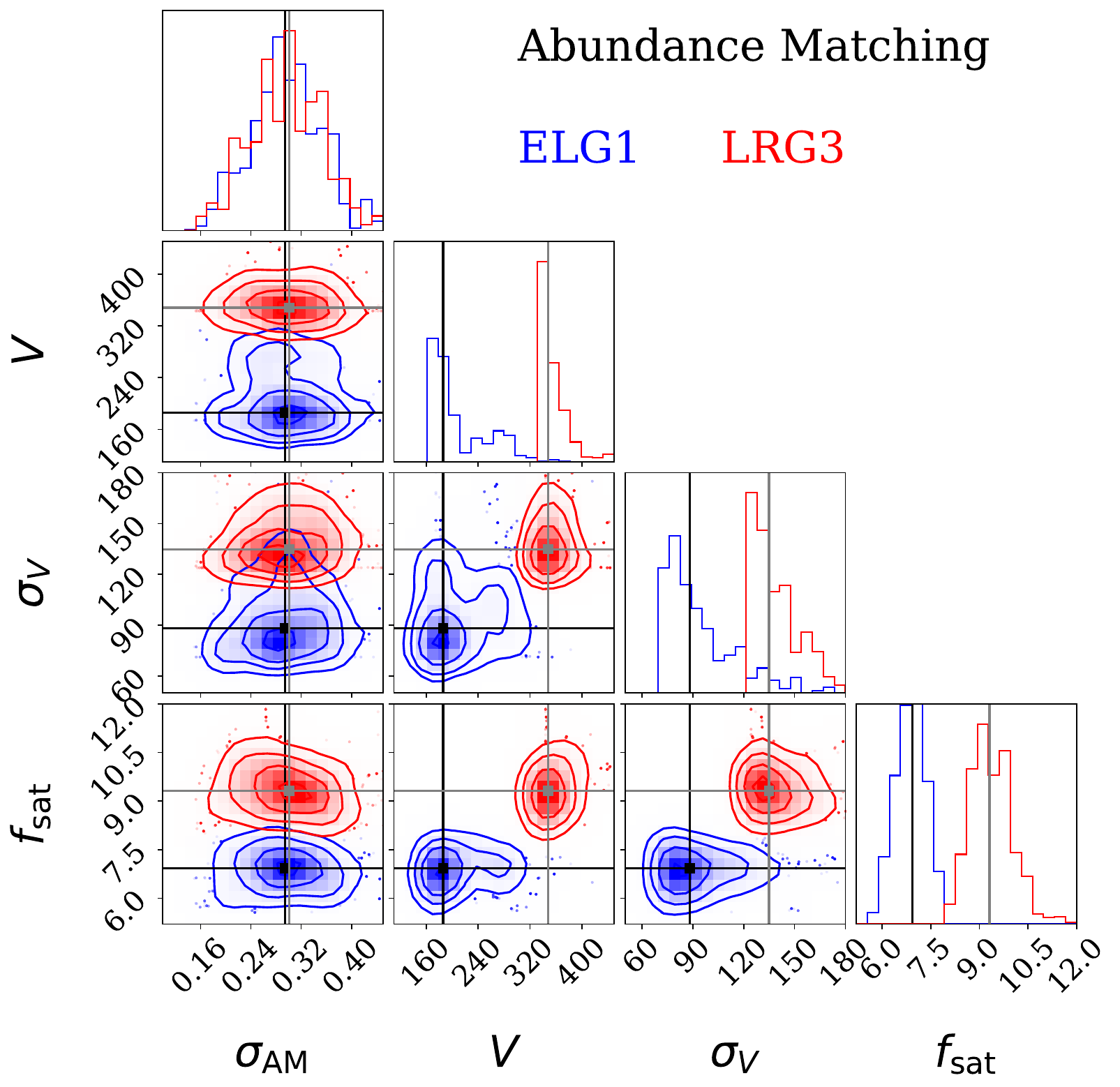}
\caption{Posterior distribution of the ELG and LRG AM parameters.}
\label{fig:cornerAM}
\end{figure*}

\begin{figure*}
\centering
\includegraphics[width=0.8\linewidth]{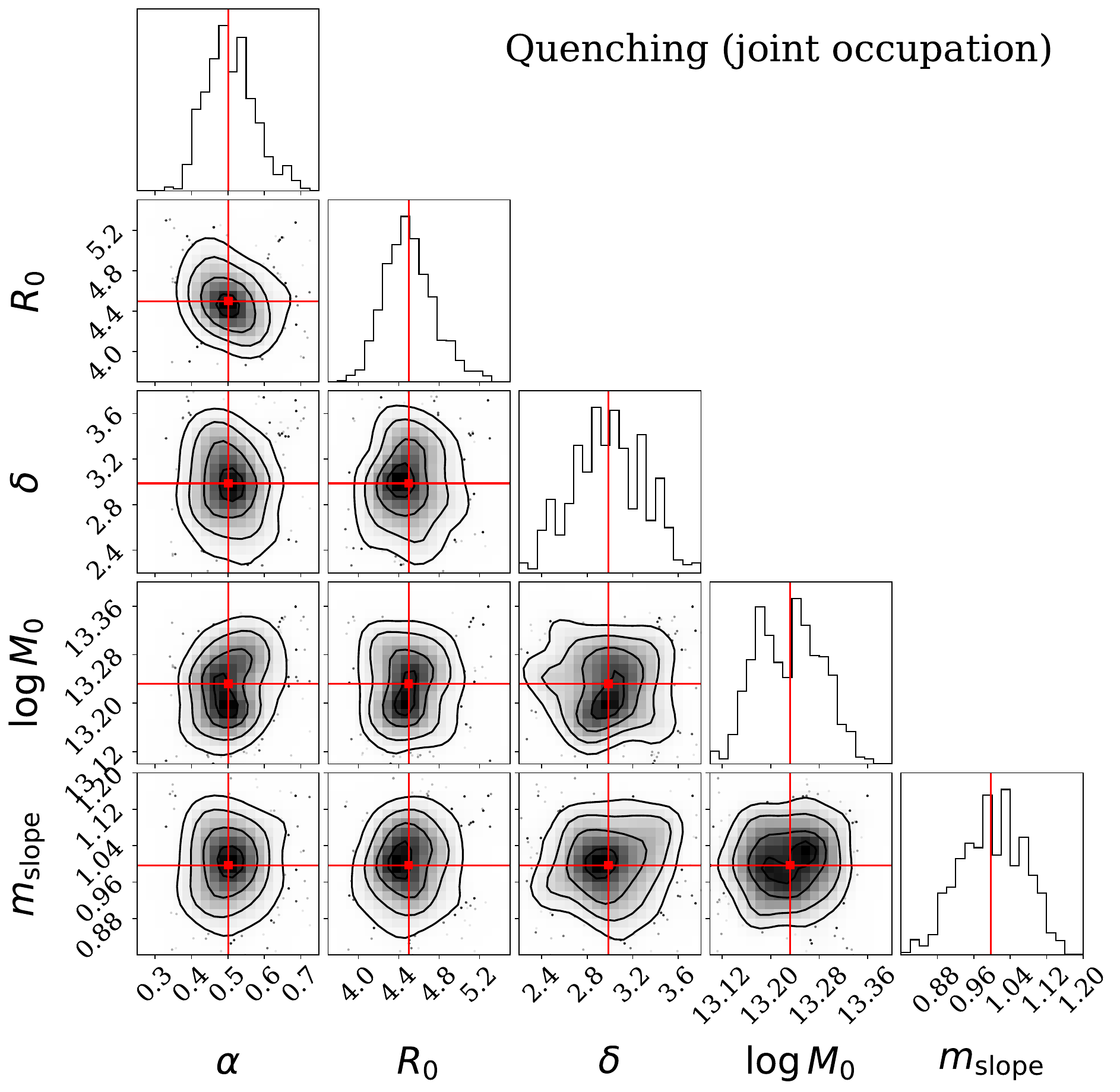}\\
\caption{Posterior distribution of the model parameters driving the joint-occupation condition that statistically emulates the effects of environmental quenching of satellite ELG in massive LRG hosts.}
\label{fig:cornerquenching}
\end{figure*}

Figures\,\ref{fig:cornerHOD}--\ref{fig:cornerquenching} present the posterior distributions, obtained from level-II inference, of the physical parameters of our model (37 in total) for DESI ELGs (blue contours and black straight lines) and LRGs (red contours and grey lines). The high dimensionality of the parameter space prevents us from representing them in unique corner plots; we split them in different panels, grouped by the physical prescription (see the captions for details). 

The posterior distributions are well contained within the adopted prior ranges and do not approach the prior boundaries, indicating that the parameter constraints are primarily driven by the likelihood rather than by the choice of prior. To verify this explicitly, we repeated the MCMC analysis using \textsc{emcee} with wider uniform priors on all parameters. The resulting posterior distributions are consistent with those obtained using Gaussian priors, showing no significant shifts in the parameter means or credible intervals.

Overall, the posterior distributions are compact and close to Gaussian for both tracers, demonstrating that the DESI two-point clustering measurements provide strong constraining power on the main ingredients of our forward model. The lack of strong non-Gaussian features or extended degeneracy tails suggests that the model parameters are well determined by the combination of real- and redshift-space clustering, and that residual parameter correlations are moderate.

We adopt the maximum likelihood estimate (MLE) as our best-fit model, which simultaneously reproduces the \texttt{ELG1} auto-, \texttt{LRG3} auto-, and \texttt{ELG1$\times$LRG3} cross-correlation functions. The best-fit parameter values are highlighted by the intersection of the straight lines and we report them in Table\,\ref{tab:clusteringresu}.

We sample \ho full posterior, for each latent realization \( \{ \boldsymbol{\theta}_i \}_{i=1}^{N} \) in Eq.\,\ref{eq:levelIposterior}, using emcee with 37 dimensions coupled with 80 walkers and 1500 steps per walker. We set the number of walkers more than 2.5 times that of the parameters, and choose the number of steps per walker based on the maximum autocorrelation time ($\tau$), so that $n_{\rm steps}> 50\,\tau_{\rm max}$, which guarantees convergence \citep{2013PASP..125..306F}.

To ensure reliable parameter inference, we apply a conservative burn-in procedure, but no thinning, to the MCMC chains. We determine the burn-in length from the integrated autocorrelation time, which we compute for each parameter using emcee with a relative tolerance of 0.01. We remove correlations with the initial conditions by discarding the first $5\,\tau_{\rm max}$ steps of each walker. We estimate $\tau_{\rm max}=24.5$, a relatively small value that ensures good mixing. This implies that: (i) the walkers move broadly across the posterior, exploring all high-probability regions; (ii) the samples are weakly correlated, i.e. the chain does not remain close to the same value for many steps; (iii) the full posterior shape is properly recovered, including any modes, tails, or correlations between parameters.

After checking the autocorrelation functions of all parameters, we kept all post–burn-in samples without thinning as this is not required for unbiased posterior estimation and only reduces the effective sample size at fixed runtime, so we adopt thin\,=\,1

The posterior distributions of the HOD parameters in Figure\,\ref{fig:cornerHOD} reveal that ELGs and LRGs occupy distinct regions of the parameter space, reflecting their different astrophysical natures. ELGs favor lower halo masses, with tight constraints around lower values of both $\log M_{\rm cut}$ and $\log M_1$, consistent with their association to younger, late-forming halos with shallow potential wells. This picture is fully consistent with the ELG (LRG) HOD inferred from DESI One-Percent data by \citet{2023JCAP...10..016R} (\citet{2024MNRAS.530..947Y}), who found that ELGs (LRGs) populate halos of $M_{\rm vir}\sim 5\times10^{11}-10^{12}\,h^{-1} {\rm M}_\odot$ ($M_{\rm vir}\sim 10^{13}\,h^{-1}{\rm M}_\odot$), with a modest (non-negligible) satellite fraction and a shallow (steep) high–mass tail.

Their occupation slope $\beta$ is similarly lower, indicating a more gradual rise of the satellite population with halo mass. In contrast, the LRG posteriors peak at significantly larger $\log{M_{\rm cut}}$ and $\log M_1$, characteristic of older, more massive and biased environments hosting quenched, early-type systems. Despite this large mass separation, both tracers exhibit well-constrained $k$ parameters---governing the satellite number normalization---with LRGs favoring a steeper increase in satellite abundance at high masses, as expected for quenched massive halos.

Finally, the constraints on $K_{\rm out}$ reinforce the complementary environmental behavior of these samples---ELGs favoring a placement slightly below the DM particle positions (cooler, more diffuse outskirts), while LRGs require satellites to be distributed somewhat beyond the average DM profile (mild over-concentration compensation). Together, these constraints demonstrate that the HOD ingredients in \ho are not only tightly determined by the observations, but also align with the established understanding of ELG and LRG formation and environment.

 The posterior distributions of the velocity–bias parameters presented in Figure\,\ref{fig:cornervbias} show that anisotropic clustering places tight constraints on satellite kinematics, directly informing how galaxy peculiar velocities deviate from those of their host dark matter halos.

For ELGs, we find $b_r,\,b_t<1$ and $f_\sigma \ll1$, corresponding to dynamically cool satellites that retain coherent infall motions and have not yet been isotropized. This interpretation is fully consistent with the observational modeling of DESI ELGs by \citet{2023JCAP...10..016R}, who likewise require sub-virial satellite motions to suppress excessive small-scale clustering power in ELG samples. Our forward model strengthens this picture by showing that such subdued kinematics arise naturally once ELGs occupy lower-mass, still–assembling halos where environmental quenching has only recently begun.

For LRGs we recover $b_r<1$ and $b_t>1$, implying suppressed radial but slightly enhanced tangential motions--- the hallmark of a population that has experienced substantial orbital decay and tidal processing in massive halos. Dynamical friction efficiently removes satellites on plunging orbits, leaving survivors on more circular paths that broaden the transverse velocity field but reduce coherent infall signatures.

These results align well with high-resolution hydrodynamical expectations \citep{2022MNRAS.510.2980A}, and with DESI HOD constraints for LRGs from \citet{2024MNRAS.530..947Y}, who similarly find sub-virial radial motions with mild tangential heating at $z\sim1$. The agreement confirms that LRG satellites are dynamically cooler than the dark matter but retain orbital anisotropy driven by their accretion histories.

Finally, the extremely small central velocity biases ($b_{\rm cen}\ll1$) ensure centrals remain nearly comoving with their halo bulk, correctly avoiding artificial FoG broadening in the 1–halo regime.

The \ho exclusion posteriors presented in Figure\,\ref{fig:cornerexclusion} show that both tracers require a probabilistic halo–halo exclusion operating at group/cluster scales, i.e. $\log{(M_{\rm excl}/(h^{-1}{\rm M}_{\odot}))}\sim13.2–13.4$ and $r_{\rm excl} \sim 3–4 \,h^{-1}$, precisely where the 1–halo to 2–halo transition is observed in the anisotropic clustering. LRGs favor a slightly higher exclusion mass and smaller exclusion radius than ELGs, consistent with their more massive and concentrated host halos, while the transition steepness ($p_{\rm excl}$) is similar for both samples. These values are fully consistent with earlier HOD treatments of halo exclusion for massive galaxies and extend that picture to the ELG population within a single, jointly–fit framework.

The posterior distributions of the abundance-matching parameters shown in Figure\,\ref{fig:cornerAM} reveal clear and physically meaningful differences between ELGs and LRGs. The stellar–to–halo connection for LRGs favors higher values of the characteristic peak circular velocity scale ($V$) and its dispersion ($\sigma_V$), consistent with these galaxies occupying deeper potential wells and tracing more massive halos. ELGs instead prefer lower values, as expected for systems residing in younger, less massive hosts where recent accretion and star formation remain more prevalent. The tight, nearly Gaussian constraints on the AM scatter parameter ($\sigma_{\rm AM}$) for both tracers imply strong sensitivity of the clustering data to how galaxies populate the halo velocity function---i.e., the rank-order mapping between $V_{\rm peak}$ and galaxy stellar mass must remain sharp to reproduce the observed two-point statistics. 

Finally, the satellite fraction parameter ($f_{\rm sat}$)---note that this input parameter differs from the \emph{realized} satellite fraction emerging, as a pure prediction, from our forward model (see Table\,\ref{tab:clusteringresu})---is well constrained and naturally separates the two samples: ELGs exhibit a much smaller intrinsic satellite abundance than LRGs, reflecting the role of environment-driven quenching that disfavors ELG satellites in massive halos. 

Taken together, these posteriors show that the galaxy–halo connection inferred by \ho for both tracers is highly informative, internally consistent, and closely aligned with the current picture of ELG and LRG formation and evolution.

The ELG$\times$LRG joint-occupation posterior in Figure\,\ref{fig:cornerquenching} confirm that the ELG$\times$LRG cross-correlation functions place strong constraints on satellite quenching physics. The model infers:
\begin{itemize}
\item A characteristic quenching mass of $\log{M_0\sim13.2}$, corresponding to LRG-host halos where ELG satellites are strongly suppressed;
\item A steep transition with $\delta\sim3$, indicating rapid quenching after infall into massive halos;
\item A radial scale $R_0\sim4-5\,h^{-1}$Mpc, showing that environmental effects extend beyond the virial radius---consistent with backsplash and pre-processing;
\item A normalization $\alpha\sim0.5$, meaning that only about half of the potential ELG satellites survive in these massive environments.
\item A steep mass dependence with $m_{\rm slope}\sim1$, implying that the quenching efficiency rises rapidly with halo mass: slightly above $M_0$, ELG satellites are only moderately suppressed, but deep in the group/cluster regime they are almost fully removed from the ELG population.
\end{itemize}

Together, these parameters describe a picture where environmental quenching is strongly mass- and radius-dependent: ELG satellites are preferentially removed in and around massive LRG halos, with efficiency that increases both with host mass and proximity to the halo center. This is fully consistent with semi-analytic results, such as \citet[][]{2018MNRAS.475.2530O} or \citet[][]{2018MNRAS.474.4024G}, which likewise predict that ELG-like galaxies avoid the most massive, quenched environments.

The tightness of the posteriors indicates that the cross-correlation is especially informative, forcing the model to accurately partition satellites into quenched (LRG-like) and star-forming (ELG-like) environments. Because \ho discovers these quenching signatures purely through forward-modeling of positions and velocities, the agreement with semi-analytic predictions provides strong validation of the approach.

\subsubsection{Impact of peculiar motions on the anisotropic clustering}
\label{sec:appvz}
In our coherent–flow prescription in Eq.\,\ref{eq:vzsat}, the two velocity–bias parameters, $b_{\rm r}$ and $b_{\rm t}$, scale the amplitude, not the direction, of the radial‑infall and tangential‑orbital components that satellites inherit from the DM particles bound to the same halo.  A value of $b_{r}=1\,(b_{t}=1)$ means that satellites on average follow the same radial (tangential) velocity field as the underlying DM.  Conversely, $b<1$ corresponds to cooler kinematics (i.e., slower infall or more circular orbits), whereas $b>1$ indicates hotter motions. 

In this framework, it is important to distinguish between the model definition of satellite motions and their observational imprint. By construction, satellites remain part of coherent flows tied to their host halos: their velocities are rescaled versions of the underlying DM radial and tangential components, preserving the dynamical coupling to the halo potential. However, when projected into redshift space, these flows do not align with the large-scale streaming motions of halos and instead appear observationally as incoherent, virial-like dispersions that drive the FoG effect. In practice, the velocity–bias parameters control the relative weight of the coherent components, while the additional random term $f_{\sigma}$ captures residual scatter. This dual description allows \ho to remain physically grounded in the halo velocity field while still reproducing the apparent incoherent small-scale anisotropy in galaxy clustering.

For \texttt{ELG1}, the model predicts $b_{\rm r}\sim0.873$ and $b_{\rm t}\sim0.802$, indicating sub-virial velocity dispersions in both the radial and tangential components, with only mild orbital anisotropy. In contrast to previous ELG HOD studies \citep[e.g.][]{2023JCAP...10..016R}, which typically infer $b_{\rm r}, b_{\rm t}\gtrsim1$, suggesting satellites dynamically hotter than the dark matter, \ho does not require such tangential heating. Instead, ELG satellites appear kinematically cooler than the host halo particles and only gently biased toward radial motions.

This behavior is physically consistent with the idea that ELGs are, on average, recently accreted satellites, which have not yet undergone strong dynamical processing (e.g., tidal stirring, harassment) nor fully experienced environmental quenching in massive halos. Their phase–space properties therefore reflect a population still transitioning from infall to virialization.

For \texttt{LRG3}, we infer $b_{\rm r}\sim0.959$ and $b_{\rm t}\sim1.180$, indicating mildly suppressed radial motions but tangential dispersions modestly hotter than those of the dark matter. This anisotropy suggests that LRG satellites have undergone significant dynamical evolution: efficient orbital decay and tidal stripping reduce their radial kinetic energy, while the preferential removal of satellites on plunging orbits leaves the surviving population on more circular, tangentially supported trajectories. Such sub-virial radial motions combined with mild tangential heating are a natural outcome of long-term dynamical friction and environmental processing within massive halos.

Importantly, while the $b_{\rm r}$ and $b_{\rm t}$ values we find still describe satellites as part of coherent flows around their hosts, the balance of those flows is shifted: radial infall is suppressed ($b_{\rm r}<1$) and tangential dispersion is enhanced ($b_{\rm t}>1$). This freedom is essential to reproduce the observed anisotropy level in the small-scale clustering. In other words, satellites in both the \texttt{ELG1} and \texttt{LRG3} samples remain dynamically coupled to their hosts, but with different anisotropic weights in the radial versus tangential components, such that the flows no longer resemble simple radial infall but instead reflect hotter, tangentially biased orbits.

These findings align with predictions from state-of-the-art hydrodynamical simulations, such as IllustrisTNG, which show that satellite velocity bias is typically suppressed below unity for massive galaxies, with only weak dependence on host halo mass and satellite properties \citep[][]{2022MNRAS.510.2980A}. They are also consistent with recent DESI HOD constraints for LRGs \citep{2024MNRAS.530..947Y}, which similarly indicate sub-virial satellite motions across the relevant redshift range. Together, these comparisons support a dynamical picture in which LRG satellites are cooler than the surrounding dark matter, consistent with long-term orbital decay and stripping within massive hosts.

Earlier semi-analytic results \citep{2018MNRAS.475.2530O} predicted $b_{\rm r}$ and $b_{\rm t}$ values closer to or above unity for LRG-like tracers, implying kinematically hotter satellites. The slight departure from those findings likely reflects differences in the physical treatments of environmental quenching and tidal processing, and emphasizes that DESI clustering adds new constraining power on satellite dynamics.

As highlighted in \S\,\ref{sec:clustering}, the satellite fraction parameters compete with the physics of peculiar motions driven by the velocity-bias parameters: increasing (decreasing) $b_{\rm r}$ or $b_{\rm t}$ amplifies (suppresses) the clustering amplitude below $10\,h^{-1}{\rm Mpc}$, making velocity anisotropy a key lever for matching DESI redshift-space observables.

For the central velocity bias, we obtain $b_{\rm cen} \sim 0.116$ and $b_{\rm cen} \sim 0.130$ for \texttt{ELG1} and \texttt{LRG3}, respectively. In our implementation (Eq.,\ref{eq:vzcen}), centrals sit at the halo center and follow the bulk peculiar velocity of their host. The parameter $b_{\rm cen}$ controls an additional residual motion relative to the halo center-of-mass, modeled as a fraction of the local dark matter velocity dispersion.

These very small $b_{\rm cen}$ values indicate that centrals are nearly at rest with respect to their host halo, as expected if they closely trace the minimum of the gravitational potential. While this residual motion has negligible impact on large scales, even a small non-zero $b_{\rm cen}$ can slightly broaden the line-of-sight distribution of pairs involving centrals. This leads to a mild suppression of the quadrupole on small, FoG-dominated scales—acting together with the satellite velocity anisotropy to match the observed redshift-space distortions.

In Eq.\,\ref{eq:vzsat}, the parameter $f_{\sigma}$ sets the amplitude of an additional stochastic LOS velocity component, expressed in units of the DM 1D velocity dispersion. This term is meant to capture small-scale, uncorrelated motions not described by the coherent radial and tangential components controlled by $b_r$ and $b_t$. We find $f_\sigma\sim0.128$ for \texttt{ELG1} (i.e., $\sim13\%$ of the DM 1D dispersion), and $0.050$ for \texttt{LRG3} ($\sim5\%$). These values introduce a modest additional random LOS velocity contribution, which enhances FoG damping on small scales. As expected, the effect of $f_\sigma$ is partially degenerate with $b_{\rm r}$ and $b_{\rm t}$, since all three parameters enter the total LOS velocity dispersion of satellites and therefore jointly shape the small-scale redshift-space anisotropy.

These results demonstrate that a realistic treatment of satellite peculiar motions is essential to reproducing the observed redshift-space anisotropy, and they highlight how peculiar velocities can serve as a sensitive probe of satellite dynamics in future cosmological analyses.

\subsubsection{\ho HOD and conformity predictions} 
\label{sec:hodpred}

\begin{figure*}
\centering
\includegraphics[width=0.8\linewidth]{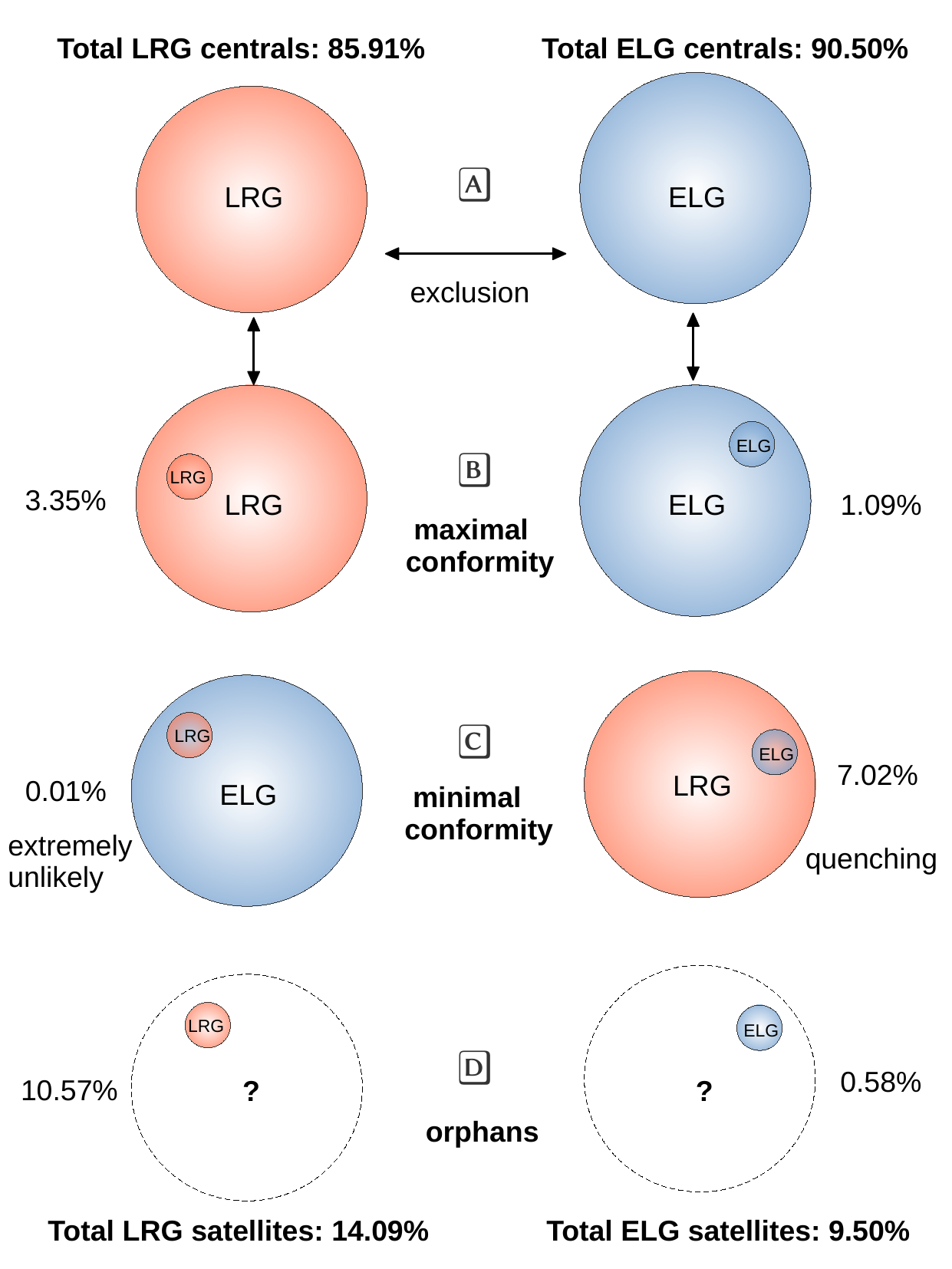}
\caption{Cartoon illustration of the \ho-inferred halo occupation configuration for the DESI Y1 \texttt{ELG1} and \texttt{LRG3} tracers. Larger (smaller) solid circles denote central (satellite) halos, while orphan satellites—i.e. satellites with no corresponding central in the final mock catalog—are shown as satellites inside dashed circles. For each tracer we identify four characteristic configurations: (A) central galaxies residing in halos with no satellites; (B) central systems hosting one or more satellites of the same tracer species, corresponding to maximal conformity; (C) central systems hosting one or more satellites of the complementary tracer, representing minimal conformity; and (D) orphan satellites with no corresponding central host in the catalog. The percentages next to each configuration are our model predictions.
}
\label{fig:cartoon}
\end{figure*}

The \textsc{Hom}e-inferred halo occupation distribution of the DESI Y1 \texttt{ELG1} and \texttt{LRG3} tracers is schematically illustrated in Figure\,\ref{fig:cartoon} and highlights both the predictive power and internal consistency of our model.

Its behavior is a direct consequence of the probabilistic class assignment based on the mixture model (\S\,\ref{sec:class}), where galaxy types are not predetermined, but stochastically sampled from a Gaussian multi-tracer PDF formulated as a function of $V_{\rm peak}$. This PDF is weighted and normalized using the input satellite fractions (Table\,\ref{tab:clusteringresu}) and the observed number densities of ELGs and LRGs (Eq.\,\ref{eq:normaliz}) to determine the relative contribution of each tracer to the clustering. This process yields a high-fidelity mock galaxy catalog simultaneously reproducing the auto- and cross-correlation functions of ELGs and LRGs with unprecedented accuracy. 

Figure \ref{fig:cartoon} shows that, for both tracers, the dominant configuration predicted by our model is that of central halos hosting no satellites---90.50\% for ELGs and 85.91\% for LRGs---while the remaining 9.50\% (ELGs) and 14.09\% (LRGs) are satellites. Among these, only 1.09\% (3.52\%) of ELG (LRG) satellites reside in halos whose central galaxy is of the same tracer type, i.e. the maximally conformal configuration. On the other hand, a substantial fraction of ELGs (7.02\%) and a negligible fraction of LRGs (0.005\%, consistent with observations) inhabit hosts with a central galaxy of the complementary type, leading to minimally conformal configurations. The remainder---0.58\% of ELGs and 10.57\% of LRG--—are orphan satellites, i.e. satellites whose parent halos do not appear as centrals in the final realization.

From these numbers, we deduce that \emph{satellite ELG strongly prefer minimally conformal configurations, while satellite LRG are mostly maximally conformity and orphan field galaxies.}

A sizable orphan contribution is expected in \ho because satellites are drawn from the full reservoir of DM particles surviving abundance–matching selection. Since LRGs are assigned first, without any conditional dependence on ELGs, it is encouraging that the model still produces a non-negligible fraction of maximally conformal LRG satellites. For ELGs, the joint-occupation (environmental quenching) condition strongly restricts their ability to remain satellites inside massive, LRG-dominated halos. This not only matches physical expectations from gas-depletion processes but also naturally reduces the orphan fraction while enhancing the cross-occupancy signal.

Therefore, it is crucial to highlight that \emph{the relative proportions of satellites in maximally/minimally conformal environments---and the orphan content of each tracer---are not imposed inputs but direct forward-model predictions that emerge self-consistently from the interplay of abundance matching, halo exclusion, and environmental quenching.}
These predictions can be validated with hydrodynamical simulations and forthcoming data, enabling direct tests of environmental quenching pathways.

\begin{figure}
\centering
\includegraphics[width=\linewidth]{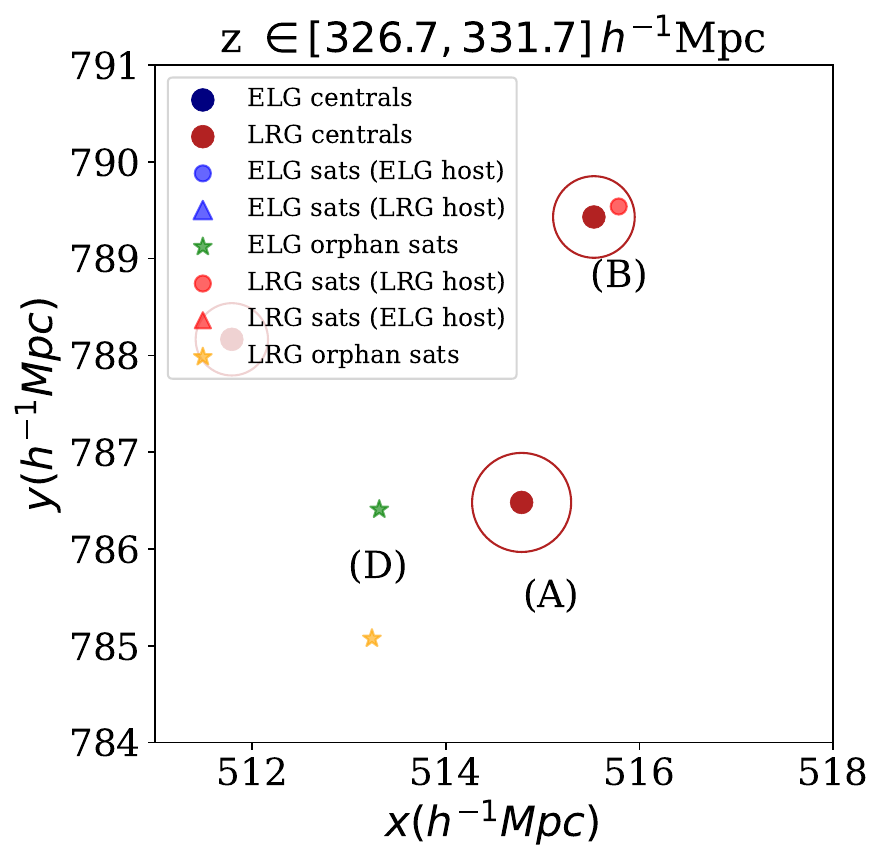}\vspace{0.4cm}
\includegraphics[width=\linewidth]{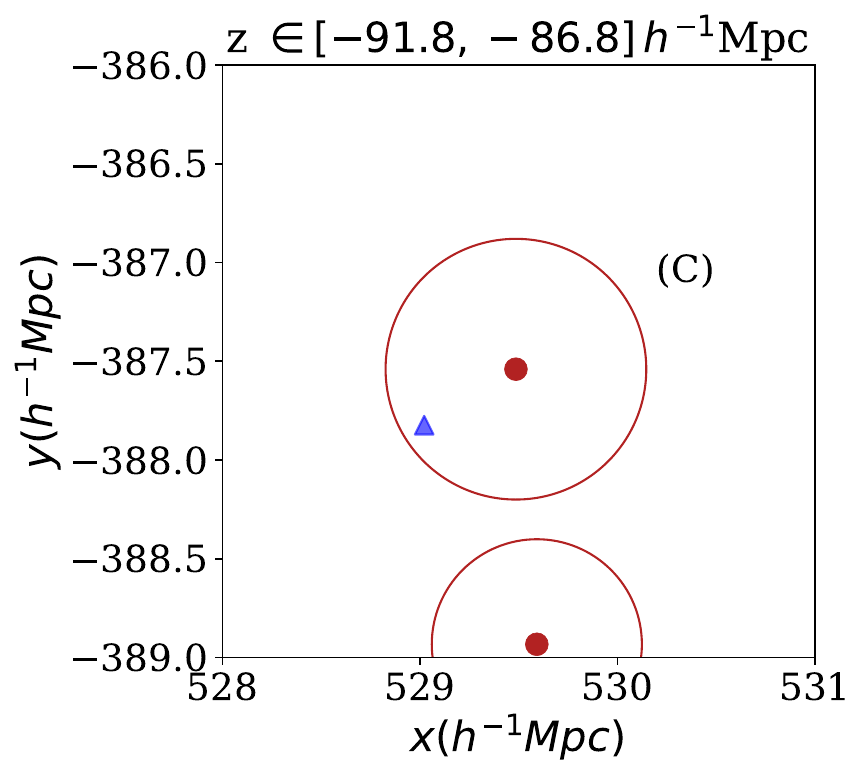}
\caption{Slices---$5\,h^{-1}$Mpc thick in $z$--- of our high-fidelity mock catalog for DESI Y1 ELGs and LRGs, showing the four configurations predicted by our forward model, which are schematically represented in Figure\,\ref{fig:cartoon}.
}
\label{fig:slices}
\end{figure}

In particular, the ELG$\times$LRG joint-occupation condition captures, in a statistical way, the environmental quenching experienced by ELG satellites as they enter the deep potential wells of massive LRG hosts---an event that our model predicts to happen in most of the ELG satellites (7.02\% out of 9.50\%). The same prescription also suppresses the occurrence of LRG satellites in ELG halos, making such configurations extremely rare (0.005\%), in line with current observational evidence.

Taken together, the physical ingredients of our model---i.e., satellite fractions, stochastic class assignment, halo exclusion, joint-occupation condition---reinforce the physical realism of \textsc{Hom}e, demonstrating that the quenching of ELG satellites in massive LRG environments and the lack of LRG satellites orbiting ELG hosts arise self-consistently from the forward model rather than being imposed by hand.

\begin{figure}
\centering
\includegraphics[width=\linewidth]{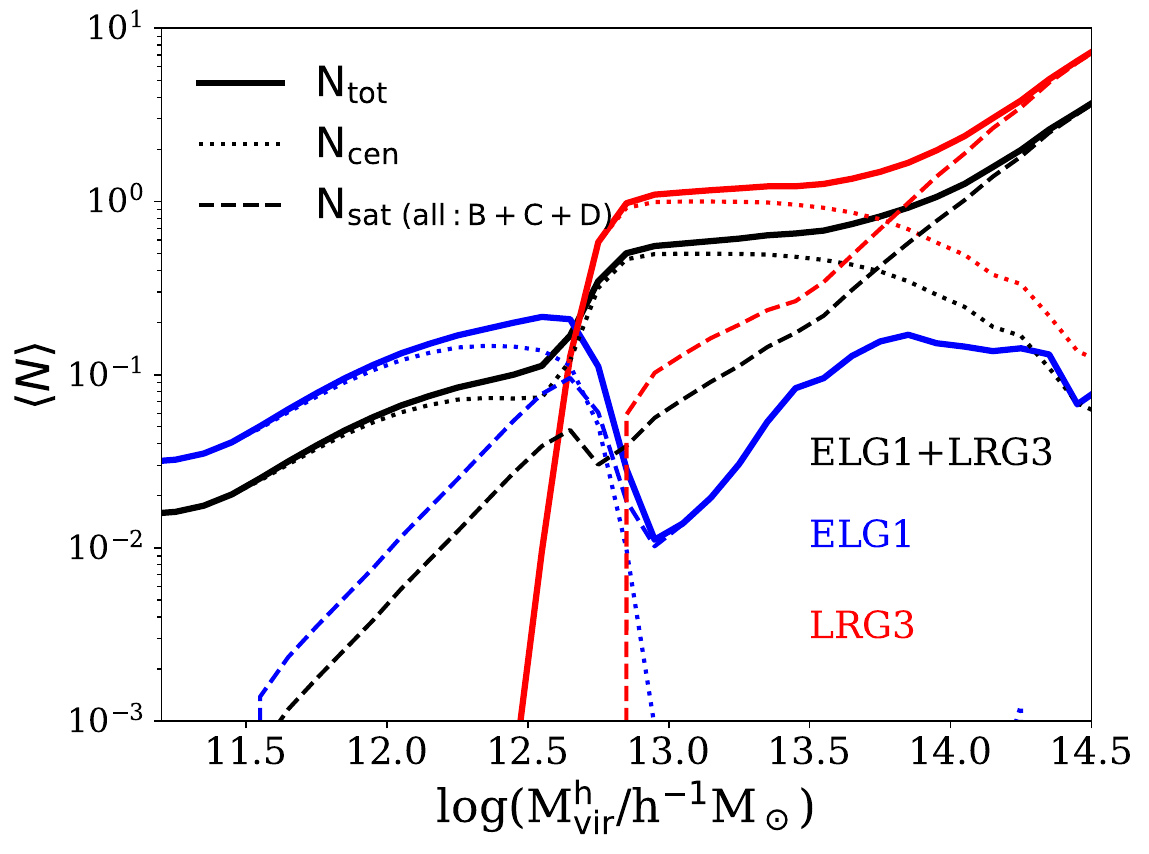}
\includegraphics[width=\linewidth]{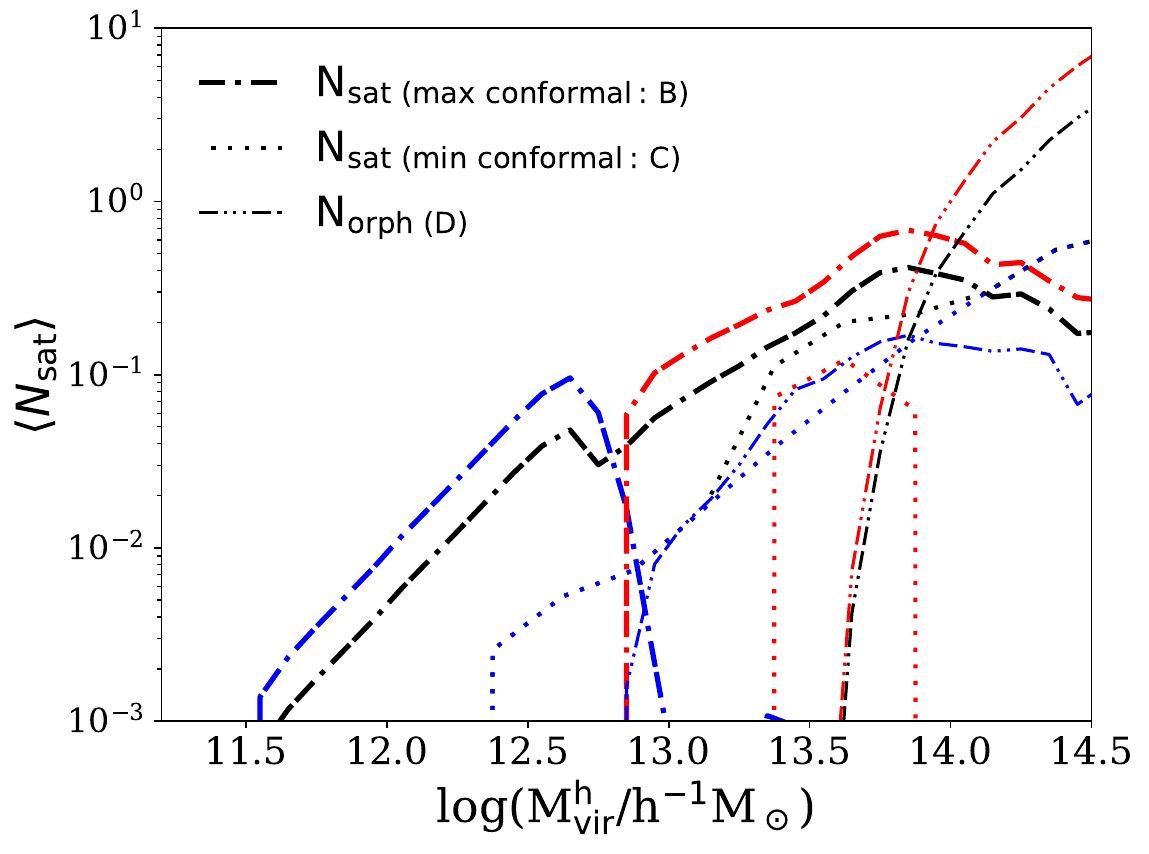}
\caption{\ho-inferred HOD as a function of the parent host halo virial mass. The contributions from \texttt{ELG1} and \texttt{LRG3} are respectively shown in blue and red; the global one is in black. The expected numbers of total, central, and satellite mocks are represented as solid, dotted and dashed lines, respectively. In the top panel the number of satellites is the total one, including the configurations (B), (C), (D) in Figure\,\ref{fig:cartoon}. In the bottom panel we separate the satellite contributions: maximally (minimally) conformal satellites are represented as dot-dashed (dotted) lines, and orphans as dot-dot-dot-dashed; these are the (B), (C), and (D) contributions which, once summed up, return the total satellite curves in the top panel. Note that we are able to isolate the contributions of orphans because we are showing the HOD as a function of the parent host halo mass, and for orphans we keep track of the parent halo mass of the DM particles in the simulation.}
\label{fig:hodglobal}
\end{figure}

Figure\,\ref{fig:slices} shows two slices of our high-fidelity \ho mock catalog, $5\,h^{-1}$Mpc thick along the $z$ coordinate, where we identify all possible configurations of the DESI  Y1 \texttt{ELG1} and \texttt{LRG3} tracers as schematically described in Figure \ref{fig:cartoon}.

The top panel in Figure\,\ref{fig:hodglobal} shows the \textsc{Hom}e-inferred HOD of the \texttt{ELG1} and \texttt{LRG3} tracers as a function of the parent host halo mass. Here, the satellite contribution is the total one including the (B), (C), and (D) configurations in Figure\,\ref{fig:cartoon}. Even though the orphans have no central host in the resulting mock catalog, we are able to represent them here as we keep track of the parent halo mass of the DM particles in the simulation. In the bottom panel we separate the contributions of maximally conformal, minimally conformal, and orphan satellites.

The central occupations inferred for \texttt{ELG1} and \texttt{LRG3} are consistent with predictions from recent HOD \citep[e.g.,][]{2018MNRAS.474.4024G, 2020MNRAS.499.5486A, 2021MNRAS.502.3599H, 2022MNRAS.510.3301Y, 2023JCAP...10..016R, 2024MNRAS.530..947Y}
and SHAM \citep[e.g.,][]{2022MNRAS.516...57Y, 2025A&A...698A.170P}
studies. As shown in Figure\,\ref{fig:hodglobal}, central ELGs (LRGs) predominantly occupy lower- (higher-)mass halos, with characteristic masses of $M_{\rm vir}\sim6.6\times10^{11}\,h^{-1}{\rm M}_\odot$ for ELGs and $M_{\rm vir}\sim1.2\times10^{13}\,h^{-1}{\rm M}_\odot$ for LRGs. Notably, the shape of the ELG central and satellite HODs exhibits an excellent match with previous HOD and SHAM analyses, in particular \citet{2018MNRAS.474.4024G}, including the location of the turnover and the high-mass suppression, reinforcing the physical robustness of our abundance–matching prescription.

For \texttt{LRG3}, the satellite component follows the steep, nearly power-law scaling with halo mass reported in earlier LRG clustering studies
\citep[e.g.,][]{2009ApJ...707..554Z, 2009ApJ...698..143R, 2024MNRAS.530..947Y}. Including orphan satellites preserves this canonical shape, whereas removing them slightly flattens the high-mass tail, although the overall mass scale and normalization remain consistent with published constraints.

At higher masses, the ELG occupations departs from a pure power law due to the combined effects of ELG satellite quenching in massive hosts---through mechanisms such as ram-pressure stripping, starvation, or tidal heating—and the halo-exclusion condition; for LRG the suppression comes from exclusion only. 

This high-mass suppression is a signature of central–satellite conformity and is essential to reproducing the observed shape and amplitude of the ELG $w_p(r_p)$ upturn around $r_p\sim10\,h^{-1}{\rm Mpc}$, which is sensitive to the interplay between ELG and LRG environments.

Orphan satellites contribute increasingly at high masses, becoming a large fraction of the already small satellite budget above $M_{\rm vir}^{\rm h}\gtrsim10^{14}\,h^{-1}{\rm M}_\odot$.

As reported in Table\,\ref{tab:nsat}, \ho predicts that the large majority of central \texttt{ELG1} and \texttt{LRG3} host individual satellites (93.17\% and 85.94\%, respectively), few of them host 2 (3.01\%; 3.86\%), while more than 2 satellites are rare. These results are in line with previous SHAM studies at lower $z$ showing that only 1.3\% of ELG hosts have more than 1 satellite \citep[][]{2016MNRAS.461.3421F}. 

\begin{deluxetable}{cccc}
\tablecaption{Percentages \label{tab:nsat} of DESI ELG and LRG tracers hosting 1, 2, and $3$ satellites, as predicted by \textsc{Hom}e.}
\tablehead{
  \colhead{$N_{\rm sat}$} & \colhead{\texttt{ELG1}}& \colhead{\texttt{LRG3}}
  }
\startdata
1 & 93.17&85.94 \\
2&3.01&3.86 \\
3&0.18&1.00 
\enddata
\end{deluxetable}

Our satellite placement scheme combines host and particle positions through Eq.,\ref{eq:outskirt}, with the parameter $K_{\rm out}$ rescaling the intrinsic dark-matter profile. We find that \texttt{ELG1} and \texttt{LRG3} favor $K_{\rm out}\sim1.028$ and $1.105$, respectively. At first glance, this might seem counterintuitive—ELGs are typically associated with more extended, diffuse satellite distributions, whereas LRG satellites are expected to concentrate toward halo centers. However, this trend is fully consistent with the host-halo selection imposed via the minimal velocity thresholds $V_{\rm peak}^{\rm min}$ (Table\,\ref{tab:clusteringresu}).

The relatively low \texttt{ELG1} threshold ($V_{\rm peak}^{\rm min}>110\,s^{-1}$km) selects late-forming halos with shallower gravitational potentials and intrinsically lower concentrations. Their satellite distributions are already extended; thus, $K_{\rm out}$ must be close to unity—and even slightly below—to avoid over-expanding satellites and inflating the small-scale clustering.

In contrast, the high threshold for \texttt{LRG3} ($V_{\rm peak}^{\rm min}>370\,s^{-1}$km) preferentially selects earlier-forming, more concentrated environments. Here, $K_{\rm out}>1$ is needed to counteract this concentration bias and reproduce the observed flattening of LRG $w_p$ at sub–Mpc scales.

Taken together, this highlights the role of $K_{\rm out}$ as a concentration-correction parameter: it compensates for assembly-bias effects introduced by the velocity-based halo selection while preserving realistic halo density profiles and maintaining agreement with observed one-halo clustering signatures.


\section{Summary and conclusions} \label{sec:summary}
We have introduced \textsc{Hom}e, a physically motivated Halo Occupation Model designed to generate high-fidelity mock galaxy catalogs for any galaxy tracer observed by any survey, using any $N$-body cosmological simulation---including those lacking resolved satellite halos. The method is embedded in a two-level hierarchical Bayesian inference framework, optimized to tightly constrain the physical parameters of the model and deliver realistic mocks with unprecedented accuracy.

The excellent performance of the method has been demonstrated by analyzing DESI Y1 ELG, LRG, and ELG$\times$LRG clustering measurements using the \textsc{AbacusSummit} $N$-body cosmological simulation products.

\ho galaxy-halo connection scheme is a hybrid mixture of abundance matching and halo occupation distribution which overcomes the lack of substructures (not tracked by the halo finder in \textsc{AbacusSummit}) by assigning satellites via DM particle positions. Our method provides reliable satellite peak circular velocities ---based on analytic prescriptions informed by external high-resolution $N$-body simulations--- to jointly model the ELG and LRG populations as complementary tracers of the same underlying dark-matter field. This enables, for the first time, a full reconstruction of their cross halo occupation distribution.

Our scheme fully accounts for the ELG and LRG intra-halo dynamics, halo exclusion, and ELG satellite quenching in LRG hosts. These are crucial ingredients to precisely model the anisotropic clustering on sub-Mpc scales, correctly shaping the 1-halo to 2-halo transition. Galaxy conformity naturally emerges as a byproduct of our forward model.
Our main findings are:
\begin{enumerate}
\item \ho predicts the DESI Y1 ELG and LRG auto- and cross-correlation functions (multipoles and projected ones) down to $200\,h^{-1}$kpc with unprecedented accuracy, precisely modeling the intra-halo dynamics and matching the observed anisotropy in the Universe. We find that satellite ELG dominate the anisotropic clustering below $4\,h^{-1}$Mpc, acting as incoherent flows with a velocity bias relative to their hosts. The \textsc{Hom}e-inferred HOD shows that: (i) 90.50\% (85.91\%) of ELG (LRG) are central galaxies with no satellites, typically residing in halos with masses of $M_{\rm vir}\sim6.6\times10^{11}\,(1.2\times10^{13})\,h^{-1}{\rm M}_\odot$; (ii) the ELG$\times$LRG cross-correlation is dominated by central–central pairs and shaped by halo exclusion on $2-5\,h^{-1}$Mpc scales; (iii) the remaining 9.50\% (14.09\%) of ELG (LRG) are satellites. Of these, 1.09\% (3.52\%) occupy a parent halo with a central galaxy of the same type (maximal conformity), 7.02\% (0.005\%) are minimally conformal living in a complementary host, and the remaining 0.58\% (10.57\%) are orphans. 

\item From the above numbers, we conclude that the ELG$\times$LRG joint occupation not only reliably emulates the quenching of ELG satellites in LRG hosts, but also prevents LRG satellites to orbit ELG centrals, matching current observational constraints \citep[e.g.,][]{2021ApJ...918...53G, 2013MNRAS.432..336W, 2014MNRAS.444.2938H, 2024ApJ...971..111R}, in line with recent HOD \citep{2020MNRAS.497..581A, 2024MNRAS.530..947Y} and SAM \citep{2018MNRAS.475.2530O, 2021MNRAS.500.4004D} studies. As a result, ELG satellites tend to prefer a minimally conformal picture, before getting quenched, while LRG satellites are either maximally conformal or orphan field galaxies.

\item The best-fit \ho model casts 9.50\% of ELGs as satellites. These dominate the anisotropic clustering signal below $4\,h^{-1}$Mpc, acting as incoherent flows with a velocity bias relative to their hosts. The interplay between halo exclusion and quenching is essential to jointly reproduce the observed clustering of all tracers at all scales considered. Galactic conformity naturally emerges from \ho at the 1-halo level, and the picture we obtain is reminiscent of the conformity built within conditional HODs \citep[e.g.,][]{2015MNRAS.454.3030P,2015MNRAS.454.1161Z,2016MNRAS.457.4360Z}, but arises here without explicit assumptions. By pushing our analysis down to $200\,h^{-1}$kpc, we reach $\sim20$ times higher resolution than previous studies \citep{2022MNRAS.516...57Y}, extracting cosmological information from two-point clustering alone while relying only on the dark matter distribution of the reference $N$-body simulation.

\item In the \ho configuration, satellite ELGs remain dynamically coupled to their host halos, with velocity-bias parameters of $b_{\rm r}\sim0.873$ and $b_{\rm t}\sim0.802$. These values indicate dispersions modestly below those of the dark matter in both components, together with a mild radial preference ($b_{\rm r}>b_{\rm t}$). Thus, our model predicts that ELG satellites are kinematically cooler than the virial expectation but not isotropic: their orbits retain slightly stronger radial coherence than tangential support. This subtle anisotropy is sufficient to shape the anisotropic clustering signal below $4\,h^{-1}$Mpc---not by decoupling satellites from their hosts, but by imprinting a modest departure from isotropic satellite motions.
In contrast, satellite LRGs occupy a different dynamical regime. We infer $b_{\rm r}\sim0.959$ and $b_{\rm t}\sim1.183$, implying that their radial motions are slightly sub-virial while their tangential dispersions are modestly enhanced relative to the dark matter. This anisotropic pattern---cooler radial infall but hotter, more circular orbits---is indicative of a dynamically evolved population in which long-term tidal stripping and dynamical friction have preferentially removed radially plunging satellites, leaving survivors on tangentially supported orbits in massive halos. Their higher satellite abundance ($f_{\rm sat}\sim14.09\%$) makes LRGs the dominant contributor to the 1-halo term. Moreover, the additional incoherent velocity term ($f_\sigma\sim0.053$, i.e. $\sim5\%$ of the DM 1-D dispersion) produces a mild broadening of the line-of-sight pairwise motions, contributing a small but non-negligible suppression of anisotropy on the smallest scales.

\item Interestingly, the sizeable orphan fraction we recover from \ho is almost entirely associated with LRGs, while ELG orphans remain negligible due to the quenching–driven joint-occupation mechanism. These LRG orphans appear as satellites without a corresponding central galaxy in the final realization, placing them in regions of the density field outside the virial extent of any identified host. Physically, such systems may correspond to backsplash galaxies, recently stripped satellites, or objects formerly bound to halos that have since been disrupted below the simulation’s resolution. Alternatively, they could reflect ejected systems that have interacted with massive environments and now linger in the diffuse outskirts of groups and filaments. Pinning down the nature of this population requires further investigation, which we plan to pursue in follow-up work using high-resolution hydrodynamical simulations, such as IllustrisTNG\footnote{\url{https://www.tng-project.org}}, where the baryonic component allows a more direct connection between galaxy quenching, stripping, and subhalo disruption. This will enable us to quantify the impact of baryonic feedback in our model. Understanding this population is key to refining halo–galaxy connection models and improving the realism of mock catalogs for future spectroscopic surveys. 

\end{enumerate}

The presence of this orphan component also motivates a thorough observational census of ELGs and LRGs using deep and wide-field galaxy surveys, such as DESI-II, Subaru PFS, or Rubin–LSST, to clarify their exact satellite fraction and its impact on the small-scale anisotropic clustering. This information, coupled with the remarkable constraining power of our method, will possibly allow us to discriminate among different cosmologies, including alternative gravity frameworks, such as $f(R)$. Our model performance will increase further by doing redshift tomography, which we plan to implement in follow-up studies.

Looking ahead, expanding our analysis to diverse $N$-body simulations with different cosmologies and incorporating additional physical constraints promises to significantly impact the calibration of the latent satellite properties and further refine the parameter space of the galaxy-halo connection framework.

\section*{Acknowledgements}
GF acknowledges F. Sinigaglia, A. Rocher and S. Saito for insightful discussions during the development of this work. She also thanks A. Carnero, A. Ross, A. de Mattia and D. Chebat for helping with the DESI data model infrastructure.

During the first stage of this work, GF has been supported by a {\em Juan de la Cierva Incorporaci\'on} grant n.\,IJC2020-044343-I. 
GF and FSK acknowledge the Spanish Ministry of Economy and Competitiveness (MINECO) for financing the \textit{Big Data of the Cosmic Web} project:
PID2020-120612GB-I00/AEI/10.13039/ 501100011033 under which this work has been conceived and carried out, and the IAC for continuous support to the \textit{Cosmology with LSS probes} project. 

The \ho method has been developed within the \textit{MUSICA (MUlti-tracer Skies for hIgh-precision Cosmological Analyses)} research line pursued at the IAC as part of the \textit{COSMIC SIGNAL (COSMIC SImulated Galaxy Networks Applied on Lightcones)} project: \url{www.cosmic-signal.org}. In this context, the reference catalogs generated with \ho will also serve to calibrate the massive production of covariance mocks within the \textit{FIRE (Field-level bayesian Inference to Reconstruct the univErse)} project, which has been recently funded by a \textit{Proyecto de Generaci\'on de Conocimiento 2024}, PID2024-160504NB-I00. 

DJE's contributions were supported by U.S. Department of Energy grant DE-SC0007881, by the National Science Foundation under Cooperative Agreement PHY-2019786 (the NSF AI Institute for Artificial Intelligence and Fundamental Interactions, \url{http://iaifi.org/}), and as a Simons Foundation Investigator.

SB is supported by the UKRI Future Leaders Fellowship [grant numbers MR/V023381/1 and UKRI2044].

Abacus development has been supported by NSF AST-1313285 and more recently by DOE-SC0013718, as well as by Simons Foundation funds and Harvard University startup funds. NM was supported as a NSF Graduate Research Fellow. The AbacusCosmos simulations were run on the El Gato supercomputer at the University of Arizona, supported by grant 1228509 from the NSF; the \textsc{AbacusSummit} simulations have been supported by OLCF projects AST135 and AST145, the latter through the Department of Energy ALCC program.
This research used resources of the Oak Ridge Leadership Computing Facility, which is a DOE Office of Science User Facility supported under Contract DE-AC05-00OR22725, and resources of the National Energy Research Scientific Computing Center (NERSC), a U.S. Department of Energy Office of Science User Facility located at Lawrence Berkeley National Laboratory, operated under Contract No. DE-AC02-05CH11231.
We would like to thank the OLCF and NERSC support teams for their expert assistance throughout this project.
We would also like to thank Stephen Bailey, Chia-Hsun Chuang, Shaun Cole, Pablo Fosalba, Salman Habib, Katrin Heitmann, Core Francisco Park, Joachim Stadel, Risa Wechsler, and Sihan Yuan for useful conversations about the \textsc{AbacusSummit} program.

\appendix

\section{Satellite placement in hosts: particle- versus halo-perspective}
\label{sec:occup}
The satellite occupation scheme in \S\,\ref{sec:positions} can be implemented either at the particle or at the halo level with subtle, yet important, differences. We summarize them here: 
\begin{itemize}
\item \emph{Particle level:} each DM particle is assigned a retention probability computed from Eq.\,\ref{eq:powerlaw}. Then a Bernoulli trial is performed for each particle to decide whether it is a satellite, based on that probability.

If not properly normalized, these probabilities tend, by construction, to overpopulate most massive hosts, skewing the 1-halo clustering term. 
In fact, by sampling a very large number of trials per halo, one tends to systematically exceed the intended mean, especially in massive halos, which dominate the total satellite count. As a result, massive halos accumulate more excess, and this shifts the overall satellite abundance upward compared to a single draw per halo. 

In principle, this excess could also be mitigated by smoothly truncating the maximum number of satellites per host, but this introduces artifacts, especially in the projected clustering. Therefore, the best is to properly normalize the expected number of satellites to the total number of particles per halo and follow from there. 

The particle-level approach has the great advantage that satellites are sampled from a non-parametric, irregular spatial distribution, i.e. the actual particle distribution, which provides the model natural anisotropy and stochasticity, especially on small scales. This leads to a more flexible and descriptive clustering model in the sub-Mpc regime, with satellites tracing different DM features of the halo shape (e.g., triaxiality, concentration, sub-structure). 

\item \emph{Host level:} each host is assigned an expected number of satellites given by Eq.\,\ref{eq:powerlaw}. Then, satellites are selected among its DM particles using a Poisson distribution with that expected number. 
This guarantees full control on the maximum satellite occupancy in massive hosts, so that normalization or truncation are not necessary. 

However, the occupation is completely locked by the host halo, independently from how many particles it has, and does not capture the halo-to-halo variation (i.e., all halos of a given mass look identical). In other words, the halo-level occupation assumes satellites are indistinguishable and not tied to any particular substructure inside the halo. As a consequence, the clustering model looses flexibility and realism in the 1-halo term.
\end{itemize}

The satellite occupation we adopt in the analysis is at the particle level with proper normalization. In this way, we maximize the control of the occupancy, as well as the model flexibility; see \S\,\ref{sec:positions} for details.

\section{Satellite core suppression}
\label{sec:core_supp}
The cuspy behavior of DM particle profiles in high-resolution simulations can be mitigated by probabilistically downsampling particles in the halo core using a retain probability defined as:
\begin{equation}
    P_{\rm keep}(r_i)=\left(\frac{r_i}{r_i+R_{{\rm h},i}^{\rm vir}}\right)^\alpha\hspace{0.5cm}{\rm with\,\,}i=1,\cdots N_{\rm p},
    \label{eq:coresupp}
\end{equation}
where $r_i$ is the 3D radial distance of each particle from its central host, $R_{{\rm h},i}^{\rm vir}$ is the host virial radius fixing the scale of the transition, and $\alpha$ determines the steepness of the suppression.

This downsampling reduces the number of satellites selected near the halo core, mimicking halo core disruption, while preserving the outer halo profile. This is especially effective for tuning the projected correlation function, leaving the anisotropic clustering mostly unaffected.

This core suppression can be combined with a radial suppression of close particle-particle (i.e., satellite-satellite) pairs in projected 2D separation, $r_{\rm p}$ based on a retain probability defined as:
\begin{equation}
    P_{\rm keep}(r_{\rm p})=\frac{s}{1+\left(\frac{r_{\rm p}}{r_{\rm cut}}\right)^2}\,,
    \label{eq:radialsupp}
\end{equation}
where $r_{\rm p}$ is the projected separation of the satellites composing the pair, $s$ is a parameter regulating the strenght of the suppression, and $r_{\rm cut}$ is the maximum radius withn which the suppression is applied. This specific shape ensures high suppression at small separations and gradual retention at larger distances.

Note that, in our analysis, we do not apply any of the above downsamplings, as we modulate the satellite profile using the $K_{\rm out}$ parameter (Eq.\ref{eq:outskirt}).

\section{Impact of the model parameters on the clustering}
\label{sec:impact_params}
We investigate how the \texttt{ELG1}, \texttt{LRG3}, and \texttt{ELG1$\times$LRG3} two-point statistics respond to $\pm10\%$ variations of the \ho physical parameters around their fiducial values reported in Table\,\ref{tab:clusteringresu}. As expected, the LRG auto-correlation functions are not affected by a change in the ELG model parameters, as the LRG selection is completely independent. On the other hand, the ELG auto-correlations do change with the HOD LRG parameters, since the ELG selection is conditional to the LRG assignment.  In other words, the LRG parameters propagate into the ELG and cross observables, while the ELG variables do not feed back into LRG clustering.

In what follows, we show the ratios of the model including the variation over the fiducial one. We represent a $+10\%$ ($-10\%$) variation in blue (cyan) for ELG, in red (orange) for LRG, and in black (grey) for their cross-correlation.

Figure\,\ref{fig:impactHOD} shows the impact of changing the ELG HOD (top) and LRG HOD (bottom) parameters. We observe that:

\begin{figure*}
\centering
\includegraphics[width=\linewidth]{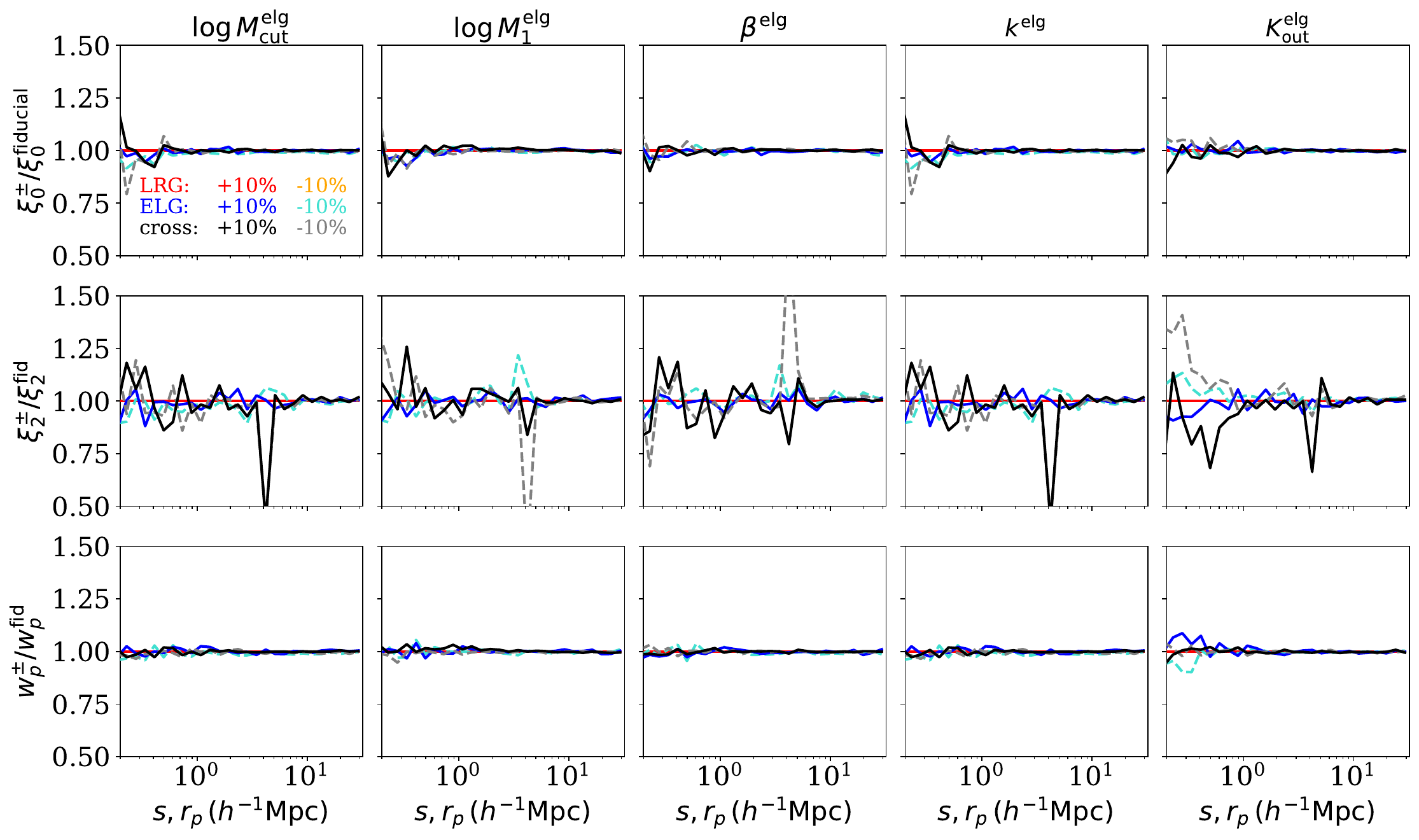}\vspace{0.1cm}
\includegraphics[width=\linewidth]{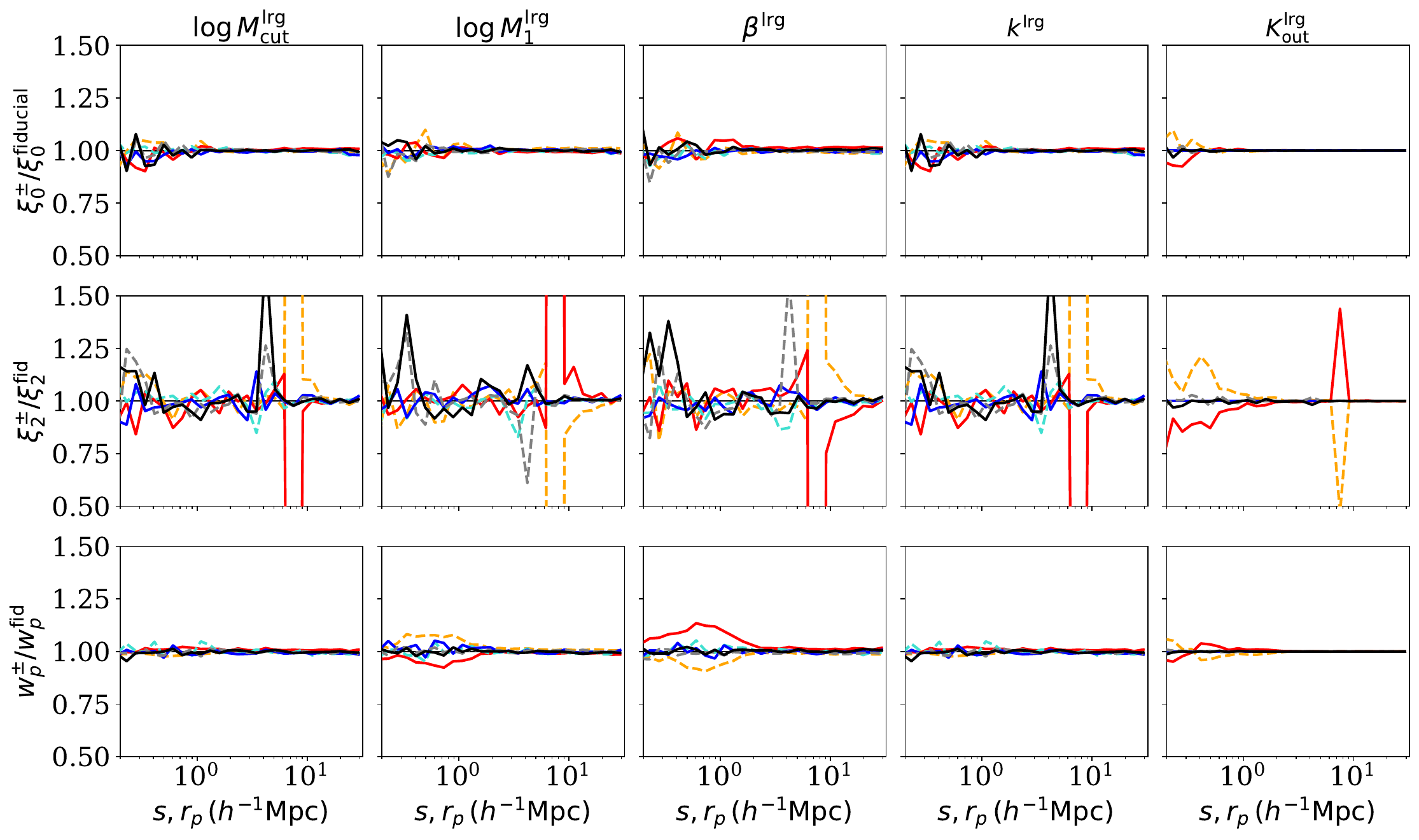}
\caption{Impact of a $\pm10\%$ variation in the HOD ELG (top paneles) and HOD LRG (bottom) fiducial best-fit parameters on the \texttt{ELG1} (blue and turquoise), \texttt{LRG3} (red and orange), and \texttt{ELG1$\times$LRG3} (black and grey) clustering. We show the ratio between the model including the variation and the fiducial one; on top of each panel we indicate the parameter that is varying.}
\label{fig:impactHOD}
\end{figure*}

\begin{itemize}
\item $\log M_{\rm cut}$:
Varying the cutoff mass has only a mild impact on the projected correlation functions, regardless of tracer, whereas the monopole and especially the quadrupole remain sensitive to it on small and intermediate scales. When the ELG cutoff mass ($\log M_{\rm cut}^{\rm elg}$) is varied (blue/turquoise curves), both the ELG and ELG×LRG clustering respond noticeably---most clearly in the quadrupole---because this parameter controls how many ELG centrals populate the low-mass end of the halo distribution. In contrast, varying the LRG cutoff mass ($\log M_{\rm cut}^{\rm lrg}$) (red/orange curves) induces a significant response not only in the LRG multipoles but also in the ELG and cross statistics (blue/turquoise and black/gray curves). This occurs because ELG assignment is conditional on the LRG catalog: shifting the LRG cutoff changes which halos are removed from the pool available to ELGs, thereby propagating LRG-side variations into all tracers.

\item $\log M_1$: Varying the satellite–onset mass scale $\log M_1$ produces strong, scale-dependent responses across all tracers, since this parameter controls the mass at which halos begin to host satellites. When the ELG parameter ($\log M_1^{\rm elg}$) is varied (blue/turquoise curves), the ELG and cross signals respond with a clear enhancement or suppression of the small-scale monopole, quadrupole, and $w_p(r_p)$, reflecting the rapid change in ELG satellite abundance. The impact on the LRG statistics (red/orange curves) is minimal, consistent with ELG variations not propagating backward into the LRG assignment.

By contrast, varying LRG parameter ($\log M_1^{\rm lrg}$; red/orange curves) induces a strong change not only in the LRG clustering but also in the ELG and cross correlations (blue/turquoise and black/gray curves). Because ELGs are assigned conditional to LRGs, shifting the LRG satellite threshold alters the distribution of halos already occupied by LRG satellites, modifying the pool of available central halos for ELG occupancy. Lowering $\log M_1^{\rm lrg}$ increases the number of LRG satellites, pushing ELGs into lower-mass halos and thereby reducing their small-scale contrast, whereas raising it has the opposite effect. As a result, the LRG $\log M_1$ parameter has a cascade effect, shaping not only the LRG 1-halo term, but also the ELG and ELG$\times$LRG small-scale amplitudes through the hierarchical selection built into \textsc{HOM}e.

\item $\beta$: Varying the satellite–occupation slope produces clear and highly scale-dependent effects on the small-scale clustering of all tracers, since this parameter regulates how rapidly the satellite number rises above 
$\log M_{\rm cut}$. When $\beta^{\rm elg}$ is varied (blue/turquoise curves), the impact is strongest in the cross monopole and quadrupole (black/gray curves) and remains visible in the ELG signal (blue/turquoise). A steeper slope quickly increases the ELG satellite fraction, amplifying the 1-halo term and deepening the quadrupole suppression from stronger virial motions. Because ELGs are selected based on LRGs, these variations have almost no effect on the LRG auto-correlation (red/orange).

Changes in $\beta^{\rm lrg}$ (red/orange curves) produce a much broader response: the LRG small-scale clustering is strongly modified—through enhanced or reduced satellite abundance---and this propagates into the ELG and cross statistics (blue/turquoise and black/gray curves). Steeper $\beta^{\rm lrg}$ values populate more satellites in high-mass halos, boosting the LRG 1-halo amplitude and simultaneously reducing the number of halos available for ELG occupation, which softens the ELG small-scale signal. In the quadrupole, increasing either $\beta^{\rm elg}$ or $\beta^{\rm lrg}$ enhances virial motions and deepens the small-scale anisotropy suppression, but the effect is strongest for LRGs due to their higher satellite fraction.

\item $\kappa$: It regulates how quickly the satellite occupation rises once the halo mass exceeds $\log M_{\rm cut}$. Its impact is weaker than that of 
$\log M_{\rm cut}$ or $\log M_1$, but it still leaves a noticeable imprint on the small- and intermediate-scale clustering. For ELGs (blue/turquoise curves), decreasing $\kappa^{\rm elg}$ shifts satellites toward lower-mass hosts and steepens the small-scale upturn in all clustering statistics. Because ELGs are assigned after LRGs, varying $\kappa^{\rm elg}$ also propagates into the cross-correlation, modifying its 1-halo term.

For LRGs (red/orange curves), the effect is milder: their satellite population already resides in massive halos where the occupation is near-saturated, so changing $\kappa^{\rm lrg}$ mainly induces small rescalings of the monopole and quadrupole. Nonetheless, the ELG signal still responds significantly, again due to the conditional ELG selection.

\item $K_{\rm out}$: It controls the radial repositioning of satellites within halos (Eq.\,\ref{eq:outskirt}). It has a strong, highly characteristic impact on all small-scale clustering statistics of both tracers.

Increasing $K_{\rm out}$ spreads satellites to larger radii, thereby suppressing the sharp small-scale rise in $w_p(r_p)$, lowering the monopole at $s\lesssim2\,h^{-1}$Mpc, and damping the quadrupole (weaker FoG elongation). Decreasing it concentrates satellites toward halo centers and enhances these signatures.

The effect is especially pronounced for LRGs (red/orange curves), whose high satellite fraction makes their 1-halo term extremely sensitive to the radial profile. As a consequence of the hierarchical assignment, varying $K_{\rm out}^{\rm lrg}$ also modifies the ELG and cross clustering (blue/turquoise and black/gray curves), because ELGs are placed after LRGs and thus respond to the altered spatial distribution of LRG hosts and satellites. Conversely, varying $K_{\rm out}^{\rm elg}$ affects only ELG and ELG$\times$LRG, leaving the LRG auto-correlation essentially unchanged.

\end{itemize}

Among the HOD ingredients, the parameters that most strongly shape the small- and intermediate-scale clustering are those that directly regulate satellite abundance and spatial distribution. The mass scale for satellite onset, $\log M_1$, is the single most influential lever: lowering it boosts the 1-halo term of all tracers, steepening the $w_p(r_p)$ upturn and enhancing the small-scale monopole and quadrupole. The satellite-occupation slope, $\beta$, provides the next most significant modulation, controlling how rapidly satellites accumulate above $\log M_{\rm cut}$ and therefore tuning the strength and steepness of the 1-halo signal. The parameter $K_{\rm out}$, which governs the radial placement of satellites, has a comparably strong effect: pushing satellites outward suppresses clustering below a few Mpc, while concentrating them enhances the FoG-induced anisotropy. In contrast, $\log M_{\rm cut}$ and $\kappa$ play secondary roles: they influence primarily the abundance of centrals in low-mass halos and only weakly affect the 1-halo term. Taken together, 
$\log M_1$, $\beta$ and $K_{\rm out}$ constitute the key physical drivers of the small-scale clustering response in both tracers and their cross-correlation.

\begin{figure*}
\centering
\includegraphics[width=0.9\linewidth]{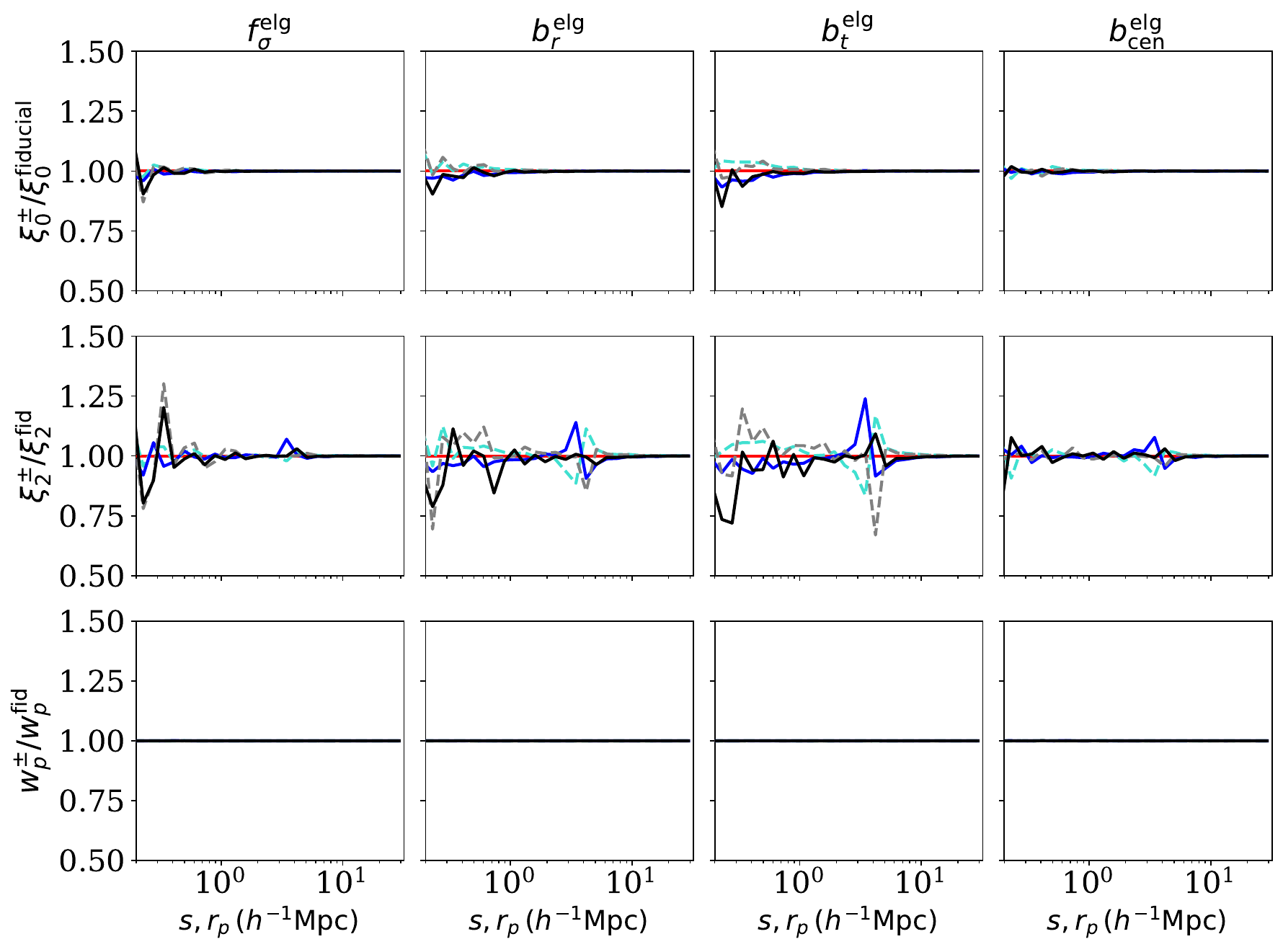}
\includegraphics[width=0.9\linewidth]{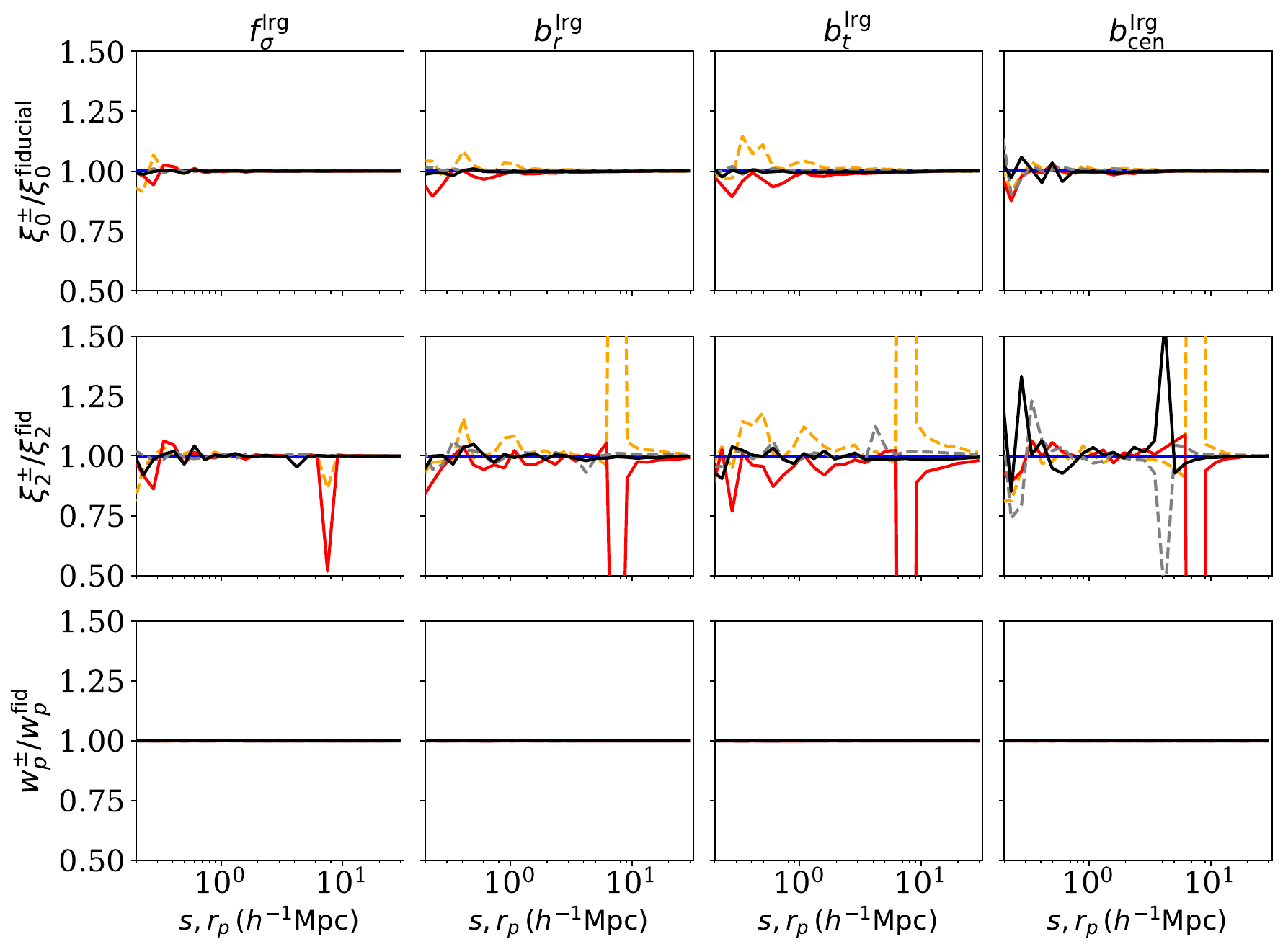}
\caption{Impact of a $\pm 10\%$ variation in the ELG (top) and LRG (bottom) velocity bias parameters on our clustering model. Lines are color-coded as in Figure\,\ref{fig:impactHOD}.}
\label{fig:impactVELB}
\end{figure*}


We now assess how the ELG and LRG velocity bias parameters affect the clustering by shaping the kinematics of satellites and centrals. As shown in Figure\,\ref{fig:impactVELB}, we find that:
\begin{itemize}
\item $f_{\sigma}$: The incoherent velocity dispersion of satellites produces strong FoG damping. Larger $f_{\sigma}$ suppresses the quadrupole substantially at small scales and reduces the monopole amplitude, especially in ELGs---where intra-halo satellite dynamics is most relevant--- and, indirectly, in ELG$\times$LRG. LRGs show a weaker response. 

\item $b_{\rm r}$: The radial velocity bias strongly impacts both the monopole and the quadrupole. Increasing $b_{\rm r}$ enhances coherent infall, making the quadrupole more negative on $s \lesssim 10\,h^{-1}{\rm Mpc}$ for all tracers, with the biggest effect in ELGs---where satellites dominate small–scale anisotropy-- and LRGs---where the satellite contribution is substantial. 

\item $b_{\rm t}$: Changing the tangential velocity bias alters the balance between coherent infall and tangential support. The biggest impact is in LRGs, followed by ELGs, where increasing $b_{\rm t}$ by $\sim10\%$ lowers the small-scale monopole (the extra tangential motion reduces the LOS pairwise compression) and quadrupole. Overall, $b_{\rm t}$ primarily redistributes small-scale anisotropy.

\item $b_{\rm cen}$: Increasing the central velocity bias adds random motions to centrals, which broadens the redshift–space distribution of the entire sample. This produces a damping of both monopole and quadrupole on small and intermediate scales. The effect is especially pronounced for LRGs and ELG$\times$LRG, where central--central pairs dominate, and milder for ELGs.
\end{itemize}

Overall, $b_{\rm r}$ and $b_{\rm t}$ regulate the anisotropic versus isotropic components of satellite motions, $f_{\sigma}$ damps coherent flows with added random scatter, and $b_{\rm cen}$ provides the strongest global control, suppressing both anisotropy and amplitude across all clustering statistics. Importantly, none of these velocity–bias parameters affects the projected correlation function, because $w_{\rm p}$ is a real–space statistic, and therefore insensitive to redshift–space distortions.

\begin{figure*}
\centering
\includegraphics[width=0.9\linewidth]{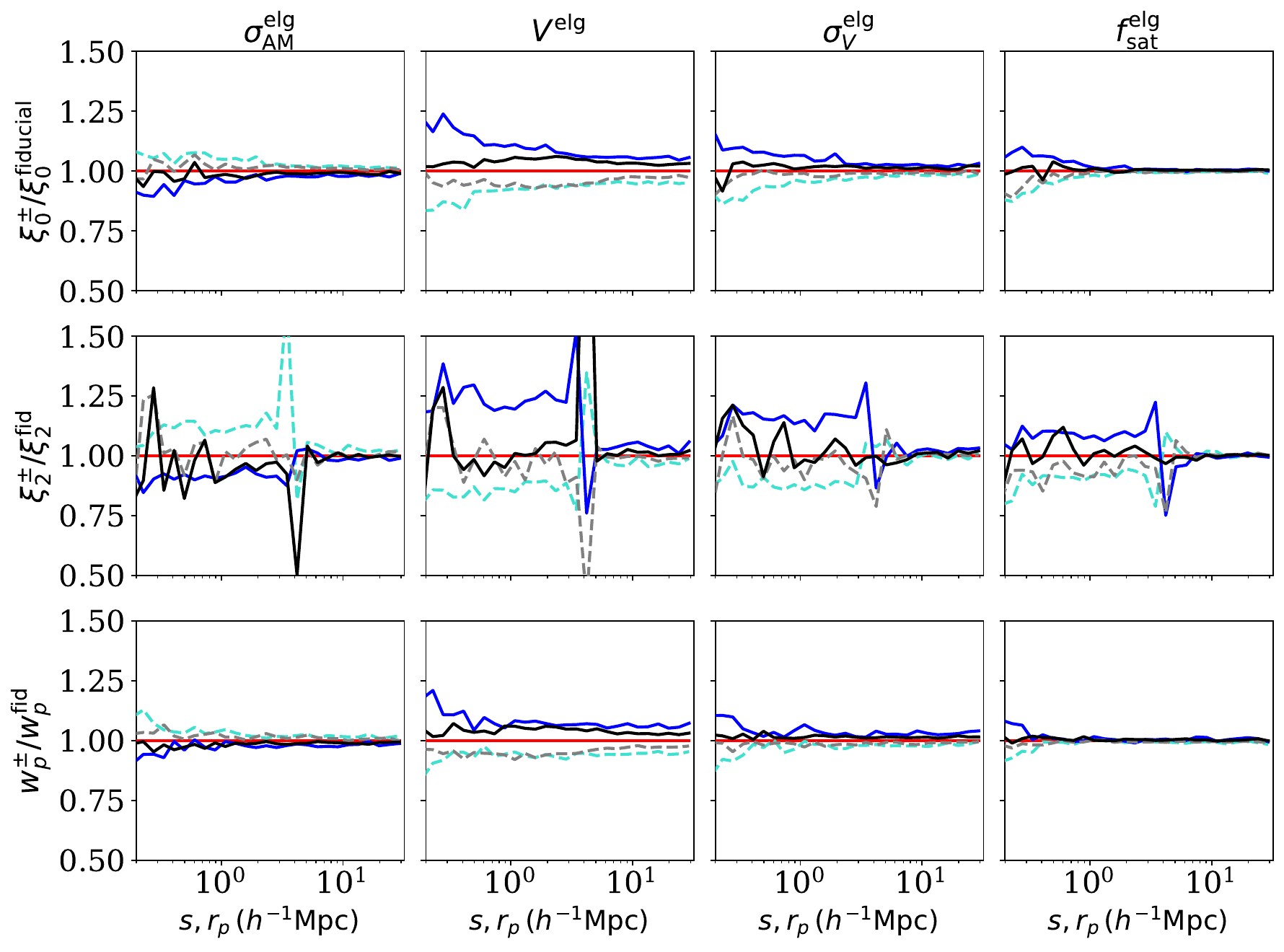}\vspace{0.1cm}
\includegraphics[width=0.9\linewidth]{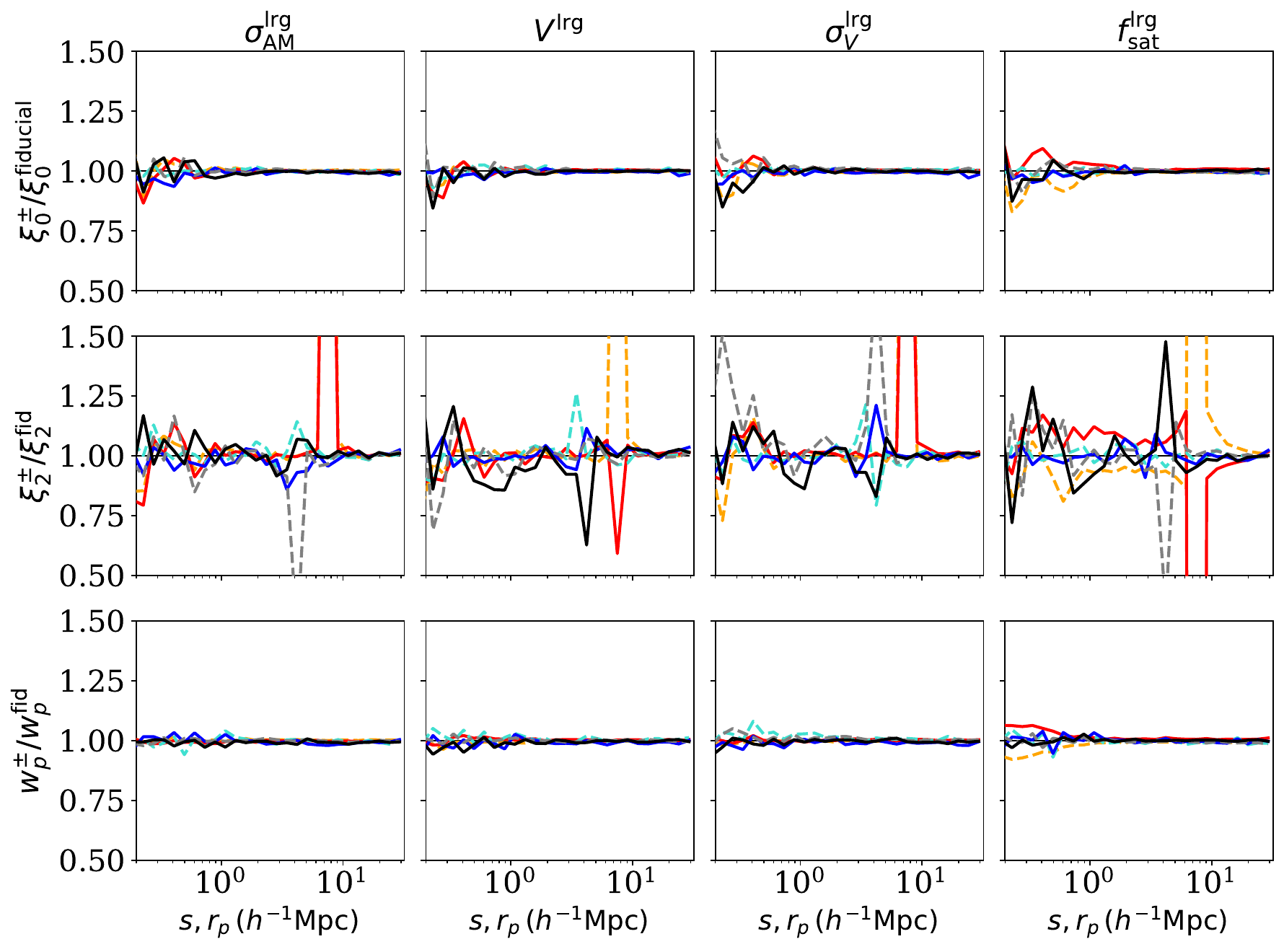}
\caption{Impact of a $\pm 10\%$ variation in the ELG (top) and LRG (bottom) abundance-matching parameters on the clustering. Lines are color-coded as in Figure\,\ref{fig:impactHOD}.}
\label{fig:impactAM}
\end{figure*}


We now examine how the abundance matching parameters affect the clustering, as shown in Figure\,\ref{fig:impactAM}:
\begin{itemize}
\item $\sigma_{\rm AM}$: Increasing the scatter between stellar mass and $V_{\rm peak}$ weakens the mapping galaxies and halos. This dilutes clustering, especially on small scales, as galaxies are redistributed into lower–bias halos. Both the monopole and $w_p(r_p)$ decrease in amplitude, while the quadrupole becomes less negative, reflecting reduced coherent infall. A tighter relation (smaller $\sigma_{\rm AM}$) has the opposite effect, boosting bias and anisotropy. The trend is strongest for ELGs and, from those, it propagates in the cross–signal.

\item $V$: These parameters set the characteristic peak circular velocities, for ELGs and LRGs, both centrals and satellites. Raising them shifts galaxies into more massive halos, increasing large–scale bias and boosting the monopole and $w_p$ amplitude, while making the quadrupole more negative on intermediate scales due to stronger Kaiser squashing. Lowering these thresholds populates lower–mass halos, reducing bias and amplitude. The impact of these parameters is strongest for ELGs, followed by ELG$\times$LRG, and LRG. Changing $V^{\rm elg}$ has stronger impact than changing $V^{\rm elg}$.

\item $\sigma_{V}$: Broadening the distribution of velocities smooths the transition in galaxy assignment, reducing sharp variations in clustering and producing smoother monopole and quadrupole shapes. Narrower distributions sharpen the transition, enhancing scale–dependent features. The effect, again, is strongest for ELGs, and milder for LRGs. The impact of changing $\sigma_{V}^{\rm elg}$ is stronger than $\sigma_{V}^{\rm elg}$.

\item $f_{\rm sat}$: The satellite fraction is a key driver of the 1–halo term. Increasing $f_{\rm sat}$ strongly boosts the monopole and quadrupole on sub–Mpc scales and produces a steep upturn in $w_p$, while lowering it suppresses all small–scale signals. This effect is pronounced both in the ELG and LRG models, underscoring the sensitivity of small–scale clustering to satellite abundance.
\end{itemize}

Overall, all AM parameters are major drivers of the two-point statistics: $\sigma_{\rm AM}$ alters the overall bias and anisotropy; $V$ shifts populations coherently across halo mass, producing large-amplitude changes; $\sigma_V$ reshapes the transfer function between halo mass and tracer type, strongly affecting all clustering signals; $f_{\rm sat}$ controls the 1-halo term and dominates small-scale anisotropy.
The four parameters act together as a tightly constrained system governing the full halo mass distribution of ELGs and LRGs, and their joint environmental statistics.


\begin{figure*}
\centering
\includegraphics[width=0.85\linewidth]{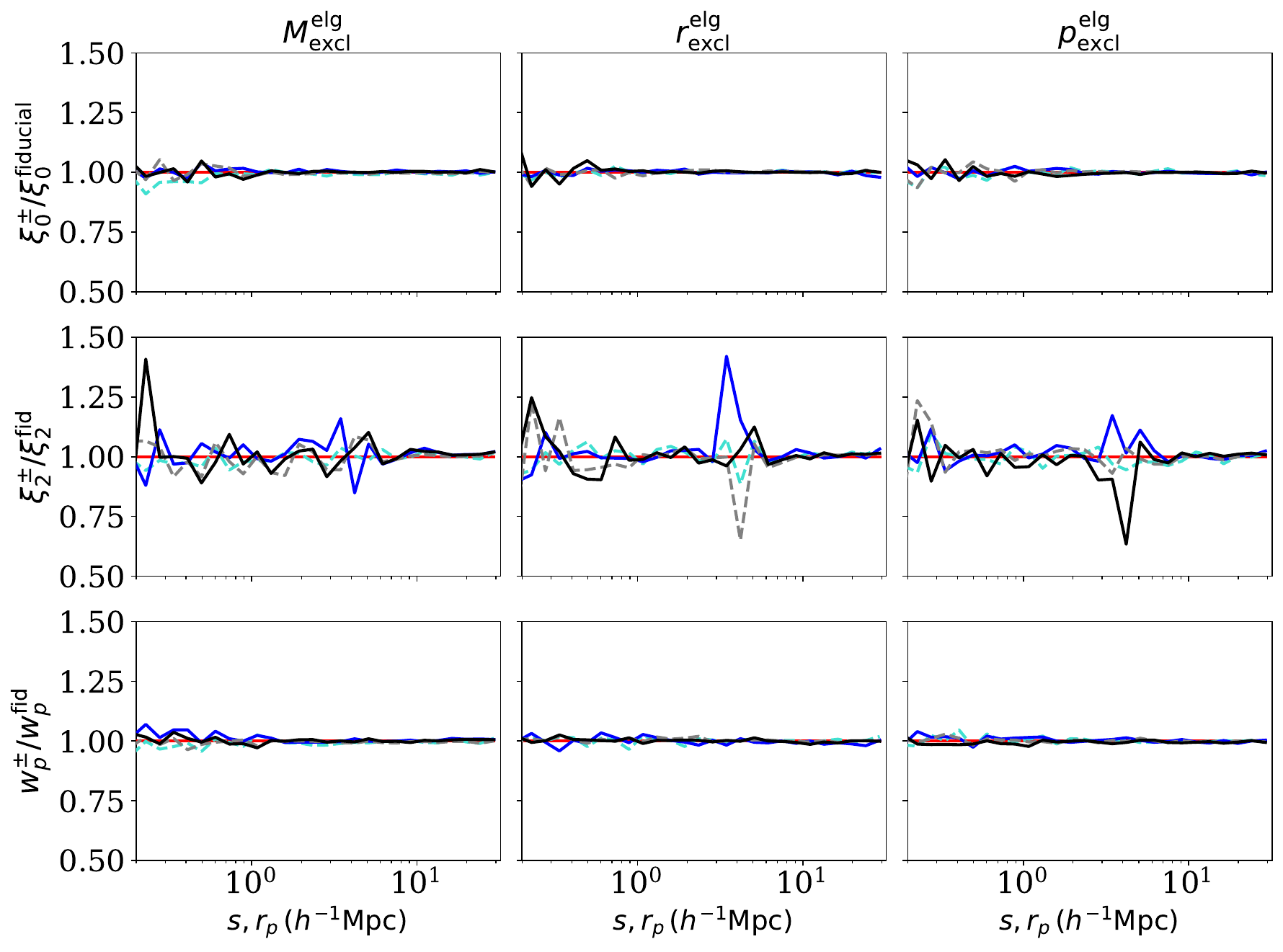}\vspace{0.1cm}
\includegraphics[width=0.85\linewidth]{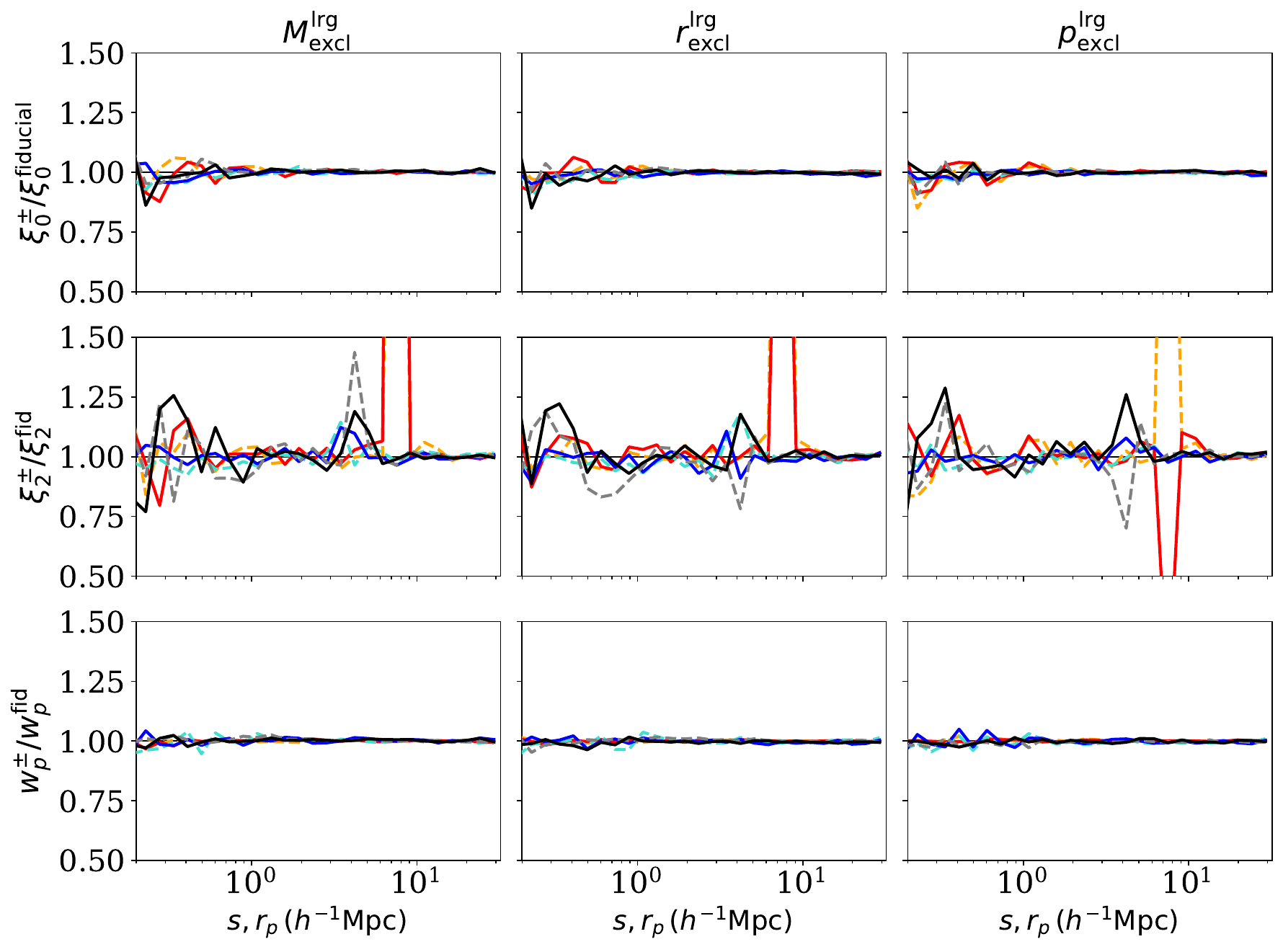}
\caption{Impact of the ELG (top) and LRG (bottom) halo exclusion parameters on the clustering. Lines are color-coded as in Figure\,\ref{fig:impactHOD}.}
\label{fig:impactEXCL}
\end{figure*}

Figure\,\ref{fig:impactEXCL} displays how halo exclusion, which regulates the minimum separation between pairs of massive halos, propagates into the clustering observables. We find that:
\begin{itemize}
\item $M_{\rm excl}$: Increasing the exclusion mass thresholds removes close halo pairs among massive hosts, and hence also their satellites. The effect is largest in the LRG and cross–correlation functions, which are dominated by more massive pairs.  The stronger the exclusion applied, the lower the 1-halo term in the monopole---since these pairs, with their satellites content, drive much of the small–scale clustering--- and the intermediate/larger scales in the multipoles and $w_p(r_p)$. The impact is milder for ELGs and stronger for LRGs and the cross-signal. Lowering $M_{\rm excl}$ has the opposite effect, steepening the rise of the monopole and $w_p$ in the cross–correlation and producing stronger quadrupole squashing.

\item $r_{\rm excl}$: Raising the exclusion radii shifts the boundary where overlap is forbidden. Larger radii strengthen the suppression at the 1–to-2–halo transition, producing a dip in the ELG$\times$LRG monopole and $w_p(r_p)$ around a few Mpc. For ELGs, the dip is present but relatively small, while for LRGs it is barely noticeable. The quadrupole also becomes less negative, as fewer close, anisotropic pairs survive. Conversely, smaller $r_{\rm excl}$ weakens exclusion, allowing halos to be packed more closely and thereby increasing clustering power on small scales.

\item $p_{\rm excl}$: The steepness of the exclusion transition is controlled by this exclusion probability parameter. Higher $p_{\rm excl}$ sharpens the cutoff, making the suppression of pairs more abrupt and visible as a sharper downturn in the monopole and $w_{p}(r_p)$. The quadrupole also reacts with a stronger transition at the exclusion scale. Lower $p_{\rm excl}$ softens the cutoff, producing a smoother and more gradual suppression, with correspondingly milder quadrupole changes. Again, the effect is strongest in the cross-correlation, milder in ELGs, and small in LRGs.
\end{itemize}

Halo exclusion primarily impacts the 1-halo to 2–halo transition regime rather than the asymptotic large scales. $M_{\rm excl}$ sets the mass scale of the halos that are impacted, $r_{\rm excl}$ fixes the scale of suppression, and $p_{\rm excl}$ controls its sharpness. Together, they tune the relative smoothness of the transition in all three clustering statistics. The effect of halo exclusion is strongest in LRGs and in the ELG$\times$LRG correlation functions, which are dominated by massive objects, and milder for ELGs.


\begin{figure*}
\centering
\includegraphics[width=0.9\linewidth]{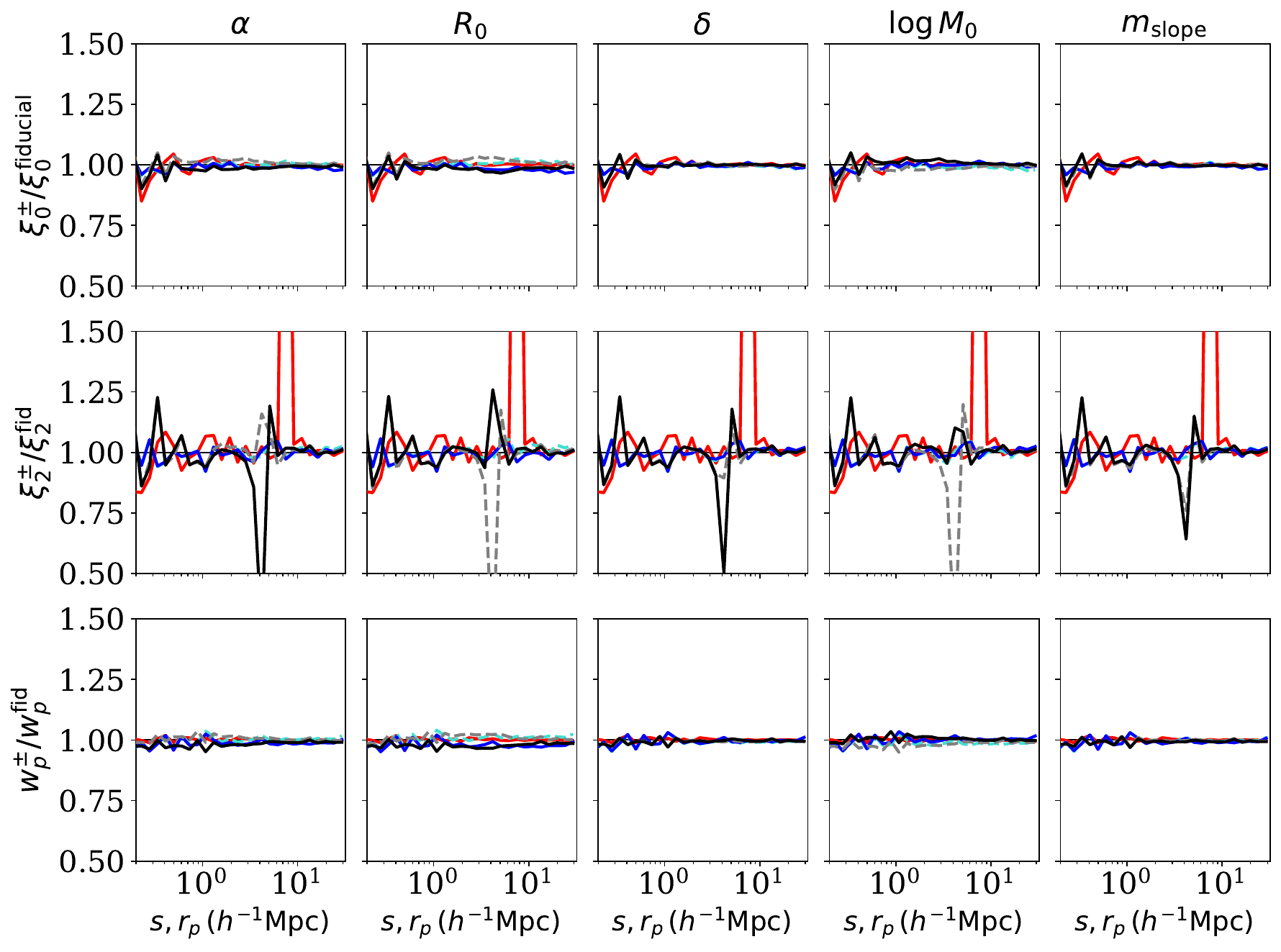}
\caption{Impact of a $\pm 10\%$ variation in the joint occupancy model parameters that emulate ELG satellites quenching in LRG hosts. Lines are color-coded as in Figure\,\ref{fig:impactHOD}.}
\label{fig:impactQ}
\end{figure*}

Finally, we study the impact on clustering of the parameters regulating the joint occupancy of ELGs and LRGs, which emulates the effect of satellite ELG quenching in massive LRG hosts from a statistical point of view. Figure\,\ref{fig:impactQ} shows that:
\begin{itemize}
\item $\alpha$: Increasing the quenching strength (larger $\alpha$) reduces the ELG satellite contribution in regions dominated by LRG hosts. This leads to a mild damping of the ELG monopole and quadrupole below $\sim 2\,h^{-1}$Mpc, a similarly small effect in the cross-correlation quadrupole, where ELG satellites are rare, but no effect on the LRG auto-correlation. The quadrupole is particularly affected, and becomes less negative across small and intermediate scales as the FoG signature weakens. Lowering $\alpha$ produces the opposite trend---satellite boost in massive halos,  steepening of the small–scale rise, and deeper negative quadrupole.---with equally modest amplitude, reflecting the fact that quenching is already saturated near the best fit. Besides reducing the 1-halo term in the clustering, quenching also flattens the satellite HOD.

\item $R_0$: Increasing it extends the spatial reach of LRG-driven suppression, allowing more LRG halos to influence nearby ELG candidates. This produces a weak reduction of small-scale power in the ELG monopole and $w_p(r_p)$, a slight weakening of the FoG signature in the ELG quadrupole, almost no effect on LRG clustering. Decreasing $R_0$ tightens the quenching to the immediate neighborhood of LRG hosts, boosting the ELG 1-halo term only marginally.

\item $\delta$: Larger $\delta$ values make the suppression more sharply localized around LRG halos.
However, because ELG satellites in LRG environments are already depleted, the clustering reacts moderatly: small adjustments in the ELG and cross-correlation multipoles below $3\,h^{-1}$Mpc, essentially no impact on $w_p(r_p)$ or on the LRG auto-correlation. Thus, $\delta$ fine-tunes only the shape of the quenching profile, not its global effect.

\item $\log{M_0}$: This parameter sets the minimal LRG halo mass capable of inducing suppression. At the best-fit point, most LRG halos already lie above this threshold, so $\pm10\%$ variations produce small adjustments to the ELG 1-halo term, and similar reactions in the cross multipoles. The clustering response remains negligible because raising or lowering $M_0$ by $\pm10\%$ hardly changes the set of halos considered massive enough to quench ELGs.

\item $m_{\rm slope}$: This exponent enhances the contribution of very massive LRG halos to the suppression field.
Increasing $m_{\rm slope}$ boosts quenched volumes around the highest-mass LRGs, which yields a faint damping of ELG small-scale anisotropies, the effect on LRG and cross statistics stays at the percent level. Since these massive LRG halos already dominate the suppression at the fiducial parameters, the model sits in a saturated regime where small changes in $m_{\rm slope}$ have limited impact.

\end{itemize}

Although the ELG$\times$LRG signal is dominated by central–central pairs (\S\,\ref{sec:clustering}), it remains sensitive to satellite quenching in massive hosts. This highlights that even a modest satellite population leaves a measurable imprint on redshift–space anisotropies, with quenching governing how this contribution evolves with halo mass.

In summary, the impact of the \ho physical parameters on the clustering signal is strongly tracer-dependent. Overall, we observe that the dominant effects derive from a change in the AM and HOD parameters, followed by the velocity bias, halo exclusion and quenching ones. In particular, quenching is a pivotal mechanism to precisely model the ELG$\times$LRG joint occupancy of halos, while halo exclusion is fundamental to shape the 1-to-2-halo transition in the cross-correlation functions. 


\section{Observed stellar mass functions}
\label{sec:MF}
In order to implement abundance matching (see \S\,\ref{sec:AM}), we adopt the \textsc{Cosmos2020} stellar mass function \citep{2023A&A...677A.184W}, which is parametrized by coadding two Schechter functions with individual normalizations ($\Phi_1^\star$, $\Phi_2^\star$) and slopes ($\alpha_1$, $\alpha_2$), and a single characteristic stellar mass ($M^\star$):
\begin{equation}
\Phi\,d\log{M}=\ln{(10)}\exp{(-10^{\log{M}-\log{M^\star}})}\times \left[ \Phi_1  \,\left(10^{\log{M}-\log{M^\star}}\right)^{\alpha_1+1}+ \Phi_2 \,\left(10^{\log{M}-\log{M^\star}}\right)^{\alpha_2+1}\right]\,d\log{M}\,.
\label{eq:schechter}
\end{equation}
The best-fit parameters to the total galaxy stellar mass function in our redshift range of interest ($0.8<z<1.1$) are: $\log{(M^\star/{\rm M}_\odot)}=10.87\pm0.07$, $\Phi_1=0.70\pm0.35\,{\rm Mpc^3\,dex^{-1}}$, $\Phi_2=0.77\pm0.34\,{\rm Mpc^3\,dex^{-1}}$, $\alpha_1=-1.35\pm0.07$, and $\alpha_2=-0.72\pm0.45$.

\begin{figure}
\centering
\includegraphics[width=0.5\linewidth]{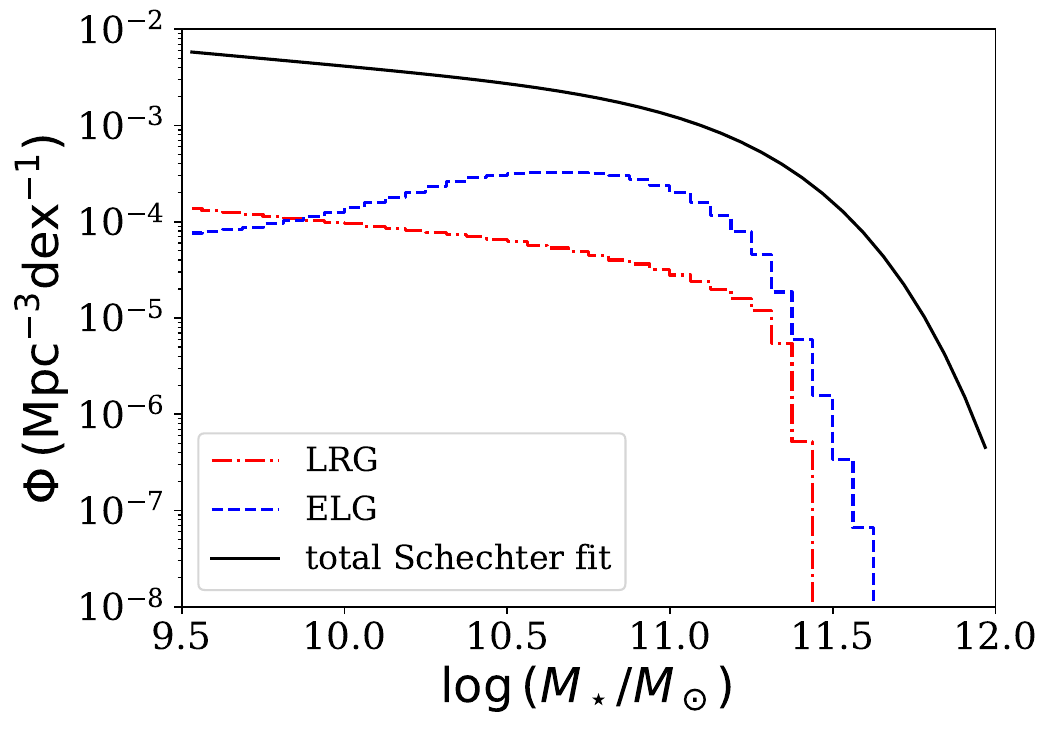}
\caption{Stellar mass functions of the ELG (blue dashed) and LRG (red dot-dashed) mock samples obtained after applying the abundance-matching down-sampling procedure (see \S\,\ref{sec:AM}). The number densities are measured from the mock catalogs in logarithmic bins of $M_\star$ and are normalized by the bin width and simulation volume. The black solid curve shows the analytic Schechter fit (Eq.\,\ref{eq:schechter}) to the observed stellar mass function of the full galaxy population from \citet{2023A&A...677A.184W}. As expected, the ELG and LRG samples lie below the total population because they represent incomplete tracer populations selected from the full galaxy catalog.}.
\label{fig:massfunc}
\end{figure}

 Figure \ref{fig:massfunc} compares the analytic fit to the observed stellar mass function of the full galaxy population with the ELG and LRG stellar mass functions obtained after applying the abundance-matching down-sampling procedure (see \S\,\ref{sec:AM}). The stellar mass functions of the tracer samples are computed by measuring the number density of galaxies in logarithmic bins of $M_\star$ and normalizing by the simulation volume and bin width.

As expected, both ELG and LRG samples lie below the full stellar mass function, reflecting their nature as incomplete subsets selected according to different criteria. The ELG population preferentially occupies intermediate stellar masses, $\log{(M_\star/M_\odot)\sim10-10.5}$, while LRGs dominate the high-mass end.

These mass functions are derived from our mock construction and are not intended to reproduce the observed DESI results \citep{2025arXiv250701593F}. Starting from the \textsc{Cosmos2020} stellar mass function, modeled with the Schechter fit of \citet{2023A&A...677A.184W}, we construct ELG and LRG samples via abundance-matching down-sampling (\S\,\ref{sec:AM}). The resulting distributions therefore reflect the mock selection, rather than the detailed DESI targeting and observational incompleteness.
A direct comparison with DESI measurements would require forward-modelling the survey selection function in the mocks, which is beyond the scope of this work. We emphasize that the stellar mass function is used only as an intermediate ingredient for abundance matching and does not constitute a constraint in the final analysis.

\bibliographystyle{aasjournal}   
\bibliography{main}

\end{document}